\documentclass[11pt]{article}
\usepackage{amsmath, amssymb, amsfonts, amsthm}

\newtheorem{theorem}{Theorem}[section]
\newtheorem{proposition}[theorem]{Proposition}
\newtheorem{corollary}[theorem]{Corollary}
\newtheorem{lemma}[theorem]{Lemma}

\newtheorem{remark}[theorem]{Remark}

\newcommand{\rd}{{\rm d}}
\newcommand{\be}{\begin{equation}}
\newcommand{\ee}{\end{equation}}
\newcommand{\bey}{\begin{eqnarray}}
\newcommand{\eey}{\end{eqnarray}}

\newcommand{\sfrac}[2]{{\textstyle \frac{#1}{#2}}}
\newcommand{\eps}{\varepsilon}

\newcommand{\bx}{{\bf x}}

\newcommand{\bA}{{\bf A}}

\renewcommand{\a}{\alpha}
\renewcommand{\b}{\beta}

\newcommand{\e}{\varepsilon}

\newcommand{\om}{{\omega}}

\newcommand{\bR}{{\mathbb R}}

\newcommand{\bN}{{\mathbb N}}

\newcommand{\tr}{\mbox{Tr}}

\newcommand{\wt}{\widetilde}
\newcommand{\wh}{\widehat}

\newcommand{\bnabla}{\mbox{\boldmath $\nabla$}}

\newcommand{\cF}{{\cal F}}
\newcommand{\cA}{{\cal A}}

\newcommand{\cE}{{\cal E}}

\newcommand{\cD}{{\cal D}}

\newcommand{\cW}{{\cal W}}
\newcommand{\cK}{{\cal K}}
\newcommand{\cH}{{\cal H}}
\newcommand{\cL}{{\cal L}}
\newcommand{\cN}{{\cal N}}

\newcommand{\B}{\left(\int \, W^2 |\nabla_m \phi|^2 + N \int  \, W^2
|\phi|^2 \right)}

\newcommand{\thi}{ \; | \!\! | \!\! | \;}
\newcommand{\supp}{\operatorname{supp}}

\input epsf

\newcommand{\fh}{{\frak h}}
\newcommand{\tfh}{\wt{\frak h}}

\newcommand{\donothing}[1]{}

\begin{document}

\title{Derivation of the Gross-Pitaevskii Hierarchy for the \\
Dynamics of Bose-Einstein Condensate}
\author{L\'aszl\'o Erd\H os${}^2$\thanks{Partially
supported by NSF grant DMS-0200235 and EU-IHP Network ``Analysis
and Quantum'' HPRN-CT-2002-0027. On leave from School of
Mathematics, GeorgiaTech, USA}\;, Benjamin Schlein${}^1$ and
Horng-Tzer Yau${}^1$\thanks{Partially supported by NSF grant
DMS-0307295 and MacArthur Fellowship.} \\
\\
Department of Mathematics, Stanford University\\ Stanford, CA 94305, USA${}^1$ \\ \\
Institute of Mathematics, University of Munich, \\
Theresienstr. 39, D-80333 Munich, Germany${}^2$
\\}

\maketitle

\begin{abstract}
Consider a system of $N$ bosons on the three dimensional unit
torus interacting via a pair potential $N^2V(N(x_i-x_j))$, where
$\bx=(x_1, \ldots, x_N)$ denotes the positions of the particles.
Suppose that the initial data $\psi_{N,0}$ satisfies the condition
\[  \langle
\psi_{N,0},  H_N^2 \psi_{N,0} \rangle \leq C N^2
\]
where $H_N$ is the Hamiltonian of the Bose system. This condition
is satisfied if $\psi_{N,0}= W_N \phi_{N,0}$ where $W_N$ is an
approximate ground state to $H_N$ and $\phi_{N,0}$ is regular. Let
$\psi_{N,t}$ denote the solution to the Schr\"odinger equation
with Hamiltonian $H_N$. Gross and Pitaevskii  proposed to model
the dynamics of such system by a nonlinear Schr\"odinger equation,
the Gross-Pitaevskii (GP) equation. The GP  hierarchy is an
infinite BBGKY hierarchy of equations so that if $u_t$ solves  the
GP equation, then the family of $k$-particle density matrices  $\{
\otimes_k u_t, k\ge 1 \}$ solves the GP hierarchy. We prove that
as $N\to \infty$ the limit points of the $k$-particle density
matrices of $\psi_{N,t}$ are solutions of the GP hierarchy. The
uniqueness of the solutions to this hierarchy remains an open
question. Our analysis requires that the $N$ boson dynamics is
described by a modified Hamiltonian which cuts off the pair
interactions whenever at least three particles come into a region
with diameter much smaller than the typical inter-particle
distance. Our proof can be extended to a modified Hamiltonian
which only forbids at least $n$ particles from coming close
together, for any fixed $n$.

\end{abstract}

\section{Introduction}\label{sec:intro}

A very simple and useful way to understand  Bose systems is to treat
all bosons as independent particles. In particular, to prove the
Bose-Einstein condensation is a simple exercise in the case of
non-interacting bosons \cite{Hu}. The true many-body problem with a
pair interaction is a much harder problem. Gross \cite{G1,G2} and
Pitaevskii \cite{P} proposed to model the many-body effect by a
nonlinear on-site self interaction of a complex order parameter (the
``condensate wave function''). The  strength of the nonlinear
interaction in this model is given by the scattering length $a_0$ of
the pair potential. The Gross-Pitaevskii (GP) equation is given by
\begin{equation}\label{nls}
i \partial_t u_t     = -\Delta u_t     +  \sigma |u_t|^2 u_t =
\frac{\delta \cE(u, \bar{u})}{\delta \bar{u}} \Big|_{u_t}, \quad
\cE(u, \bar u)= \int \rd x \Big [ \, |\nabla u|^2 + \frac {\sigma} 2
|u|^4 \Big ]\; ,
\end{equation}
where $\cE$ is the Gross-Pitaevskii energy functional and $\sigma= 8
\pi a_0$. The Gross-Pitaevskii equation was considered as a
mean-field model;   some corrections were obtained in \cite{LHY,
LeeYang} and more recently, e.g., in  \cite{TTH}.

The many-body effects can be analyzed in much more details  by the
Bogoliubov transformation. The Bogoliubov's theory, it should be
emphasized, postulates that the ratio between the non-condensate and
the condensate is small. Apart from this assumption, the coupling
constant $\sigma$ derived from the Bogoliubov's theory is the
semiclassical approximation to the scattering length. One can
perform some higher order diagrammatic re-summation to recover the
scattering length. In the case of hard core potential, the
traditional theory relies on the non-rigorous pseudo-potential model
\cite{HY}.

The first rigorous result concerning the many-body effects of the
Bose gas was Dyson's estimate of the ground state energy. Dyson
\cite{Dy} proved the correct leading upper bound to the energy and a
lower bound off by a factor around $10$.  The matching lower bound
was obtained  by Lieb-Yngvason \cite{LY1,LY2} forty years later. The
last result has inspired many subsequent works, including a proof
\cite{LSY1,LY1,LY2} that the Gross-Pitaevskii energy functional
correctly describes the ground state in a scaling limit to be
specified later.

The experiments on the Bose-Einstein condensation were conducted by
observing the dynamics of the condensate when the confining traps
are switched off. It is remarkable that the Gross-Pitaevskii
equation, despite being a mean-field equation, has provided a very
good description  for the dynamics of the condensate. The validity
of the Gross-Pitaevskii equation asserts that the fundamental
assumption of the Gross-Pitaevskii theory (i.e., that the many-body
effects can be modelled by a nonlinear on-site self interaction of
the order parameter) applies not only to the ground states, but to
certain excited states and their subsequent time evolution. This
remarkable and fundamental property of the Bose gas has mostly been
taken for granted and has not been treated rigorously in the
literature. In order to explain the issues involved, we now
introduce some notations and set up the scaling for the
Gross-Pitaevskii theory.

Let $V\ge 0$ be a fixed  nonnegative, spherically symmetric, smooth
potential with compact support. The zero energy scattering solution
$f$ satisfies the equation
\begin{equation}\label{ze}
\Big [ \, -\Delta + \frac 12  \,  {V (x)}  \, \Big ] f (x) = 0 \, .
\end{equation}
If we fix  the normalization $\lim_{|x| \to \infty} f (x) =1$ and
write $f (r) = q(r)/r$ (where $r = |x|$), the scattering length
$a_0$ of $V$ is defined by \be a_0 = \lim_{r \to \infty} r - q(r)
\ee and thus for $x$ large, \be\label{fb} f(x) \sim  1 - a_0/|x| \,
. \ee {F}rom the zero energy equation,  we also have the identity
\be\label{13} \int \rd x V(x) f(x)= 8 \pi a_0 \, .\ee

Let $\Lambda$ be the three-dimensional torus of unit side-length.
Denote the position of $N$ bosons in $\Lambda$ by $\bx=(x_1, x_2,
\ldots , x_N)$, $x_j\in \Lambda$. Lieb, Seiringer and Yngvason
\cite{LSY1} pointed out that, in order to obtain the GP functional,
the  length scale of the pair potential should be of order  $1/N$.
Notice that the density of the system is $N$ and the typical
inter-particle distance is $N^{-1/3}$, which is much bigger than the
length scale of the potential. The system is really a dilute gas on
the scale of the range of the interaction, but it is scaled in such
a way that the size of the total system is of order one.

The Hamiltonian of the Bose system on the torus $\Lambda$ is given
by
\begin{equation}\label{eq:ham1}
H = - \sum_{j=1}^N \Delta_j + \sum_{j < k}^N N^2 V (N(x_j - x_k))\;
.
\end{equation}
By scaling, the scattering length of the potential $N^2 V (Nx)$ is
$a:= a_0/N$. We shall from now on use the notation
\begin{equation}
V_a (x): = N^2 \, V \left( N x\right)\; , \qquad a= a_0/N \, .
\label{eq:Vrescale}
\end{equation}
Notice that the pair potential in \eqref{eq:ham1} is an
approximation to the mean field Dirac delta interaction:
$$
\frac {b_0} N \sum_{j < k}^N \delta(x_j-x_k), \qquad b_0= \int \rd x
V(x)\, .
$$
The Schr\"odinger equation is given by
\begin{equation}\label{eq:SC}
i\partial_t \psi_t =  H \psi_t, \quad \mbox{or} \quad i\partial_t
\gamma_N = [H , \gamma_N ] \; ,
\end{equation}
where $\gamma_N$ is the $N$-particle density matrix. For a pure
state, $\gamma_{\psi}: = |\psi \rangle \langle \psi |$  is the
orthogonal projection onto $\psi$.

Introduce the shorthand notation
$$
\bx:=(x_1, x_2, \ldots, x_N), \quad \bx_k:=(x_1, \ldots , x_k),
\quad \bx_{N-k}:=(x_{k+1}, \ldots , x_N)
$$
and similarly for the primed variables, $\bx'_k:=(x_1', \ldots ,
x_k')$. The $k-$particle density matrix is given by
\begin{equation}
\gamma_N^{(k)} (\bx_k; \bx_k'): = \int \rd \bx_{N-k} \gamma_N \left(
\bx_k, \bx_{N-k}; \bx_k', \bx_{N-k} \right) \; . \label{eq:marginal}
\end{equation}
Our normalization implies that  \( \tr \; \gamma_N^{(k)} =1 \) for
all $k=1,\dots, N$.

The density matrix $\gamma_{N,t}^{(1)}$ satisfies  the following
equation
\begin{equation}\label{1.1}
\begin{split}
i\partial_t \gamma_{N,t}^{(1)} ({x}_1 ; {x}'_1) = \; &(-\Delta_{x_1}
+ \Delta_{x'_1}) \gamma_{N,t}^{(1)} ({x}_1; {x}'_1) \\& + (N -1)
\int \rd x_{2} (V_a (x_1 - x_{2}) - V_a (x'_1 - x_{2}))
\gamma_{N,t}^{(2)} ({x}_1 , x_{2};{x}'_1 , x_{2}),
\end{split}
\end{equation}
and similar equations hold for $\gamma^{(k)}_{N, t}$ for $k\ge 1$.
To close this equation, one needs to assume some relation between
$\gamma_{N,t}^{(2)}$ and $\gamma_{N,t}^{(1)}$. The simplest
assumption would be  the factorization property, i.e., \be
\gamma^{(2)}_{N, t} (x_1, x_2; x_1', x'_2) = \gamma^{(1)}_{N,t}
(x_1; x_1') \gamma^{(1)}_{N,t} (x_2; x'_2) \, . \ee This does not
hold for finite $N$, but it may hold for a limit point
$\gamma^{(k)}_{t}$ of $\gamma^{(k)}_{N, t}$ as $N\to \infty$, i.e.,
\be\label{g2} \gamma^{(2)}_{t} (x_1, x_2; x_1', x'_2) =
\gamma^{(1)}_{t} (x_1; x_1') \gamma^{(1)}_{t} (x_2; x'_2) \, . \ee
Under this assumption, $\gamma_{t}^{(1)}$ satisfies the limiting
equation
\begin{equation}\label{1.2}
i\partial_t \gamma_{t}^{(1)} ({x}_1; {x}'_1) = \; (-\Delta_{x_1} +
\Delta_{x'_1}) \gamma_{t}^{(1)} ({x}_1; {x}'_1) +  (Q_t(x_1)
-Q_t(x'_1)) \gamma_{t}^{(1)} ({x}_1; {x}'_1)
\end{equation}
where \be\label{b0} Q_t(x) = \lim_{N \to \infty} N \int \rd y
V_a(x-y) \rho_t(y), \quad \rho_t(x) = \gamma_{t}^{(1)} ({x}; {x})\,
. \ee

If, instead of $V_a(x)$, we use the unscaled mean field potential
$V(x)/N$, then $Q_t$ is the convolution of $V$ with the density
$\rho_t(x)$. The equation \eqref{1.2} becomes the Hartree equation.
For Bose systems with mean field interaction and product initial
wave function, the factorization assumption \eqref{g2} can be proved
for a general class of potentials. See the work of Hepp \cite{H} and
Spohn \cite{Sp} for bounded potentials and \cite{BGM,EY} for
potentials with Coulomb type singularity. For certain quasi-free
initial data, Ginibre and Velo \cite{GV} can handle all integrable
singularities. We note that in one dimension the convergence to the
GP hierarchy \eqref{eq:GPH2} for the delta potential was established
by Adami, Bardos, Golse and Teta in \cite{ABGT}.

In our setting, $V_a(x)$ lives on scale $1/N$ and is much more
singular than all the cases considered previously.  If $\rho_t(x)$
is continuous, then $Q_t$ is given by
$$
Q_t(x) = b_0 \rho_t(x) .
$$
Thus (\ref{1.2})
 gives the GP equation with the {\it incorrect} coupling constant
$\sigma = b_0$ instead of $\sigma = 8\pi a_0$.
 It is known that $b_0/8 \pi $ is the first Born
approximation to the scattering length $a_0$ and the following
inequality holds: \be\label{slbound} a_0 \le \frac {b_0}{8 \pi}=
\frac 1 {4 \pi} \int_{\bR^3} \,  \frac  1 2 \, V (x) \, dx \, .\ee
To go beyond the Born approximation, we need to understand the short
scale correlations of the ground state which we now review.

The ground state of a dilute Bose system with interaction potential
$V_a$ is believed  to be very close to  the form \be
 W (\bx) :=  \prod_{i<j} f(N(x_i-x_j))
\ee where $f$ is the zero-energy solution  \eqref{ze}. We remark
that Dyson \cite{Dy} took a different function which was not
symmetric, but the short distance behavior was the same as in $W$.
Since in the experiments the initial states were prepared with a
trapping potential, living on a scale of  order one, the ground
state for such a trapped gas is of the form $\psi (\bx) = W(\bx)
\phi (\bx)$ \cite{LSY1, LSY2} where $\phi(\bx)$ is close to a
product function. Thus we shall consider initial data of the form
$W(\bx) \phi (\bx)$.

We assume for the moment that the ansatz,  $\psi_t (\bx) = W(\bx)
\phi_t (\bx)$ with $\phi_t$ a product function, holds  for all time.
The reduced density matrices for $\psi_t (\bx) $ satisfy
\be\label{g3} \gamma^{(2)}_t (x_1, x_2; x_1', x_2')\sim
f(N(x_1-x_2))f(N(x'_1-x_2')) \gamma_{t}^{(1)} ({x}_1;
{x}'_1)\gamma_t^{(1)}(x_2; x_2')\,. \ee Together with \eqref{13} and
the assumption that $\rho_t$ is smooth, we have \be\label{g4}
\lim_{N \to \infty}  N \int \rd x_{2} V_a (x_1 - x_{2})
\gamma_{N,t}^{(2)} ({x}_1 , x_{2} ; {x}'_1 , x_{2}) = 8 \pi a_0
\gamma_{t}^{(1)} ({x}_1; {x}'_1) \rho_t(x_1) \, . \ee We have used
that $\lim_{|x| \to \infty} f (x) =1$ and the last equation is valid
for $|x_1-x'_1|\gg 1/N$. This gives the GP equation with the correct
dependence on the scattering length.

Notice that the relations \eqref{g3} and \eqref{g4} are very subtle.
The correlation in $\gamma^{(2)}$ occurs at the scale $1/N$. Testing
the relation \eqref{g3} in a weak limit, all correlations at the
scale $1/N$ disappear and the product relation \eqref{g2} will hold.
However, this short distance correlation shows up in the GP equation
due to the singular potential $N V_a (x_1 - x_{2})$.  Therefore, a
rigorous justification of the GP equation requires a proof that the
relation \eqref{g3} holds with such a precision that \eqref{g4} is
valid---a formidable task.

We would like to divide this task into two parts. The first question
is whether the short scale structure of $\gamma^{(2)}_{N,t}$ is
given by $f(N(x_1-x_2))f(N(x'_1-x_2'))$, i.e., whether, for large
$N$, \be\label{g5} \gamma^{(2)}_{N,t} (x_1, x_2; x_1', x_2')\sim
f(N(x_1-x_2))f(N(x'_1-x_2')) \gamma^{(2)}_t (x_1, x_2; x_1', x_2')
\ee where $\gamma_{t}^{(2)}$ is a weak limit of $\gamma_{N,t}^{(2)}$
(the short distance correlations given by $f(N(x_1 - x_2)) f(N(x_1'
-x_2'))$ vanish when the weak limit is taken) living on a scale of
order one so that \be\label{g6} \lim_{N \to \infty} \int \rd x_{2}
V_a (x_1 - x_{2}) \gamma_{N,t}^{(2)} ({x}_1 , x_{2};{x}'_1 , x_{2})
= 8 \pi a_0 \gamma_{t}^{(2)} ({x}_1 , x_{1};{x}'_1 , x_{1})\,. \ee
The second question is whether $\gamma^{(2)}_t$ factorizes, i.e.,
whether \be\label{Ufact} \gamma^{(2)}_{t} (x_1, x_2; x_1', x'_2) =
\gamma^{(1)}_{t} (x_1; x_1') \gamma^{(1)}_{t} (x_2; x'_2). \ee

If we replace the Hamiltonian \eqref{eq:ham1} by a modified
Hamiltonian $\wt H$, in which we remove the pair interaction when
three or more particles come close together in a very short
distance, then we can prove a certain version of \eqref{g5} and
\eqref{g6}. Let $\ell$ denote a distance much  smaller than the
typical inter-particle distance, say, $\ell=N^{-1/3-\delta}$ for
some $\delta> 0$. The modified Hamiltonian is approximately of the
form
\begin{equation}\label{eq:modham1}
\wt H \sim  - \sum_{j=1}^N \Delta_j + \sum_{j < k}^N \Big [ N^2 V
(N(x_j - x_k)) \prod_{i \not = j, k} 1(|x_i-x_k|\ge \ell) \Big ]\,.
\end{equation}
The precise definition  will be given in (\ref{eq:modham}). This
cutoff modification changes the Hamiltonian for events that are
rare with respect to the expected typical particle distribution,
therefore it should have little effect on the dynamics.
Unfortunately we cannot control this effect rigorously. (In
principle, the original unmodified dynamics may introduce local
clustering of particles, despite that it is unfavorable for the
local energy.) In the computation of the ground state energy by
Lieb-Yngvason \cite{LY1}, no such modification was needed since the
positivity of the contribution from these rare events could be
exploited. Our method is based on the conservation of $\wt H^2$
along the dynamics and an  inequality of the form:
\be\label{eq:keyesti0} \sum_{i<j} \int \, W^2 |\nabla_i \nabla_j
\phi_t|^2 \le C \int \, | (\wt H+N)  \psi_t|^2 = C \int |(\wt H +
N)\psi_0|^2  \leq C N^2,\ee with $\psi_t= W \phi_t$. The idea of
using the conservation of higher power of the Hamiltonian was
introduced in \cite{EY}. Clearly, the computation of $\wt H^2$
involves derivatives of the pair potential which have no definite
sign.  If we use the original Hamiltonian $H$ instead of $\wt H$,
these terms cannot be controlled by the kinetic energy operator in
the rare situation when many particles come close together. The
modified Hamiltonian \eqref{eq:modham1} removes this technical
obstacle.

Using (\ref{eq:keyesti0}) we will prove that weak limit points
$\gamma_{t}^{(k)}$ of the $k$-particle density $\gamma_{N,t}^{(k)}$
(whose time evolution is generated by the modified Hamiltonian $\wt
H$) satisfy the following infinite BBGKY hierarchy, which will be
called GP-hierarchy:
\begin{equation}\label{eq:GPH2}
\begin{split}
i\partial_t \gamma_{ t}^{(k)} ({\bf x}_k ;{\bf x}'_k) =\; &
\sum_{j=1}^k (-\Delta_{x_j} + \Delta_{x'_j}) \gamma_{t}^{(k)} ({\bf
x}_k ;{\bf x}'_k) \\ & + \sigma \sum_{j=1}^k \int \rd x_{k+1}
(\delta (x_j - x_{k+1}) - \delta (x'_j - x_{k+1})) \gamma_{
t}^{(k+1)} ({\bf x}_k,x_{k+1};{\bf x}'_k,x_{k+1}) \, .
\end{split}
\end{equation}
with the correct coupling constant $\sigma=8 \pi a_0$. Notice that
if $u(t, x)$ is a solution of the GP equation \eqref{nls}, then \be
\gamma_t^{(k)}(\bx_k; \bx'_k) = \prod_{j=1}^k u(t, x_j) \bar u(t,
x'_j) \label{prodgp} \ee is a solution of the hierarchy
\eqref{eq:GPH2}. To conclude  the factorization property
(\ref{Ufact}), it remains to answer the following three open
questions. The first one is to rigorously justify the cutoff
modification of the Hamiltonian by controlling the clustering of
particles. The second one is the uniqueness of the solution of the
hierarchy in a certain space. The third one is to prove a-priori
bounds on the density matrices of the Schr\"odinger equation so that
their limits fall into the space needed in the uniqueness theorem.
Recently the uniqueness problem was solved in  \cite{ESY}.
Furthermore, the required a-priori estimates  were established for
certain mean field Hamiltonians without the cutoff modification (see
Section \ref{sec:MF} for more details). However, the a-priori
estimates and the removal of the cutoff modification for the GP
scaling remain interesting open problems.

\section{Definition of the Model and the Main Result}
\setcounter{equation}{0}

We now define the modified Hamiltonian and state the main result.
Throughout the paper we use the notation that $a \ll b$, if $a/b =
O(N^{-\a})$ for some $\a >0$. The notation $W^{k,p}$ will stand for
the standard Sobolev spaces.

\subsection{The Two-Body Problem}\label{sec:2body}

We consider the problem of a single  particle in the field of the
scaled potential \[ V_a (x)/2 = (1/2) N^2 V (Nx).\] Here we assume
$V (x)$ to be a smooth, spherically symmetric, and compactly
supported potential. For $0<\kappa\ll 1$ let $e_\kappa$ and
$(1-w^\kappa)$ be the lowest Neumann eigenvalue and eigenfunction on
the ball $\{y : |y|\leq \kappa \}$, \be (-\Delta +  \sfrac 1 2 V_a)
(1- w^\kappa ) = e_\kappa (1- w^\kappa ) \ee with normalization
$w^\kappa(|x|=\kappa)=0$. We can extend $w^{\kappa}$ to be
identically zero for $|x| \geq \kappa$ so that $w^{\kappa}$
satisfies \be (-\Delta + \sfrac 1 2  V_a) (1- w^\kappa ) = e_\kappa
(1- w^\kappa ) \chi_\kappa(x), \ee where
$\chi_\kappa(x):=\chi(|x|\leq\kappa)$ is the characteristic function
of the ball of radius $\kappa$. We will prove in Lemma~\ref{lm:E&g}
that
$$
    e_\kappa = 3a\kappa^{-3} (1 + o(1)), \qquad
$$
as long as $a\ll \kappa \ll 1$.

Let $\ell_1$ be a scale to be fixed later on with $a\ll \ell_1\ll 1
$. Let $\mu(d\kappa): = g(\kappa) d\kappa$ be a probability measure
supported on $[\ell_1/2, 3\ell_1/2]$ with smooth density $g$. Define
$w$ by
\begin{equation}\label{eq:w}
    1-w:= \int (1- w^\kappa) \mu (d\kappa)\,.
\end{equation}
One can check that $w$ satisfies the equation \be
    (-\Delta + \sfrac 1 2  V_a)(1-w) = q (1-w)
\label{eq:neumann1} \ee where
\begin{equation}\label{eq:q}
     q(x) : = \frac{
     \int e_\kappa 1(|x|\leq \kappa)(1-w^\kappa(x))
     \mu(d\kappa)}{ \int(1-w^\kappa(x))\mu(d\kappa)} \; .
\end{equation}
Some properties of the functions $w(x)$ and $q(x)$, which will be
used in the rest of the paper, are collected in Appendix
\ref{app:2body} (see, in particular, Lemma \ref{lm:w&q}).

\subsection{Removal of triples}
\label{sec:rem} We introduce a new length-scale $\ell$ with $\ell_1
\ll \ell\ll 1$.
 The following procedure excludes
configurations with more than three particles within a distance
$\ell$ from each other.

 Let $h$ be the exponential
cutoff at scale $\ell$ defined by
$$
   h(x): = e^{-\sqrt{x^2+\ell^2}/\ell}\; , \qquad x\in \Lambda \; .
$$
For a configuration $\bx=(x_1, \ldots x_N)$ let $\cN_{ij}(\bx)$ be
the number of particles other than $i$ and $j$ that are within
distance $\ell$ to either $i$ or $j$:
$$ \cN_{ij}(\bx):=\sum_{k \not = i, j} [ h(x_k-x_i)+h(x_k-x_j) ]\; ,
\qquad i\neq j \; .
$$
Let $0< \e < \frac{1}{10}$ be fixed. Define
$$
F(u): = e^{- u/\ell^\e }, \quad u \ge 0
$$
and
$$
F_{ij}(\bx):= F(\cN_{ij}(\bx)) \; .
$$
Thus $F_{ij} (\bx)$ is exponentially small if $|x_k - x_i| \lesssim
\ell$ or $|x_k -x_j| \lesssim \ell$ for some $k \neq i,j$. The
functions $F_{kj}$, modulo an exponentially small error, forbid
particle triples within a distance $\ell$ and provide a strong
non-overlap property. This fact and some other important properties
of the functions $F_{ij}$ will be presented in Appendix
\ref{app:cutoff}.

Introduce the notations
$$
\chi_{jk}= \chi_{jk}(\bx):=1( |x_j-x_k|\leq \ell_1) \qquad \mbox{if}
\quad k\neq j \; ,
$$
$$
\wt\chi_{jk}= \wt\chi_{jk}(\bx):=1( |x_j-x_k|\leq \ell) \qquad
\mbox{if} \quad k\neq j\; ,
$$
and $\chi_{jk}\equiv \wt\chi_{jk}\equiv 0$ if $j=k$. Analogously, we
will freely use the shorthand notation
$$
    V_{ij} = V_a(x_i-x_j), \quad w_{ij} = w(x_i-x_j), \quad
    h_{ij} = h(x_i-x_j)\; , \quad q_{ij}= q(x_i-x_j)\; ,
$$
and similarly for their derivatives. In particular $w_{ii}=0$,
$(\nabla w)_{ii}=0$.

We also fix a smooth function $\theta \in C_0^{\infty} (\bR)$ with
$\theta (s) = 1$ for $s \leq 1$ and $\theta (s)=0$ for $s
>2$. For some $K>0$ we introduce the notation
\be
   \theta_{kj}: =  \theta \Big( \frac{|x_k -x_j|}{K\ell|\log\ell|} \Big) \;  .
\label{def:theta} \ee The constant $K$ will be chosen sufficiently
large but independent of $N$ at the end of the proof. The
$K$-dependence will be omitted from the notation.

\subsection{The Modified Hamiltonian}\label{sec:modham}

Consider the function $w$ defined in (\ref{eq:w}) for the  length
scale $\ell_1$.  We define \be G_i(\bx):= 1 - \sum_{j \neq i } w
(x_i - x_j) F_{i j}(\bx) \; , \quad \mbox{and}\quad
    W:= \prod_{i=1}^N \sqrt{G_i} \; .
\label{def:G} \ee The function $W$ is our approximation to the wave
function of the ground state.  Due to the cutoffs it differs
slightly from the form $\prod_{i<j} f(x_i-x_j)$ mentioned in Section
\ref{sec:intro}. It is also slightly different from Dyson's
construction \cite{Dy} of the ground state, which is non-symmetric.
By Lemma \ref{lm:w&q}, $w \le d_0 < 1$ is separated away from 1. We
will prove, in Section \ref{sec:approx} (see also Lemma
\ref{lm:pos}), that there exists a constant $c_1 >0$ such that $ c_1
\leq G_i \leq 1$.

Introduce the notation $M_{kj}$ by \be
     M_{kj}:= \frac{F_{kj}}{2}( G_j^{-1} + G_k^{-1}) \; .
\label{eq:mkj} \ee The modified Hamiltonian $\wt H$ is defined by
\begin{equation}\label{eq:modham}
\wt H : = H - \sum_{k\neq j}( \sfrac 1 2  V_{kj} -q_{kj}) [ 1-
(1-w_{kj})M_{kj}]  \; .
\end{equation}
Since $V$ and $q$ have compact support of size at most of order
$\ell_1$ (Lemma \ref{lm:w&q} in  Appendix \ref{app:2body}), the
factor $(1/2)V_{kj}-q_{kj}$ vanishes  for any $k\not = j$ unless
$|x_k-x_j| \leq O(\ell_1)$. Suppose now $|x_k-x_j|\le O(\ell_1)$. If
there is no third particle at $x_m$ with $m\neq k, j$,  such that
$|x_k-x_m|\lesssim \ell /|\log\ell|$ then $G_j\approx G_k\approx
1-w_{kj}$ and $F_{kj}\approx 1$ with exponential precision. Thus $ [
1-(1-w_{kj})M_{kj}]$ is exponentially small. This shows that the
difference between $H$ and $\wt H$ is exponential small unless three
or more particles are closer than $\ell$ to each other. Since we
will choose $\ell = N^{-2/5 -\kappa}$, for some $\kappa >0$, the
probability for this to happen is, for large $N$, very small.

The reason for this modification will be clear in Section
\ref{sec:energy}; without the subtraction of the three-body terms in
(\ref{eq:modham}) we are not able to prove the a priori estimate for
$H^2$ (part ii) of Proposition \ref{prop:energy2}.

Our methods can be easily generalized to remove only $n$-body
collisions instead of removing triples collisions, for any $n \geq
3$. More precisely, let $f \geq 0$ be a smooth function such that
$f(x) =1$ for $x \leq 0$, $f(x) \simeq e^{-x}$ for $x \gg 1$. We
define \[ F^{(n)} (u) := f \left( \frac{u-(n-3)}{\ell^{\eps}}
\right) \] and
\[ F^{(n)}_{ij} (\bx) := F^{(n)} ( \cN_{ij} (\bx)). \] Then, apart
from exponentially small errors, $F_{ij}^{(n)} (\bx) =1$, unless
there are at least $n-2$ other particles in the $\ell$-vicinity of
$x_i$ and $x_j$ (and $F_{ij}^{(n)} \simeq 0$ if this is the case).
The modified Hamiltonian \begin{equation}\label{eq:Hn} \wt H^{(n)} =
H - \sum_{k\neq j}( \sfrac 1 2 V_{kj} -q_{kj}) [ 1-
(1-w_{kj})M^{(n)}_{kj}] \end{equation} with $M^{(n)}_{kj}$ defined
through \eqref{eq:mkj}, with $F_{ij}$ replaced by $F^{(n)}_{ij}$.
Note that $\wt H^{(n)}$ now differs from $H$ by a term which is
exponentially small unless there are $n$ or more particles closer
that $\ell = N^{-2/5 -\kappa}$ to each other.

\bigskip

Since we are dealing with bosons, the Hamiltonian (\ref{eq:modham})
acts on the Hilbert space $\cH^{\otimes_s N}:=L^2 (\Lambda , \rd
x)^{\otimes_s N}$ of functions of $3N$ variables which are symmetric
with respect to any permutation of the $N$ particles, i.e., $\psi
\in \cH^{\otimes_s N}$ if and only if $\psi \in \cH^{\otimes N}$ and
$R_\pi\psi = \psi$ for all permutation $\pi \in S_N$, where
\begin{equation}\label{eq:Rpi}
(R_{\pi} \psi) (x_1 ,\dots, x_N ) := \psi (x_{\pi 1}, \dots , x_{\pi
N}) \; .
\end{equation}
The dynamics is given by the Schr\"odinger equation
\begin{equation}\label{eq:schr2}
i\partial_t \psi_t =  \wt H \psi_t, \quad \mbox{or} \quad
i\partial_t \gamma_N = [\wt H , \gamma_N ] \; ,
\end{equation}
for a density matrix $\gamma_N$.

\subsection{Weak* Topology for the Kernel of Density Matrices}
\label{sec:top}

Let $\cH := L^2 (\Lambda, \rd x)$ be the one particle Hilbert space,
and let $\cH^{\otimes_s n}$ be the symmetric subspace of the
$n$-fold tensor product $\cH^{\otimes n}$. We denote by $\cL^1_n :=
\cL^1 (\cH^{\otimes n})$  the space of trace class operators  on the
Hilbert space $\cH^{\otimes n}$. We will work with families of trace
class operators  $\Gamma =\{ \gamma^{(n)}\}_{n\geq 1}$ with
$\gamma^{(n)} \in \cL^1_n$ for all $n \geq 1$.

In this section we consider the operator kernels
$\gamma^{(n)}(\bx_n; \bx'_n)$ as elements of $L^2 ( \Lambda^{n}
\times \Lambda^{n})$, with the norm \be \| \gamma^{(n)} \|_2 = \int
\rd \bx_n \rd \bx'_n \, |\gamma^{(n)} (\bx_n ; \bx'_n)|^2 \, .\ee

For $\Gamma= \{\gamma^{(n)} \}_{n \ge 1}  \in \bigoplus_{n \geq 1}
L^2 (\Lambda^{n} \times \Lambda^{n})$, and any fixed $\nu
> 1$ we define the norm
\be \| \Gamma \|_{H^{(\nu)}_-} := \sum_{n=1}^{\infty} \nu^{-n} \|
\gamma^{(n)} \|_{2} \ee and we define the Banach space
 $H^{(\nu)}_- := \{ \Gamma \in \bigoplus_{n \geq 1} L^2 (\Lambda^{n} \times \Lambda^{n}): \|
\Gamma \|_{H^{(\nu)}_-} < \infty \}$. We also define $H^{(\nu)}_+ :=
\{ \Gamma \in \bigoplus_{n \geq 1} L^2 (\Lambda^{n} \times
\Lambda^{n}) : \lim_{n \to \infty} \nu^n \| \gamma^{(n)} \|_2 = 0
\}$. We equip $H^{(\nu)}_+$ with the norm
\[
\| \Gamma \|_{H^{(\nu)}_+} = \sup_{n \geq 1} \nu^n \| \gamma^{(n)}
\|_{2} \, .
\]
Similarly to the standard proof of the duality $\ell^1 = c_0^*$
between scalar valued summable sequences and vanishing sequences,
one readily checks the duality $H_-^{(\nu)} = \big( H_+^{(\nu)}
\big)^*$.
 By  the
Banach-Alaoglu Theorem, the unit ball $B^{(\nu)}_-$  in
$H^{(\nu)}_-$ is compact with respect to  the weak$*$ topology.

We shall fix a time $T>0$ for the rest of this paper. Denote by $C
([0,T] , H_-^{(\nu)})$ the space of weak* continuous functions from
$t \in [0,T]$ to $H_-^{(\nu)}$. Since the space $H^{(\nu)}_+$ is
separable, we can fix a dense countable subset in the unit ball of
$H^{(\nu)}_+$, denoted by $\{J_i\}_{i \ge 1}$. Define the metric on
$H^{(\nu)}_-$ by
\begin{equation}\label{eq:rho}
\rho (\Gamma, \wt \Gamma) : = \sum_{i=1}^\infty 2^{-i} \Big |
\sum_{n=1}^\infty \int \rd \bx_n \rd \bx'_n \overline{J_i^{(n)}
(\bx_n ; \bx'_k)} \big [ \gamma^{(n)}(\bx_n ; \bx'_n)- \wt
\gamma^{(n)}(\bx_n; \bx'_n) \big ] \Big | \; .
\end{equation}
Note that the topology induced by $\rho (.,.)$ and the  weak*
topology are equivalent on the unit ball $B^{(\nu)}_-$. We  equip $C
([0,T] , H_-^{(\nu)})$  with the metric \be \wh \rho  (\Gamma, \wt
\Gamma) := \sup_{0\le t \le T}  \; \rho\,(\Gamma(t), \wt
\Gamma(t))\; . \label{def:varrho} \ee

\subsection{The Main Result}\label{sec:main}

Consider the rescaled potential $V_a (x) = (a_0 /a)^2 V ((a_0/a) x)$
(see \eqref{eq:Vrescale}) whose scattering length $a$ is chosen such
that $Na=a_0$. Let $\ell, \ell_1$ be two lengthscales satisfying
\be\label{scales} a \ll \ell_1 \ll \ell \ll 1, \quad a \ell_1 \ll
\ell^{4}, \quad a \ll \ell_1^{3/2}, \quad \ell_1 \ll \ell^{3/2},
\quad \ell_1^3 \ll a \ell^{9/4}, \quad \ell^5 \ll a^2 \ee (for
example $\ell_1 = N^{-2/3 +\kappa}$ and $\ell = N^{-2/5 - \kappa}$
for $\kappa>0$ small enough).

Let $\wt H$ be the modified Hamiltonian given in (\ref{eq:modham}).
Let $\psi_{N,t}$ be a normalized solution of the $N$-body
Schr\"odinger equation
\begin{equation}
i\partial_t \psi_{N,t} = \wt H\psi_{N,t} \; . \label{eq:schr3}
\end{equation}
and let  $\gamma^{(k)}_{N,t}(\bx_k ; \bx_k')$ be its $k$-particle
marginal densities defined in (\ref{eq:marginal}). These density
matrices satisfy the normalization \be
   \tr \; \gamma^{(k)}_{N,t} =1
\label{gammanorm} \ee for all $N, k$ and time $t$. The main result
of this paper is the following theorem.

\begin{theorem}\label{thm:main} Assume the unscaled pair
potential $V$ is positive, spherically symmetric, smooth and
compactly supported and assume that $\varrho = (8\pi)^{-1} (\|V\|_1
+ \|V\|_{\infty})$ is sufficiently small (of order one). Let $T>0$
be fixed. Suppose that the normalized initial wave function
$\psi_{N,0}$ satisfies the energy bounds \be\label{mainassump}
\langle \psi_{N,0}, \wt H \psi_{N,0} \rangle \leq C_1 N \quad
\text{and } \quad \langle \psi_{N,0}, \wt H^2 \psi_{N,0} \rangle
\leq C_2 N^2 . \ee Then the sequence $\gamma^{(k)}_{N,t}$ has at
least one non-trivial limit point and any limit point  satisfies
the infinite Gross-Pitaevskii Hierarchy (\ref{eq:GPH2}) in a weak
sense with $\sigma= 8 \pi a_0$. More precisely:
\begin{itemize}
\item[i)] For sufficiently large $\nu$, the sequence $\Gamma_{N,t}
= \{ \gamma_{N,t}^{(k)} \}_{k \geq 1} \subset C ([0,T],
B_-^{(\nu)})$ is  compact w.r.t. the topology induced by the $\wh
\rho$ metric. \item[ii)] Let $\Gamma_{\infty,t} = \{
\gamma_{\infty,t}^{(k)} \}_{k \geq 1} \in C ([0,T], B_-^{(\nu)})$ be
any limit point of $\Gamma_{N,t}$ in the $\wh \rho$ metric. Then
$\Gamma_{\infty,t}$ satisfies
\begin{equation}
\tr \, (1 - \Delta_i) ( 1-\Delta_j) \gamma^{(k)}_{\infty ,t} < C^k
\end{equation}
for all $i \neq j$, $i,j = 1, \dots , k$, for all $t \in [0,T]$, and
all $k \geq 1$. \item[iii)] Any limit point $\Gamma_{\infty,t}$ of
the sequence $\Gamma_{N,t}$ is non-trivial. In particular
\begin{equation}
\tr \; \gamma^{(1)}_{\infty ,t} = 1 \, \quad \quad \text{and }\quad
\quad \tr \; \gamma^{(2)}_{\infty,t} = 1 \,. \label{nontriv}
\end{equation}
\item[iv)] Let $\Gamma_{\infty,t}$ be a limit point of
$\Gamma_{N,t}$ and assume $h_r (x) = (3/4\pi) \, r^{-3} h(x/r)$ for
any $h\in C_0^\infty(\Lambda)$ with $\int_\Lambda h =1$. Then, for
any $k \geq 1$ and $t \in [0,T]$, the limit
\begin{equation}\label{eq:limrr'}
\begin{split}
\lim_{r,r' \to 0} \int \rd x'_{k+1} \rd x_{k+1} \, h_r (x'_{k+1} -
x_{k+1}) h_{r'} &(x_{k+1} - x_j)
\gamma^{(k)}_{\infty,t} (\bx_k , x_{k+1} ; \bx'_k , x'_{k+1}) \\
&= : \gamma^{(k)}_{\infty,t} (\bx_k, x_j ; \bx'_k , x_j)
\end{split}
\end{equation}
exists in the weak $W^{-1,1} (\Lambda^k \times \Lambda^k)$-sense and
defines $\gamma^{(k)} (\bx_k, x_j ; \bx'_k , x_j)$ as a distribution
of $2k$ variables. \item[v)] For any $k \ge 1$ and any regular test
function $J^{(k)} ({\bf x}_k ;{\bf x}'_k)$ in the Sobolev space
$W^{2, \infty}(\Lambda^k\times \Lambda^k)$, we have, for any $t \in
[0,T]$,
\begin{equation}\label{eq:GPH3}
\begin{split}
\int \rd {\bf x_k} \rd {\bf x}_k' J^{(k)} ({\bf x}_k ; {\bf x}_k')
\gamma_{\infty,t}^{(k)} &({\bf x}_k ; {\bf x}_k') = \int \rd {\bf
x}_k \rd {\bf x}_k' J^{(k)}
({\bf x}_k ; {\bf x}_k') \gamma_{\infty,0}^{(k)} ({\bf x}_k ;{\bf x}_k') \\
-i \sum_{j=1}^k \int_0^t \rd &s \, \int \rd {\bf x}_k \rd {\bf x}_k'
\, J^{(k)} ({\bf x}_k; {\bf x}_k') \, ( -\Delta_j + \Delta_j')
\gamma^{(k)}_{\infty,s} ({\bf x}_k; {\bf x}_k')
\\ -8\pi i a_0 \sum_{j=1}^k &\int_0^t \rd s \int \rd {\bf x}_k \rd {\bf x}_k' \rd
x_{k+1} \, J^{(k)} ({\bf x}_k ;{\bf x}_k')
\\ &\times (\delta (x_j - x_{k+1})
- \delta (x_j' - x_{k+1})) \gamma^{(k+1)}_{\infty,s} ({\bf x}_k ,
x_{k+1}; {\bf x}_k',x_{k+1}) \; .
\end{split}
\end{equation}
\end{itemize}
\end{theorem}

\begin{remark}   The main assumption \eqref{mainassump} is satisfied
for $\psi_{N,0}= W \phi_{N}$ with $\phi_N$ sufficiently regular. See
Lemma \ref{lm:h2bound} in the Appendix for a proof.
\end{remark}

For simplicity, we state the theorem for  initial conditions that
are pure states. The same result holds for initial conditions
$\gamma_{N,0} \in \cL^1(\cH^{\otimes_s N})$, $\tr \;
\gamma_{N,0}=1$,  with the bounds
\[
\tr \; \wt H \gamma_{N,0} \leq C_1 N \quad \text{and } \quad \tr \;
\wt H^2 \gamma_{N,0} \leq C_2 N^2 . \]

As pointed out in the introduction, the factorization of the limit
point $\gamma_{\infty,t}^{(k)}$ depends on the uniqueness of the
solution to the infinite GP hierarchy (\ref{eq:GPH3}), which is an
open problem.

A similar proof holds if we replace the Hamiltonian $\wt H$
(\ref{eq:modham}) with $\wt H^{(n)}$ (\ref{eq:Hn}) for any fixed $n
\geq 3$, if $\varrho$ is sufficiently small, depending on $n$.

The structure of the rest of the paper is the following. In Section
\ref{idea} we explain some of the main ideas behind the proof of the
main theorem. In Section \ref{sec:approxi} we define an
approximation $U_{N,t}$ of $\Gamma_{N,t}$. In Section
\ref{sec:energy} we prove energy estimates for $\wt H$ and $\wt
H^2$, which rely on some Hardy type and Sobolev type inequalities,
which are stated and proved in Section \ref{sec:hardy}. The energy
estimates are turned into a-priori estimates for the approximation
$U_{N,t}$ and for any limit point $U_{\infty,t}$ of $U_{N,t}$ in
Section \ref{sec:apriori}. In Section \ref{sec:approx} we show how
these a-priori estimates can be used to regularize the
delta-function appearing in the limiting GP-Hierarchy. Finally, in
Section \ref{sec:mainproof}, we prove Theorem \ref{thm:main}. Some
technical estimates are collected in Section \ref{sec:control} and
in the Appendices.

\section{Idea of the Proof}\label{idea}
\setcounter{equation}{0}

We write the solution of the Schr\"odinger equation in the form
$\psi_{N,t} = W \phi_{N,t}$, where $W$, defined by (\ref{def:G}), is
an approximation of the wave function of the ground state of the $N$
boson system. We shall implement a general idea that consists of two
steps. A similar idea in a quite different context lies behind the
relative entropy method \cite{Y}.

\begin{itemize}
\item[(i)] Construct an approximate solution (in our case, $W$)
to the dynamics and factor out this approximation from the full
solution $\psi$.

\item[(ii)]  Derive an effective inequality governing the
remaining part $\phi$.  In general, one can prove a stronger
estimate on $\phi$ than on $\psi$. This estimate often involves
Dirichlet form with respect to the approximate solution.
\end{itemize}

\medskip

\noindent In our case, we obtain  the following key estimate on
$\phi_{N,t}$ (Corollary \ref{cor:energy2}):
\begin{equation}
\int W^{2} |\nabla_i \nabla_j \phi_{N,t}|^2 \leq C \label{apriori1}
\end{equation}
for any fixed $i \neq j$. We define $U_{N,t}^{(k)}$ to be roughly
the $k$-particle density matrix corresponding to the wave function
$\phi_{N,t}$ (the precise definition of $U_{N,t}^{(k)}$ will be
given in Section \ref{sec:approxi}). We are going to use
$U_{N,t}^{(k)}$ as an approximation of the $k$-particle density
$\gamma_{N,t}^{(k)}$ (they will coincide in the weak $N \to \infty$
limit). Then (\ref{apriori1}) implies that \be \label{eq:UDelta} \tr
\, (1 - \Delta_i) ( 1 - \Delta_j) U^{(k)}_{N,t} \leq C^k \; .\ee Let
now $U_{\infty,t}^{(k)}$ be a weak limit point of $U_{N,t}^{(k)}$
which also satisfies (\ref{eq:UDelta}). Then, using this bound, we
can show, by Proposition \ref{prop:delta}, that $U^{(k)}_{\infty, t}
(\bx_k,x_j; \bx'_k, x_j)$ is well-defined. Moreover we will show
(Lemma~\ref{lm:compGamma}) that the limit points of
$\gamma_{N,t}^{(k)}$ and of $U_{N,t}^{(k)}$ coincide, i.e. that
$\gamma_{\infty,t}^{(k)} = U_{\infty,t}^{(k)}$. Therefore
(\ref{eq:limrr'}) holds true, and the right hand side of
\eqref{eq:GPH3} is well-defined. Note that \eqref{eq:limrr'} and
\eqref{eq:GPH3} cannot be proven directly, since we do not have
estimates for $\tr \, (1 -\Delta_i) (1 -\Delta_j)
\gamma_{N,t}^{(k)}$; this is why we need to introduce the auxiliary
densities $U^{(k)}_{N,t}$.

A more refined calculation is needed to obtain $\sigma=8 \pi a_0$.
The idea of this calculation was explained in the introduction; it
can be made rigorous (Section \ref{sec:mainproof}) via the key
estimate \eqref{apriori1} and certain generalizations of the Hardy
and Poincar{\'e} inequalities (Section \ref{sec:hardy}). To explain
\eqref{apriori1} in more details, we first review the related mean
field models.

\subsection{The mean-field model}\label{sec:MF}

The mean field Hamiltonian is defined by
\begin{equation}\label{eq:ham-mean}
H_N = - \sum_{j=1}^N \Delta_j + N^{-1}\sum_{j < k}^N  V^{(\tau)}
(x_j - x_k)\;
\end{equation}
where we have introduced a scale parameter $\tau$ and $V^{(\tau)}
(x)= \tau^{-3} V(x/\tau)$. If we fix $\tau$ and take the limit $N
\to \infty$ (the mean field limit), then with a product initial wave
function Hepp \cite{H} and Spohn \cite{Sp} proved that the one
particle density matrix of the solution of the Schr\"odinger
equation \eqref{eq:SC} converges to the Hartree equation
\be\label{har} i
\partial_t u_t     = -\Delta u_t     +  (V^{(\tau)}\ast|u_t|^2)
u_t \, .\ee If we take $\tau\to 0$ in this equation,
 we recover the GP equation \eqref{nls} but
with the so-called mean-field interaction constant $\sigma=b_0=\int
V$ instead of $\sigma= 8\pi a_0$ involving $a_0$, the scattering
length of $V$. To  investigate the simultaneous limit $\tau\to 0$,
$N\to \infty$ in the many body problem,
 we set $\tau=N^{-\beta}$. If $0<\beta<1$, then the scattering length of
the interaction potential, $N^{-1} V^{(\tau)}$, is much smaller than
the range of the potential.  So we are in the mean-field regime and
the one particle density matrix of the solution of the Schr\"odinger
equation \eqref{eq:SC} is expected to converge to the GP equation
with $\sigma=b_0$. If $\beta=1$, then
 the Hamiltonian in
\eqref{eq:ham-mean} recovers the Hamiltonian in the GP scaling
\eqref{eq:ham1} and  the limit is expected to be the GP equation
with $\sigma = 8\pi a_0$.

In a joint work with A. Elgart we prove that, {\sl for product
initial data and $0<\beta < {2/3}$, the density matrices of the
Schr\"odinger equation \eqref{eq:SC} converge to a solution of the
GP hierarchy \eqref{eq:GPH2} with $\sigma=b_0$}. In \cite{ESY}, we
also prove that the GP hierarchy has a unique solution in an
appropriate Sobolev space. Moreover, we show in \cite{ESY} that for
$0<\beta < 1/2$  any limit of the solution to the Schr\"odinger
equation with product initial data satisfies the a-priori Sobolev
bound. Hence, in this case, we prove the uniqueness as well, and
therefore we complete the derivation of the GP equation with
$\sigma=b_0$ in a mean-field scaling. The same result is expected to
hold also for $1/2\leq \beta <1$, but the a-priori bound in this
regime is open.

The outline of the proof in \cite{EESY} is similar to this paper's.
The first step is to prove the estimate
\begin{equation}
\sum_{i, j}\int  |\nabla_i \nabla_j \psi_{N,t}|^2 \leq C N^2\,.
\label{apriori2}
\end{equation}
The proof of the convergence to the hierarchy makes use of arguments
similar to the ones used in Section \ref{sec:approx} and Section
\ref{sec:mainproof}. Since the proof of the estimate
\eqref{apriori2} is instructive, we reproduce it here.

\begin{proposition}\label{prop2/3}
Suppose $\psi(\bx)$ is a smooth normalized function. Then the
following inequality holds
\begin{equation}\label{eq:H^2}
\begin{split}
(\psi, H_N^2 \psi) \geq \; &C_1 \left(1 -\frac{1}{N
\tau^{3/2}}\right) \sum_{j,\ell=1}^N \int \rd{\bf x} |\nabla_j
\nabla_{\ell} \psi ({\bf x})|^2 \\ &- \frac{C_2 N}{N \tau^{3/2}}
\sum_{j=1}^N \int \rd {\bf x} |\nabla_j \psi ({\bf x})|^2 -\frac{C_3
N^2}{N \tau^{3/2}} \int \rd{\bf x} | \psi ({\bf x})|^2 .
\end{split}
\end{equation}
In particular, for $\tau=N^{-\beta}$ with $0\le \beta < 2/3$, we
have
\begin{equation}\label{eq:H^2.1}
 \sum_{j, \ell=1}^N \int \rd{\bf x} |\nabla_j
\nabla_{\ell} \psi ({\bf x})|^2 \le C (\psi, H_N^2 \psi)  + C N^2 \,
.
\end{equation}
\end{proposition}

\begin{proof} By expanding the square of the energy operator and dropping
some positive diagonal terms, we have
\begin{equation}\label{eq:prop1}
\begin{split}
\| H_N \psi \|^2 \geq \; &\sum_{j,\ell} \int \rd{\bf x} \Delta_j
\bar \psi ({\bf x}) \Delta_{\ell} \psi ({\bf x}) \\ &+ \frac{1}{N}
\sum_{j=1}^N \sum_{\ell\neq m} \int \rd{\bf x} (-\Delta_j \bar
\psi ({\bf x})) V^{(\tau)} (x_{\ell} - x_{m}) \psi ({\bf x}) \\
&+ \frac{1}{N} \sum_{j=1}^N \sum_{\ell \neq m} \int \rd {\bf x}
(-\Delta_j  \psi ({\bf x})) V^{(\tau)} (x_{\ell} -x_m) \bar \psi
({\bf x}) \,.
\end{split}
\end{equation}
{F}rom integration by parts, the second term on the right hand side
of the previous equation equals
\begin{equation*}
\begin{split}
\frac{1}{N} \sum_{j=1}^N \sum_{\ell\neq m} \int \rd{\bf x}
(-\Delta_j \bar \psi ({\bf x})) V^{(\tau)} (x_{\ell} - x_{m}) \psi
({\bf x})  =\; & \frac{1}{N} \sum_{j=1}^N \sum_{\ell\neq m} \int
\rd{\bf x} |\nabla_j \psi ({\bf x}))|^2 V^{(\tau)} (x_{\ell} -
x_{m}) \\ &+ \frac{2}{N} \sum_{j \neq \ell}^N  \int \rd{\bf x} \big
( \nabla_j \bar \psi \big )({\bf x}) \psi ({\bf x}) \nabla_j
V^{(\tau)} (x_j -x_{\ell}) \, .
\end{split}
\end{equation*}
{F}rom the Schwarz inequality, the last term is bounded by
\begin{equation*}
\begin{split}
\int \rd{\bf x} \nabla_j \bar \psi ({\bf x}) \psi ({\bf x}) \nabla_j
V^{(\tau)} (x_j -x_{\ell})   \geq \; &- \frac{1}{\tau^{3/2}} \int
\rd{\bf x} \, |\nabla_j \psi (\bx)|^2 \frac{1}{\tau^2} \Big |\nabla
V \left( \sfrac{x_j -x_{\ell}}{\tau}\right)\Big | \\ &-
\frac{1}{\tau^{3/2}} \int \rd{\bf x} |\psi ({\bf x})|^2 \,
\frac{1}{\tau^3} \Big | \nabla V \left( \sfrac{x_j
-x_{\ell}}{\tau}\right)\Big | \, .
\end{split}
\end{equation*}
By Lemma \ref{lm:sob}, the last two terms are bounded below by
$$
-\frac{C_1}{ \tau^{3/2}}  \sum_{j,\ell =1}^N \int \rd{\bf x} \left(
| \nabla_{\ell} \nabla_j \psi ({\bf x})|^2 + |\nabla_j \psi (\bx)|^2
+ |\psi (\bx)|^2 \right) \, .
$$
The last term in \eqref{eq:prop1} can be bounded in a similar way.
Combining these equations, we  have proved Proposition
\ref{prop2/3}.
\end{proof}

To understand the restriction $\tau \gg N^{-2/3}$, consider a single
particle in the potential $N^{-1} V^{(\tau)}$. Define the operator
$\fh$ in $\bR^3$ by
$$
\fh := -\Delta_x  + N^{-1} V^{(\tau)}(x) \, .
$$
Clearly, $\langle f, \fh^2 f \rangle$ is given by \be \langle f,
\fh^2 f \rangle =  \Big \langle \,  f, \; \Big \{ \Delta\Delta -
\Delta N^{-1}  V^{(\tau)} - N^{-1}  V^{(\tau)} \Delta + N^{-2}
(V^{(\tau)})^2 \Big \}  \, f \, \Big \rangle \, . \ee Notice that
for a general smooth function $f$ living on scale one, the last term
$\langle f, N^{-2}  (V^{(\tau)})^2  f \rangle$ is of order $N^{-2}
\tau^{-3}$. This term diverges in the large $N$ limit if $\tau \ll
N^{-2/3}$. The middle terms, on the other hand, are bounded by $C
N^{-1}$. Hence for smooth functions living on scale one, $\langle f,
\fh^2 f \rangle$ typically diverges. Therefore, functions with
$\langle f, \fh^2 f \rangle$ finite (in particular, the eigenstates
of $\fh$) should have short scale structure. For such functions, the
divergent parts of $N^{-2} (V^{(\tau)})^2$ (and that of
$\Delta\Delta$) are cancelled by the middle terms $\Delta N^{-1}
V^{(\tau)} + N^{-1} V^{(\tau)} \Delta$. This cancellation can be
understood using  the following idea of ``resolution of
singularities" with the help of an approximate solution $W$. We
explain the method in the setting of our modified Hamiltonian $\wt
H$.

\subsection{Resolution of singularities}

In this section we explain the idea  behind the estimate (see
\eqref{eq:keyesti0})
\begin{equation}\label{eq:HWpp}
\sum_{i, j} \int \, W^2 |\nabla_i \nabla_j \phi|^2  \le  C \langle W
\phi,  (\wt H^2 + N^2)  W \phi \rangle = C \langle \psi,  (\wt H^2 +
N^2)  \psi \rangle \, .
\end{equation}
Since the right side is a constant of motion, this implies the
a-priori bound \eqref{apriori1}.
 Recall
that  $\wt H$ contains the singular potential $N^2V(Nx)$. The
difficulty of the singular potential was resolved in the work of
Lieb-Yngvason \cite{LY1} for  the ground state energy problem. The
basic idea was to replace the singular potential by some flattened
out potential and to use the variational principle  and the
positivity of the potential. Our aim, instead,  is to  estimate the
Sobolev norm of $\phi$ in terms of $\wt H^2$. Therefore, there is no
variational principle and the usage of positivity of the potential
is limited.  Our strategy is to factor out the approximate ground
state $W$. The resulting effective Hamiltonian contains a flattened
interaction potential, similar to the one in \cite{LY1}. In this
way, we also maintain the almost perfect cancellation between the
potential and kinetic energy operators, a critical property
explained in the previous section.

The action of the Hamiltonian $\wt H$ on a product function $\psi=W
\phi$ can be written in a more convenient form. Let  $B$ be the
function \be
   B:= W^{-1}\wt H W \, .
\label{def:B1} \ee Define the operator \be
 L : = \sum_k\Big( -\Delta_k  - 2 \nabla_k (\log W) \nabla_k \Big)\; .
\label{def:L1} \ee The operator $L$ is  self-adjoint with respect to
$W^2$, since
\begin{equation*}
\int  W^2 \bar\phi_1 (L \phi_2) = \sum_{k=1}^N\int  W^2 (\nabla_k
\bar\phi_1)( \nabla_k \phi_2) =\int W^2 (L\bar\phi_1)\phi_2 \;  .
\end{equation*}
Then we have
\begin{equation*}
W^{-1} \wt H \, ( W \phi) =  L \phi + B \phi, \quad \qquad \int \,
\bar \psi \wt H \psi = \sum_k \int \, W^2 |\nabla_k  \phi|^2 + \int
\, W^2 B |\phi|^2 \; ,
\end{equation*}
and the following identity
\begin{equation*}
\int \, |\wt H W \phi|^2 = \int \, W^2 |L \phi|^2 + \int \, W^2 B
\bar\phi L \phi + \int \, W^2 B\phi L\bar\phi + \int \, |\phi|^2
(\wt H W)^2 \; .
\end{equation*}
Last equation can be rewritten in the following more convenient
form:
\begin{equation}\label{eq:HWphi1}
\begin{split}
\int \, |\wt H W \phi|^2 = \; &\sum_{i, j} \int \, W^2 |\nabla_i
\nabla_j \phi|^2 - 2 \sum_{i, j}\int \,
W^2 (\nabla_i \nabla_j \log W ) \, \nabla_i\bar \phi \nabla_j \phi  \\
& +2 \sum_{m} \int \, W^2 B |\nabla_m \phi|^2 + \int \, W^2
(\nabla_m B) \left( \bar\phi \nabla_m \phi + \phi \nabla_m \bar \phi
\right) + \int \, B^2 |\phi|^2 W^2 .
\end{split}
\end{equation}
For more details see (\ref{eq:HWphi})-(\ref{eq:Lphi2}). If $W$ is
the true ground state of $\wt H$, then $B$ is a positive constant
and the middle term in the second line of the last equation
vanishes. {F}rom the Schwarz inequality, we have
\begin{equation}\label{eq:HWp}
\int \, |\wt H W \phi|^2 \geq  \sum_{i, j} \int \, W^2 |\nabla_i
\nabla_j \phi|^2 - 2 \sum_{i, j} \int \, W^2 |\nabla_i \nabla_j \log
W|  \, |\nabla_i \phi|^2 .
\end{equation}
The singularities of the ground state wave function are of the form
$1- Ca/|x_i-x_j|$ for some constant $C$ and for $r \leq |x_i-x_j|\ll
1$, where $r$ is the radius of the support of $V_a$ ($r$ is of order
$a$). The second derivative of the ground state wave function is
bounded by $Ca / |x_i-x_j|^3 \leq C\varrho/ |x_i-x_j|^2$ in this
regime, since $a \leq C\int V_a \leq  C a^{-2} \| V \|_\infty r^3
\leq C r\varrho$. Recall that $\varrho = (8\pi)^{-1} (\|V\|_1 + \|
V\|_{\infty})$. The same estimate also holds for $|x_i-x_j|\leq r$
(see Appendix A). Thus $|\nabla_i\nabla_j \log W| \leq
C\varrho/|x_i-x_j|^2$. Applying the Hardy inequality, the last term
can be bounded by
\begin{equation*}
\int \, W^2 |\nabla_i \nabla_j \log W|  \, |\nabla_i \phi|^2 \leq C
\varrho \sum_{i,j} \int W^2 |\nabla_i \nabla_j \phi|^2 \,.
\end{equation*}
Assuming that $\varrho$ is small enough (of order one), together
with (\ref{eq:HWp}) we get the key estimate \eqref{eq:HWpp}.

In practice, we do not know the exact ground state function and its
approximation will be used. When $x_i$ is near $x_j$, the
approximate ground state behaves like the ground state of the
Neumann boundary problem given in \eqref{eq:neumann1}. Thus the
singularity of $B$ for $|x_i-x_j|\ll 1$ behaves like $q(x_i-x_j)$
where $q$, defined in \eqref{eq:q}, lives on scale $\ell_1 \gg 1/N$.
This procedure, roughly speaking, replaces the singular potential
$V_a$ by an effective potential living on a bigger scale $\ell_1$.
The price to pay is the two middle terms in \eqref{eq:HWphi1}. Since
the singularities of $W$ and $q$ are milder than that of  $V_a$,
this procedure in a sense resolves the singularity of $V_a$. This is
the key idea behind the proof of the estimate \eqref{apriori1}.

\section{Construction of the approximate solution}\label{sec:approxi}
\setcounter{equation}{0}

Recall the definition of the functions $G_i$ and $W$, \be G_i(x):= 1
- \sum_{j \neq i } w (x_i - x_j) F_{i j}(x) \; , \quad
\mbox{and}\quad
    W:= \prod_{i=1}^N \sqrt{G_i} \; .
\ee In the next lemma we prove that $G_i$ is separated away from
zero.

\begin{lemma}\label{lm:pos}
There is a positive constant $c_1$ depending only on $V$ such that
for sufficiently large $N$ \be
    c_1 \leq G_i \leq 1 \, .
\label{eq:Gsep} \ee
\end{lemma}
\begin{proof}
This can easily be seen because, for fixed $i$, the sum $\sum_{j
\neq i} w_{ij} F_{ij}$ equals zero, if $|x_i -x_j|
>\ell$ for all $j \neq i$, while, if $|x_i - x_m| \leq \ell$ for
some $m \neq i$, it is bounded by
\begin{equation*}
\chi (|x_i -x_m| \leq \ell) \sum_{j \neq i} w_{ij} F_{ij} \leq
w_{im} F_{im} + N e^{-c\ell^{-\e}} \leq c_1 + Ne^{-c\ell^{-\e}}
\end{equation*}
where we applied Lemma \ref{lemma:noover}. So if $N$ is large enough
we obtain (\ref{eq:Gsep}) with some $c_1\in (c_0, 1)$.
\end{proof}

We will also need versions of $G_i$ and $W$ which are independent of
some of the variables $x_1,\dots x_N$. We define, for $i,m \neq
k_1,\dots, k_\a$ fixed, a version of $F_{im}$ independent of
$x_{k_1}, \cdots, x_{k_\a}$:
$$
    F^{(k_1 \ldots k_\a)}_{im}: = F( \cN_{im}^{(k_1 \ldots k_\a)}), \qquad
    \cN_{im}^{(k_1 \ldots k_\a)} : = \sum_{u\neq i,m,k_1,\ldots,k_\a} [ h_{iu} +h_{mu}] \; .
$$
Moreover, for $i \neq k_1,\dots, k_\a$, we define
$$
    G_i^{(k_1\ldots  k_\a)}: = \left(1- \sum_{m \neq i,k_1 , \ldots , k_\a}
    w_{im} F^{(k_1\ldots k_\a)}_{im}\right)
$$
and \be
      W^{(k_1 \ldots k_\a)} : = \prod_{i \neq k_1, \ldots , k_\a} \sqrt{G_i^{(k_1 \ldots   k_\a)}} \; .
\label{def:Wmanyk} \ee Thus $W^{(k_1 \ldots k_\a)}$ is independent
of $x_{k_1}, \ldots , x_{k_\a}$. When $\alpha=1$, we shall drop the
label $\alpha=1$ and denote the function by $W^{(k)}$. Notice that
we have
$$
    G_i^{(k)} = G_i + w_{ik} F_{ik} - \sum_{m\neq k,i} w_{im}(F^{(k)}_{im}
    -F_{im})\, .
$$

We shall also use  the notations $W^{[k]}:= W^{(1 \ldots k)}$,
$F^{[k]}_{im}: = F^{(1 \ldots k)}_{im}$ for any $i,m
>k$, and $G^{[k]}_i: = G^{(1 \ldots k)}_i$, for $i > k$. In
Appendix \ref{app:remove} we shall prove the existence of a constant
$C$, such that, for any choice of the indices $k_1, .. k_{\a}$,
$$
C^{-\alpha} \leq W^{(k_1 .. k_{\a})} / W \leq C^{\a} \; .
$$
Since $W \neq 0$ (by Lemma \ref{lm:pos}), we can write the solution
of the Schr\"odinger equation (\ref{eq:schr3}) as \be \psi_{N,t}  =
W \phi_{N,t} .\ee

We define a modified $k$-particle marginal distributions by
\begin{equation}\label{eq:UNt}
\begin{split}
U^{(k)}_{N,t} (x_1, \dots, x_k ; x_1', \dots, x_k') : =  \int \rd
x_{k+1} \dots\rd x_N (W^{[k]} &(x_{k+1} , \dots, x_N))^2
\\ \times \phi_{N,t} (x_1,, \dots, x_k, x_{k+1}, \dots,x_N) &\overline{\phi}_{N,t}
(x_1', \dots, x_k', x_{k+1}, \dots, x_N) \; .
\end{split}
\end{equation}
These density matrices will satisfy the requirements of the Main
Theorem \ref{thm:main}.

Clearly  $U^{(k)}_{N,t}\ge 0$ as an operator on $L^2 (\Lambda, \rd x
)^{\otimes_s k}$. Furthermore,  by (\ref{eq:WkW2}), we have
\begin{equation*}
\begin{split}
\tr \; U^{(k)}_{N,t} &= \int \rd x_1 \dots \rd x_N (W^{[k]} (x_{k+1}
,\dots, x_N))^2 |\phi_{N,t} (x_1, \ldots x_k, x_{k+1} ..x_N)|^2 \\
&\leq C^k_0 \int \rd x_1 \ldots \rd x_N \, W (x)^2 |\phi (x)|^2 \leq
C^k \int \rd x |\psi_{N,t} (x)|^2 =C^k_0 \; .
\end{split}
\end{equation*}
It is  instructive to compare the approximate densities
$U_{N,t}^{(k)}$ with the true marginal densities
$\gamma_{N,t}^{(k)}$, which are given by
\begin{equation*}
\begin{split}
\gamma^{(k)}_{N,t} (x_1, \dots, x_k ; x_1', \dots, &x_k')  : = \int
\rd x_{k+1} \dots\rd x_N \;  W (x_1 , \dots, x_N)  W (x_1' , \dots,
x_N')
\\ &\times \phi_{N,t} (x_1,, \dots, x_k, x_{k+1}, \dots,x_N)
\overline{\phi}_{N,t} (x_1', \dots, x_k', x_{k+1}, \dots, x_N) \; .
\end{split}
\end{equation*}
A simple computation shows that $\gamma_{N,t}^{(2)}$ and
$U_{N,t}^{(2)}$ satisfy the relation \eqref{g5}.

\section{Hardy type and Sobolev type inequalities}\label{sec:hardy}
\setcounter{equation}{0}

We need the following finite volume version of the usual Hardy
inequality \be
     \int_{\bR^3}\frac{|f|^2}{|x|^2}\leq C \int_{\bR^3} |\nabla f|^2 \; .
\label{eq:hardyold} \ee

\begin{lemma}[Hardy Type Inequality]\label{lm:hardy}
Let $0\leq U(x)\leq c|x|^{-2}$ on $B: = \{ |x|\leq \ell\}\subset
\bR^3$, and let
$$
     \langle f \rangle: = |B|^{-1}\int_B f
$$
be the average of $f$ on $B$ ($|B|=(4/3)\pi \ell^3$ is the volume of
$B$). Then \be
    \langle U|f|^2 \rangle \leq C \langle|\nabla f|^2\rangle
     + C \langle U\rangle
    \langle|f|^2\rangle \; .
\label{eq:hardy} \ee
\end{lemma}

\begin{proof}
Let $\bar f: = \langle f \rangle$, then
$$
\langle U \; |f|^2 \rangle  \leq 2\langle U\; |f-\bar f|^2 \rangle +
2 \bar f^2 \langle U\rangle \; .
$$
By the usual Hardy inequality
$$
\langle U|f-\bar f|^2 \rangle \leq C \langle |\nabla f|^2\rangle +
C\ell^{-2} \langle |f-\bar f|^2\rangle \; .
$$
By the Neumann spectral gap we have
$$
\ell^{-2} \langle |f-\bar f|^2\rangle \leq C\langle |\nabla
f|^2\rangle \, .
$$
The lemma now follows from
$$
    |\bar f|^2 = |\langle f \rangle| ^2 \leq\langle |f|^2\rangle .
$$
\end{proof}

We also need some well-known inequalities, collected in the
following lemma.
\begin{lemma}[Sobolev Type Inequalities] \label{lm:sob}
The following two inequalities are standard.
\begin{itemize}
\item[i)] Suppose $U \in L^{3/2} (\Lambda)$, $U \geq 0$, and $\psi
\in W^{1,2} (\Lambda)$. Then
\begin{equation}
\int_\Lambda \,|\psi |^2  U \leq C \| U\|_{L^{3/2}(\Lambda)} \,
\int_\Lambda  \, \Big[ | \nabla \psi |^2 + |\psi |^2\Big]
\,.\label{eq:sob1}
\end{equation}
\item[ii)] Suppose $U \in L^1 (\Lambda)$. Then, considering
$U(x-y)$ as an operator on the Hilbert space $L^2 (\Lambda , \rd x)
\otimes L^2 (\Lambda , \rd y)$ we have the operator inequality
\begin{equation}
U(x-y) \leq C \| U \|_{L^1} \, (1 - \Delta_x) ( 1-\Delta_y) \; .
\label{eq:2delta}
\end{equation}
\end{itemize}
\end{lemma}

\begin{proof}
i) Applying the H\"older inequality we find
\begin{equation*}
\int U \, |\psi|^2   \leq \left( \int_\Lambda  \, U^{3/2}
\right)^{2/3} \left( \int_\Lambda  |\psi|^{6} \right)^{1/3} \; .
\end{equation*}
By the Sobolev inequality we have
\begin{equation*}
\left( \int_\Lambda |\psi |^{6} \right)^{1/3} \leq  C \int_\Lambda
\Big[ |\nabla \psi |^2 + |\psi |^2 \Big] \; ,
\end{equation*}
which implies part i). The proof of part ii) can be found in
\cite{EY}.
\end{proof}

Let
\begin{equation}\label{eq:lambda-sigma}
   \lambda(x): = \frac{\chi \{ |x|\leq 3\ell_1/2 \}}{|x|^2}
    \quad \text{and } \quad \sigma (x) := \frac{\chi
   \{ |x|\leq 3\ell_1 /2\}}{|x|^3 + a^3} \qquad x\in
   \Lambda\; ,
\end{equation}
and, as usual, $\lambda_{ij}: = \lambda(x_i-x_j)$, $\sigma_{ij} :=
\sigma (x_i -x_j)$. We note the following inequalities, whose proof
is given in Appendix \ref{app:2body} \be
\begin{split}
   w\leq \wt\chi, \qquad w\leq Ca\ell_1 \lambda, \qquad  |\nabla w|\leq Ca^{-1}, \qquad
    |\nabla w|\leq Ca\lambda , \\  |\nabla^2 w|\leq  C
    \varrho\lambda, \qquad
    |\nabla^2 w|\leq  C a \sigma, \qquad |\nabla w|^2 \leq
    C a \sigma, \qquad w^2 \leq Ca^2\lambda ,\;  \label{eq:wlambda}
\end{split}
\ee where $\varrho = (8\pi)^{-1} (\|V\|_{\infty} + \| V \|_1)$.

In the sequel we will use the convention that integrals without
specified measure and domain will always refer to Lebesgue
integration on $\Lambda^N$:
$$
    \int \Big(\ldots\Big): = \int_{\Lambda^{N}} \Big(\ldots\Big) \rd \bx \; .
$$
We will use the Hardy-type and the Sobolev type inequalities in the
following form:

\begin{lemma}\label{lemma:combined} Let $\phi(\bx)\in L^2(\Lambda^N)$.
For any $q>0$ and for fixed $k,j$ indices and fixed $x_1, .. ,\wh
x_k, .. ,  x_N$ we have \be
   \int \rd x_k W^2 F_{kj}^q \lambda_{kj} |\phi |^2
    \le C_q\int \rd x_k  W^2 F_{kj}^{q/4} \wt \chi_{kj} |\nabla_k\phi|^2
    + C_q\ell_1\ell^{-3}\int  \rd x_k W^2 F_{kj}^{q/4} \wt \chi_{kj} |\phi|^2
\label{eq:nabw} \ee for sufficiently large $N$. In particular \be
   \sum_{kj} \int  W^2 F_{kj}^q\lambda_{kj} |\phi |^2
    \le C_q \\ \int   W^2  \Big[ \sum_k |\nabla_k\phi|^2
    + N\ell_1\ell^{-3}\int   |\phi|^2\Big]
\label{eq:nabwsum} \ee by (\ref{eq:overlapsum}). Moreover we have
\be \label{eq:sigmaest} \sum_{kj} \int W^2  \sigma_{kj} |\phi|^2
\leq C (\log N) \, \int
 W^2 \left( \sum_{jk} |\nabla_j \nabla_k \phi|^2 + N \sum_{j} |\nabla_j \phi|^2 +
 N^2 |\phi|^2 \right)\; .
\ee The same estimates hold if we set $W\equiv 1$ in all integrals.
\end{lemma}\begin{proof}
We define
\begin{equation*}
F_{j}^{(k)} := e^{-\ell^{-\e} \sum_{m \neq k,j} h (x_j -x_m)} .
\end{equation*}
Note that $F_{j}^{(k)}$ is independent of the variable $x_k$, and
that, trivially, $ F_{kj}\leq F_{j}^{(k)}$. Moreover we have
\begin{equation}
\begin{split}
\wt \chi_{kj} \, F_{kj}^q  &= \wt \chi_{kj} \, e^{-q \ell^{-\e}
\sum_{m \neq k,j} (h (x_j -x_m) + h (x_k -x_m))} \\ &\geq \wt
\chi_{kj} \, e^{-q \ell^{-\e} \sum_{m \neq k,j} h (x_j -x_m)\left( 1
+ e^{|x_j - x_k|/\ell}\right)} \\ &\geq \wt \chi_{kj} \,
[F_{j}^{(k)}]^{4q},
\end{split}
\label{eq:wtF}\end{equation} because $1 + e \leq 4$. Here we used
that $((x_k -x_m)^2 + \ell^2)^{1/2} \leq ((x_j -x_m)^2 +
\ell^2)^{1/2} + |x_j -x_k|$. Using (\ref{eq:Wk}) to replace $W^2$ by
$[W^{(k)}]^2$, which is independent of $x_k$, applying
(\ref{eq:hardy}) for $x_k$, and finally changing the measure back,
we find:
\begin{equation*}
\begin{split}
\int W^2 F_{kj}^q \lambda_{kj}|\phi |^2  \leq & \; C_0^2 \int
[W^{(k)}]^2  (F_{j}^{(k)})^q
\lambda_{kj} |\phi |^2 \\
\leq &\; C  \int \rd x_1\ldots \widehat{\rd x_k} \ldots \rd x_N \,
[W^{(k)}]^2  (F_{j}^{(k)})^q  \, 
\int \rd x_k
\Big[ |\nabla_k\phi|^2
+ \ell_1\ell^{-3} |\phi|^2 \Big] \wt\chi_{kj} \\
\leq &\; C \int W^2  (F_{j}^{(k)})^q \,
\wt\chi_{kj}\Big[|\nabla_k\phi|^2 + \ell_1\ell^{-3}|\phi|^2 \Big] \;
.
\end{split}
\end{equation*}
The estimate (\ref{eq:nabw})  follows now from (\ref{eq:wtF}).
Similarly the estimate (\ref{eq:sigmaest}) follows from
(\ref{eq:2delta}), after replacing $W$ by $W^{(k,j)}$. It is clear
that setting $W\equiv 1$ would not alter the validity of the proof.
\end{proof}

\section{Energy Estimate} \label{sec:energy}
\setcounter{equation}{0}

We begin by computing the action of the Hamiltonian $H$ defined in
(\ref{eq:ham1}) on the wave function $W$, defined in Section
\ref{sec:modham}. For any fixed index $k$ we have
$$
W^{-1} (-\Delta_k)     W= \frac {-\Delta_k  G_j } { 2G_j }-
 1(i\neq j)\frac{1}{4}
 \frac { \nabla_k  G_i }{ G_i }
 \frac {\nabla_k G_j }{ G_j } + \frac{1}{4} \left(\frac{\nabla_k
 G_j}{G_j}\right)^2 \; .
$$
Here, and in the rest of the paper, we use the summation convention:
all unspecified indices are summed up; in this case the $j$ and $i$
indices on the right hand side.

Direct computation shows that for any fixed $k$
$$
     -\Delta_k G_k = (\Delta w)_{kj} F_{kj} + \Omega_k\; ,
$$
where \be
    \Omega_k: = 2(\nabla w)_{kj} \nabla_k F_{kj} +w_{kj}\Delta_k
    F_{kj}
\label{def:Omegak} \ee and, if $k\neq j$ are fixed, then
$$
   -\Delta_k G_j = (\Delta w)_{kj} F_{kj} + \Omega_{kj}
$$
with
$$
   \Omega_{kj} := 2(\nabla w)_{kj}\nabla_k F_{kj} +w_{ij}
   \Delta_k F_{ij} \; .
$$

{F}rom (\ref{eq:neumann1}) it follows that
$$
    \Delta w = - (1/2) V_a (1-w) +(1-w)q \; .
$$
Then we have
\begin{equation}\label{eq:WHW}
W^{-1}HW = ((1/2) V_{kj} - q_{kj}) [ 1- (1-w_{kj})M_{kj}] + q_{kj} +
\Omega
\end{equation}
with
\begin{equation}\label{eq:Omega}
   \Omega
   := \frac{\Omega_k}{2G_k} +\frac{\Omega_{kj}}{2G_j}
    -\frac{1}{4}\frac{\nabla_kG_i}{G_i}\frac{\nabla_k G_j}{G_j}1(i\neq
    j)+ \frac{1}{4} \left( \frac{\nabla_k G_i}{G_i}\right)^2
\end{equation}
and, for any $k,j$ (see \ref{eq:mkj}) \be
     M_{kj}:= \frac{F_{kj}}{2}( G_j^{-1} + G_k^{-1}) \; .
\ee The action of the modified Hamiltonian (\ref{eq:modham}) on the
wave function $W$ is then simply given by
\begin{equation*}
W^{-1}\wt HW = q_{kj} + \Omega \, .
\end{equation*}
It follows that the action of $\wt H$ on the product $W \phi$ is
given by
\begin{equation}\label{eq:LB}
W^{-1} \wt H W \phi =  L \phi + B \phi
\end{equation}
where we defined the operator \be
 L : = \sum_k\Big( -\Delta_k  - 2 \nabla_k (\log W) \nabla_k \Big)\; ,
\label{def:L} \ee and the function \be
   B:= W^{-1}\wt H W = q_{kj} +\Omega \; .
\label{def:B} \ee Note that the operator $L$ has the
self-adjointness property that (with the summation convention)
\begin{equation}\label{eq:L}
\int  W^2 \bar\psi (L \phi) = \int  W^2 (\nabla_k \bar\psi)(
\nabla_k \phi) =\int W^2 (L\bar\psi)\phi \;  .
\end{equation}

Next we compute the last two terms in $\Omega$ (see
(\ref{eq:Omega})). For fixed $i,k$ we have
\begin{equation*}
\nabla_k G_i =  (\nabla w)_{ik} F_{ik} - (\nabla w)_{k\ell}
F_{k\ell} \delta_{ki} -w_{i\ell} \nabla_k F_{i\ell} \, .
\end{equation*}
Hence we obtain
\begin{equation*}
\begin{split}
\left( \nabla_k G_i \right)^2 = \; &(\nabla w)_{ik}^2 F_{ik}^2 +
(\nabla w)_{k\ell} (\nabla w)_{ks} F_{k\ell} F_{ks} \delta_{ki}
-2(\nabla w)_{ik} w_{i\ell} F_{ik} \nabla_k F_{i\ell} \\ &+ 2
(\nabla w)_{k\ell}  w_{ks} F_{k\ell} \nabla_k F_{ks} \delta_{ki} +
w_{i\ell} w_{is} \nabla_k F_{i\ell} \nabla_k F_{is} \, .
\end{split}
\end{equation*}
Using Lemma \ref{lemma:noover} we note that the second term on the
r.h.s. is exponentially small unless $\ell =s$. Analogously the
third term is exponentially small unless $k = \ell$, while the
fourth and fifth terms are exponentially small unless $\ell =s$.
Thus summing over $i,k$ on the left and the right side of the
equation, we obtain (changing the name of the indices in the second
and fourth term)
\begin{equation}\label{eq:GiGi}
\begin{split}
\frac{ (\nabla_k G_i )^2}{4G_i^2} =  \; &\frac{(\nabla w)_{ik}^2
F_{ik}^2}{2 (1 - w_{ik} F_{ik})^2} - \frac{(\nabla w)_{ik} w_{ik}
F_{ik} (\nabla_k F_{ik} - \nabla_i F_{ik})}{2 G_i^2} +
\frac{w_{i\ell}^2 (\nabla_k F_{i\ell})^2}{4G_i^2} + O \left(
e^{-c\ell^{-\e}}\right) \, .
\end{split}
\end{equation}
Here we also used that, on the support of $(\nabla w)_{ik}$, we have
$ G_i= 1- w_{ik}F_{ik} + O(e^{-c/\ell^\e})$. Analogously, for $i\neq
j$ we have
\begin{equation}\label{eq:GiGj}
\begin{split}
\frac{\nabla_k G_i \nabla_k G_j}{4G_i G_j} = \; &\frac{(\nabla
w)_{ik}^2 F_{ik}^2}{2 (1 - w_{ik} F_{ik})^2} - \frac{(\nabla w)_{ik}
w_{j\ell} F_{ik} (\nabla_k F_{j\ell} - \nabla_i F_{j \ell})}{2G_i
G_j} + \frac{w_{i\ell} w_{js} \nabla_k F_{i\ell} \nabla_k
F_{js}}{4G_i G_j} + O \left( e^{-c\ell^{-\e}}\right)
\end{split}
\end{equation} where we are summing over all $k$ and all $i
\neq j$ on the left and the right side of the equation. We see that
there are some cancellations between (\ref{eq:GiGi}) and
(\ref{eq:GiGj}). It follows that
\begin{equation}\label{eq:wtOm}
\Omega = \wt \Omega + O \left( e^{-c\ell^{-\e}}\right) \; ,
\end{equation}
where
\begin{equation}
\wt \Omega = \frac{\Omega_k}{2G_k} +\frac{\Omega_{kj}}{2G_j} +
\Gamma \label{def:Omtilde}
\end{equation}
and
\begin{equation*}
\Gamma = s_{ij} \left( \frac{(\nabla w)_{ik} w_{j\ell} F_{ik}
(\nabla_k F_{j\ell} - \nabla_i F_{j \ell})}{2G_i G_j} -
\frac{w_{i\ell} w_{js} \nabla_k F_{i\ell} \nabla_k F_{js}}{4G_i
G_j}\right).
\end{equation*}
Here $s_{ij} = -1$ if $i=j$ and $s_{ij} = 1$ if $i\neq j$ (and sum
over all indices is understood). A more careful analysis also shows
that, analogously to (\ref{eq:wtOm}), for any fixed $m$,
\begin{equation}\label{eq:nablaOm}
\nabla_m \Omega = \nabla_m \wt \Omega + O \left( e^{-c\ell^{-\e}}
\right) \; .
\end{equation}
This fact will be used in the proof of the next proposition, which
is the key energy estimate.

\begin{proposition}\label{prop:energy2}
Assume $a= a_0 N^{-1}$, $a \ll \ell_1 \ll \ell \ll 1$, $a\ell_1
 \ll \ell^{4}$, $a \ll \ell_1^{3/2}$ and $a \ll \ell^2$ (for example $\ell_1 =
N^{-2/3 + \kappa}$ and $\ell = N^{-2/5 - \kappa}$, for $\kappa>0$
sufficiently small). Put $\varrho = (8\pi)^{-1} ( \| V \|_{\infty} +
\| V \|_1)$.
\begin{itemize}
\item[i)] Then we have
\begin{equation*}
( W \phi , \wt H W \phi) = \int \, W \bar{\phi} (\wt H W \phi) \geq
(1 -o(1)) \int \, W^{2} \sum_k |\nabla_k \phi|^2 - o(N) \int \, W^2
|\phi|^2 \, ,
\end{equation*}
for $N \to \infty$. \item[ii)] Moreover, if $\varrho$ is
sufficiently small, there is $C >0$ such that
\begin{equation*}
\begin{split}
\int |\tilde{H} W \phi|^2  \geq \; & (C - o (1)) \int \,
W^2\sum_{ij} | \nabla_i \nabla_j \phi|^2  - o(1) \left( N \int \,
W^2 \sum_i |\nabla_i \phi|^2 + N^2 \int \, W^2 |\phi|^2 \right)
\end{split}
\end{equation*} as $N\to \infty$.
\end{itemize}
\end{proposition}

This Proposition and the conservation of $\wt H$ and $\wt H^2$ along
the solution of the  Schr\"odinger equation (\ref{eq:schr3})
immediately implies the following

\begin{corollary}\label{cor:energy2}
Let $\psi_{N,t} (\bx) = W(\bx) \phi_{N,t} (\bx)$ be a symmetric
(with respect to permutations of the particles) wave function
solving the Schr\"odinger equation (\ref{eq:schr3}), with
 initial data $\psi_{N,0}$ satisfying the bounds
$$
(\psi_{N,0}, \wt H \psi_{N,0}) \leq C_1 N \; , \quad \mbox{and}\quad
(\psi_{N,0}, \wt H^2 \psi_{N,0}) \leq C_2 N^2\; .
$$
Suppose moreover that the assumptions of Proposition
\ref{prop:energy2} are satisfied. Then for any time $t$ and for any
fixed indices $i \neq j$, we have
\begin{equation}
\int W^{2} |\nabla_i  \phi_{N,t}|^2 \leq C,\qquad \int W^{2}
|\nabla_i \nabla_j \phi_{N,t}|^2 \leq C \label{apriori}
\end{equation}
for some constant $C$, independent of $t$ and of the indices $i,j$.
\end{corollary}

\begin{proof}[Proof of Proposition \ref{prop:energy2}]
{F}rom (\ref{eq:LB}), (\ref{eq:L})  we have (recall the summation
convention)
\begin{equation*}
\int \, W \bar{\phi} (\wt H W \phi) = \int \, W^2
|\nabla_k \phi|^2 + \int \, W^2 B |\phi|^2 \\
\geq \int \, W^2 |\nabla_k \phi|^2 - \int \, W^2 | \Omega|
|\phi|^2\; .
\end{equation*}
Here we used that $B = q_{kj} + \Omega \geq \Omega$, because $q_{kj}
\geq 0$. Part i) thus follows immediately from (\ref{eq:wtOm}) and
Lemma \ref{lm:error1} below.

As for part ii) we note that, from (\ref{eq:LB}),
\begin{equation}\label{eq:HWphi}
\int \, |\wt H W \phi|^2 = \int \, W^2 |L \phi|^2 + \int \, W^2 B
\bar\phi L \phi + \int \, W^2 B\phi L\bar\phi + \int \, |\phi|^2
(\wt H W)^2 \; .
\end{equation}
Using (\ref{eq:L}), we can rewrite the second and third term in the
last equation as
\begin{equation}\label{eq:BOmega}
\begin{split}
\int  \, W^2 B \bar\phi  L \phi + \int \, W^2 B \phi L \bar\phi &= 2
\int \, W^2 B |\nabla_m \phi|^2 + \int
\, W^2 (\nabla_m B) \left( \bar\phi \nabla_m \phi + \phi \nabla_m \bar \phi \right) \\
&\geq 2 \int \, W^2 \Omega |\nabla_m \phi|^2 - 2 \int \, W^2
|\nabla_m B| |\phi||\nabla_m \phi| .
\end{split}
\end{equation}
Here we used again the positivity of $q$.
 {F}rom Eqs. (\ref{eq:wtOm}) and
(\ref{eq:nablaOm}), we find
\begin{equation*}
\begin{split}
\int  \, W^2 B &\bar\phi  L \phi + \int \, W^2 B \phi L \bar\phi
\\ \geq \; &- 2 \int \, W^2 |\wt \Omega| |\nabla_m \phi|^2 - 2
\int \, W^2 |\nabla_m q_{kj}| |\phi| |\nabla_m \phi| \\ &- 2 \int \,
W^2 |\nabla_m \wt \Omega| |\phi| |\nabla_m \phi| + O
\left(e^{-c\ell^{-\e}}\right) \B \, .
\end{split}
\end{equation*}
Analogously, the first term on the r.h.s. of (\ref{eq:HWphi}) can be
written as
\begin{equation}\label{eq:Lphi2}
\begin{split}
\int \, W^2 |L \phi|^2 &= \int \, W^2 \nabla_j (L \bar\phi) \nabla_j
\phi \\ &= \int \, W^2 (L \nabla_j \bar \phi) \nabla_j \phi + \int
\, W^2 \Big( [\nabla_j ,  L] \bar\phi\Big)
\nabla_j \phi \\
&= \int \, W^2 \nabla_i \nabla_j  \bar\phi \nabla_i \nabla_j \phi -
2 \int \, W^2 \nabla_j {\Big (
\frac {\nabla_i W} W \Big ) } \nabla_i\bar \phi \nabla_j \phi \\
&= \int \, W^2 |\nabla_i \nabla_j \phi|^2 - 2 \int \, W^2 (\nabla_i
\nabla_j \log W ) \, \nabla_i\bar \phi \nabla_j \phi \; .
\end{split}
\end{equation}
Next we note that, for any fixed $i.j$,
\begin{equation*}
\nabla_i \nabla_j \log W = \frac{1}{2} \sum_{\ell} \nabla_i \nabla_j
\log G_{\ell} = \frac{1}{2} \sum_{\ell} \frac{\nabla_i \nabla_j
G_{\ell}}{G_{\ell}} - \frac{\nabla_i G_{\ell} \nabla_j
G_{\ell}}{G_{\ell}^2} \; .
\end{equation*}
The second term gives a positive contribution. Hence we have
\begin{equation*}
\int \, W^2 |L \phi|^2 \geq \int \, W^2 |\nabla_i \nabla_j \phi|^2 -
\int \,  W^2  \frac{\nabla_i \nabla_j G_{\ell}}{ G_{\ell}}
\nabla_i\bar \phi \nabla_j \phi .
\end{equation*}
Thus, using $G_{\ell} \geq c_1 >0$, we obtain
\begin{equation}\label{eq:HWphi2}
\begin{split}
\int \, |\wt H W \phi|^2 \ge \;&\int \, W^2 |\nabla_i \nabla_j
\phi|^2 - (1/c_1) \int \, W^2 |\nabla_i \nabla_j G_{\ell}| \,
|\nabla_i \phi| |\nabla_j \phi| \\ & -2 \int \, W^2 |\wt \Omega|
|\nabla_m \phi|^2 - 2 \int \, W^2 |\nabla_m q_{kj}| |\phi| |\nabla_m
\phi| \\ &- 2 \int \, W^2 |\nabla_m \wt \Omega| |\phi| |\nabla_m
\phi| + O \left(e^{-c\ell^{-\e}}\right)   \B \; .
\end{split}
\end{equation}
The first term on the r.h.s. is the positive contribution we are
interested in, all the other terms are errors terms which we
estimate separately in Lemmas \ref{lm:error1} (note the remark after
Lemma \ref{lm:error1}), \ref{lm:error2}, \ref{lm:error3} and
\ref{lm:error4}. The proposition follows then from last equation and
from the results of these lemmas.
\end{proof}

In the rest of this section we use the following notations for
brevity
$$
   \cE(\phi):= \int  W^2 \sum_k |\nabla_k \phi|^2 + N \int  W^2|\phi|^2
$$
$$
  \cF(\phi): =  \int  W^2 \sum_{kj}  |\nabla_k\nabla_j \phi|^2 + N \int  W^2 \sum_k |\nabla_k \phi|^2 + N^2
  \int  W^2|\phi|^2 \; ,
$$
where the dependence on $N$ is omitted.

\begin{lemma}\label{lm:error1}
Assume $a\ll\ell_1\ll\ell\ll 1$ and $a\ell_1\ll \ell^4$. Then
\begin{equation*}
\int  W^2 |\wt \Omega| |\phi|^2 \leq o(1) \cE(\phi)
\end{equation*} as $N\to \infty$.
\end{lemma}

{\it Remark. } Replacing $\phi$ by $\nabla_m \phi$, and summing over
$m$, it also follows from Lemma \ref{lm:error1} that
\begin{equation*}
\int  W^2 |\wt \Omega| \sum_m |\nabla_m \phi|^2 \leq o(1) \cF(\phi)
\; .
\end{equation*}

\begin{proof} Using $G_j\ge c_1>0$,
 we have from (\ref{def:Omtilde}) (using the summation convention)
\begin{equation*}
\begin{split}
\int \, W^2 |\wt \Omega| |\phi|^2 \leq & \;  C\int \, W^2\Big[
|(\nabla w)_{kj}| |\nabla_k F_{kj}|+w_{ij}
|\Delta_kF_{ij}| \Big]|\phi|^2 \\
&+ C\int \, W^2 \Big[ |(\nabla w)_{ik}| w_{jr} F_{ik} |\nabla_k
F_{jr}| + w_{im} w_{jr} |\nabla_k F_{im}|
|\nabla_k F_{jr}| \Big] |\phi|^2 \\
\leq & \; C\int W^2(a\ell^{-1} + a\ell_1\ell^{-2}) \lambda_{kj}
F_{kj}^{1/2} |\phi|^2\; ,
\end{split}
\end{equation*}
where we used the estimates $w_{jr}\leq \wt\chi_{jr}$,  $w_{ij} \leq
Ca\ell_1\lambda_{ij}$ and $|\nabla w_{ij}|\leq Ca\lambda_{ij}$ from
(\ref{eq:wlambda}), and we also used the formulas (\ref{eq:Fest}),
(\ref{eq:kfix}) to sum up indices. Using $\ell_1\leq \ell$ and
(\ref{eq:nabwsum}), we obtain
$$
  \int \, W^2 |\wt \Omega| |\phi|^2 \leq Ca\ell^{-1}\int W^2 \Big[
|\nabla_k\phi|^2
  + N\ell_1\ell^{-3}|\phi|^2\Big] = o(1)\cE(\phi) \; .
$$
\end{proof}

\begin{lemma}\label{lm:error2}
Assume $a \ll \ell_1 \ll \ell$ and $a\ell_1 \ll \ell^4$. Then there
is a constant $C$ such that
\begin{equation*}
\int \, W^2 \sum_{i,j,m}|\nabla_i \nabla_j G_m | |\nabla_i \phi|
|\nabla_j \phi| \leq \; C  \varrho \int \, W^2 \sum_{i,j}|\nabla_i
\nabla_j \phi|^2 + o(1) \cE(\phi)
\end{equation*}
for $N \to \infty$. Recall that $\varrho = (8 \pi)^{-1} (\| V
\|_{\infty} + \| V \|_1)$.
\end{lemma}
\begin{proof} A direct computation of
$\nabla_i \nabla_j G_m$ and an application of Schwarz inequality (to
separate $|\nabla_j \phi|$ and $|\nabla_i \phi|$) show that (with
the summation convention)
\begin{equation}\label{eq:Gij}
\begin{split}
\int  \, W^2 |\nabla_i \nabla_j G_m | &|\nabla_i \phi| |\nabla_j
\phi| \\ \leq &\; 4\int  \, W^2 |(\nabla^2 w)_{ij}| F_{ij} |\nabla_i
\phi|^2 + C \int  \, W^2 |(\nabla w)_{k j}| |\nabla_i F_{k j}|
|\nabla_i \phi|^2 \\& + C \int  \, W^2 |(\nabla w)_{k j}| |\nabla_i
F_{k j}| |\nabla_j \phi|^2 + \int  \, W^2 w_{k m} | \nabla_i
\nabla_j F_{k m}| |\nabla_i \phi|^2 \; .
\end{split}
\end{equation}
Applying $|\nabla w|\leq Ca\lambda$ (see (\ref{eq:wlambda})),
(\ref{eq:Fest}), and (\ref{eq:nabwsum}), the second term is bounded
by
\begin{equation}
\int \, W^2 |(\nabla w)_{k j}| |\nabla_i F_{k j}| |\nabla_i \phi|^2
\leq C a\ell^{-1} \int \, W^2 \lambda_{k j} F_{k j}^{1/2} |\nabla_i
\phi|^2
 \leq o(1) \cF(\phi)
\label{eq:secterm}
\end{equation}
if $a \ell_1 \ll \ell^4$, and $a \ll \ell$. The third term on the
r.h.s. of (\ref{eq:Gij}) can be bounded analogously. As for the
fourth term in (\ref{eq:Gij}), we note that, using (\ref{eq:Fest})
and $w\leq Ca\ell_1\lambda$ (by (\ref{eq:wlambda})), it is bounded
by
\begin{equation*}
\int  \, W^2 w_{k m} | \nabla_i \nabla_j F_{k m}||\nabla_i \phi|^2
\leq C a\ell_1 \ell^{-2} \int  \, W^2 \lambda_{k m} F_{km}^{1/2}
|\nabla_i \phi|^2
\end{equation*}
and the same estimate  holds as in (\ref{eq:secterm}) since $a\ell_1
\ell^{-2}\leq a\ell^{-1}$.

Finally we consider the first term on the r.h.s. of (\ref{eq:Gij}).
Using $|\nabla^2\om|\leq C\varrho\lambda$ from (\ref{eq:wlambda})
and the estimate (\ref{eq:nabw}) we obtain
\begin{equation*}
\begin{split}
\int  \, W^2 &|(\nabla^2 w)_{ij}| F_{ij} |\nabla_i \phi|^2 \leq C
\varrho \int  \, W^2 \lambda_{ij} F_{ij} |\nabla_i \phi|^2 \\ \leq
\; &C  \varrho \int  W^2\wt \chi_{ij} F_{ij}^{1/4} |\nabla_i
\nabla_j \phi|^2 + C \varrho \ell_1 \ell^{-3}
\int  \, W^2 \wt \chi_{ij} F_{ij}^{1/4} |\nabla_i \phi|^2 \\
\leq \; &C  \varrho \int  \, W^2 |\nabla_i \nabla_j \phi|^2 + C
\frac{\ell_1}{ N \ell^{3}} \, N \int  \, W^2  |\nabla_i \phi|^2 \; ,
\end{split}
\end{equation*}
where we also used (\ref{eq:overlapsum}). This completes the proof
because $\ell_1 \ell^{-3} N^{-1} =o(1)$, as $N\to \infty$.
\end{proof}

\begin{lemma}\label{lm:error3}
Assume $a \ll \ell_1 \ll \ell$ and $a \ll \ell_1^{3/2}$. Then
\begin{equation*}
\int \, W^2 \sum_{m,k,j}|\nabla_m q_{kj}| |\phi||\nabla_m \phi| \leq
o(1) \cF(\phi) \; ,
\end{equation*}
for $N \to \infty$.
\end{lemma}

\begin{proof}
By part iii) of Lemma \ref{lm:w&q} we have (with the summation
convention)
\begin{equation*}
\int \, W^2 |\nabla_m q_{kj}| |\phi||\nabla_m \phi| \leq C a
\ell_1^{-3} \int  \, W^2 \, \frac{\chi (|x_m -x_j| \leq \sfrac{3}{2}
\ell_1)}{|x_m - x_j|}  | \phi| |\nabla_m \phi| \, .
\end{equation*}
Next we apply a weighted Schwarz inequality:
\begin{equation*}
\begin{split}
C a \ell_1^{-3} \int  \, W^2 \, \frac{\chi (|x_m -x_j| \leq
\sfrac{3}{2}\ell_1 )}{|x_m -x_j|} & | \phi| |\nabla_m \phi| \\
\leq \; &C \a a \ell_1^{-3} \int  \, W^2 \, \frac{\chi (|x_m -x_j|
\leq \sfrac{3}{2}\ell_1)}{|x_m -x_j|} |\phi|^2 \\ &+ C \a^{-1} a
\ell_1^{-3} \int  \, W^2 \, \frac{\chi (|x_m -x_j| \leq
\sfrac{3}{2}\ell_1)}{|x_m - x_j|} |\nabla_m \phi|^2 \, .
\end{split}
\end{equation*}
In the first term we use (\ref{eq:2delta}) from Lemma \ref{lm:sob},
in the second one we estimate $\chi (|x_m -x_j| \leq
\sfrac{3}{2}\ell_1)\leq C\ell_1 |x_m-x_j|^{-1}$ and apply the usual
Hardy inequality (\ref{eq:hardyold}). In both cases we first have to
change $W$ to $W^{(j)}$ by (\ref{eq:Wk}) then back again to $W$ to
make the weight function $W$ independent of the $x_j$ variable. The
result is
\begin{equation*}
C a \ell_1^{-3} \int  \, W^2 \chi (|x_m -x_j| \leq
\sfrac{3}{2}\ell_1 )
 | \phi| |\nabla_m \phi| \leq
 C \left(\frac{\a a \ell_1^2}{\ell_1^3} + \frac{a \ell_1}{\ell_1^3 \a} \right) \cF(\phi) \; .
\end{equation*}
We choose $\a = \ell_1^{-1/2}$. Then
$$
    C a \ell_1^{-4} \int  \, W^2  \chi (|x_m -x_j| \leq \sfrac{3}{2}
    \ell_1)|\phi||\nabla_m \phi| \leq o(1) \cF(\phi)
$$
provided $a \ll \ell_1^{3/2}$.
\end{proof}

\begin{lemma}\label{lm:error4}
Assume $a \ll \ell_1  \ll \ell \ll 1$, $a \ll \ell^2$ and $a \ell_1
\ll \ell^4$. Then
\begin{equation*}
\int  \, W^2 \sum_m |\nabla_m \wt \Omega| |\phi||\nabla_m \phi| \leq
o(1) \cF(\phi)
\end{equation*}
as $N \to \infty$.
\end{lemma}

In order to prove this lemma we need the following estimates (we
recall the definition of $\theta_{km}$ from  (\ref{def:theta})).

\begin{lemma}\label{lm:err5}
Let $a \ll \ell_1 \ll \ell \ll 1$ and $a \ll \ell^2$, then for any
$q>0$ \be  \sum_{m,k} \int W^2 \sigma_{km} \, F_{km}^q
|\phi||\nabla_m\phi| \leq o(N \ell) \cF(\phi) \; , \label{mkphim}
\ee \be
  \sum_{m,k,j} \int W^2 \lambda_{kj} F^q_{kj} \theta_{km}
  |\phi||\nabla_m\phi| \leq o(1)\cF(\phi) \; .
\label{mkjphim} \ee
\end{lemma}
\begin{proof}
With a Schwarz inequality we obtain (using the summation convention)
\begin{equation*}
\begin{split}
    \int W^2 \sigma_{km} F^q_{km} |\phi||\nabla_m\phi| &\leq
    \int W^2 \sigma_{km} F^q_{km} \Big( \a |\phi|^2 +
     \a^{-1} |\nabla_m \phi|^2\Big)\\
    &\leq  \a \int W^2 \sigma_{km} |\phi|^2 + \a^{-1} a^{-1} \int W^2
    \lambda_{km} F_{km}^q |\nabla_m \phi|^2
\end{split}
\end{equation*}
where we used that $\sigma_{km} \leq a^{-1} \lambda_{km}$. Now we
apply (\ref{eq:sigmaest}) and (\ref{eq:nabw})  from Lemma
\ref{lemma:combined} to the r.h.s. of the last equation. We find
\begin{equation*}
\int W^2 \sigma_{km} F^q_{km} |\phi||\nabla_m\phi| \leq C_q\left( \a
\log N  + \a^{-1} a^{-1}+ \a^{-1} \ell_1 \ell^{-3} \right) \cF
(\phi) \; .
\end{equation*}
Eq. (\ref{mkphim}) follows choosing $\a = \ell^{-1-\delta}$ for
sufficiently small $\delta > 0$.

As for (\ref{mkjphim}) we have, from (\ref{eq:nabw}),
\begin{equation*}
\begin{split}
\int W^2 \lambda_{kj} F^q_{kj} \theta_{km} |\phi||\nabla_m\phi| \leq
\; & \int W^2 \lambda_{kj} F^q_{kj} \theta_{km}
  (\a^{-1} |\nabla_m\phi|^2 + \a |\phi|^2) \\ \leq \; &\int W^2 \wt \chi_{kj}
  F^{q/4}_{kj} \theta_{km} \\ &\times \left( \a^{-1} |\nabla_j \nabla_m \phi|^2 + \a^{-1} \ell_1
  \ell^{-3} |\nabla_m \phi|^2 + \a |\nabla_j \phi|^2 + \a \ell_1
  \ell^{-3} |\phi|^2 \right) \; .
\end{split}
\end{equation*}
Applying \eqref{eq:sob1} for the second and third terms, and
(\ref{eq:2delta}) for the last term, we get
\begin{equation*}
\begin{split}
\int W^2 &\lambda_{kj} F^q_{kj} \theta_{km}  |\phi||\nabla_m\phi|
\\ &\leq C \left(\a^{-1} + \a^{-1} \ell_1 \ell^{-3} \ell^2 |\log
\ell|^2 + \a \ell^2 |\log \ell|^2 + \a \ell_1  \ell^{-3} \ell^3
|\log \ell|^3 \right) \cF (\phi) \; .
\end{split}
\end{equation*}
Choosing for example $\a = \ell^{-1}$ we find (\ref{mkjphim}).
\end{proof}

\begin{proof}[Proof of Lemma \ref{lm:error4}]
By the definition of $\wt \Omega$ we find (using the summation
convention)
\begin{equation}\label{eq:nablaOmega}
\begin{split}
\int  \, W^2 |\nabla_m \wt \Omega| | \phi| |\nabla_m \phi| \leq \;
&\int \, W^2 \Big |\nabla_m \frac{\Omega_k}{2G_k}\Big | |\phi|
|\nabla_m \phi| \\ &+ \int  \, W^2 \Big |\nabla_m
\frac{\Omega_{jk}}{2G_j}\Big | |\phi| |\nabla_m \phi| + \int \, W^2
\Big |\nabla_m \Gamma \Big | |\phi| |\nabla_m \phi| \; .
\end{split}
\end{equation}
We begin considering the first term on the r.h.s. of
(\ref{eq:nablaOmega}).
\begin{equation}\label{eq:nablaOmegak}
\int \, W^2 {\Big |} \nabla_m \frac{\Omega_k}{2G_k}\Big| |\phi|
|\nabla_m \phi| \leq C \int \, W^2 \left( |\nabla_m \Omega_k | +
|\Omega_k | |\nabla_m G_k| \right) |\phi| |\nabla_m \phi| \; .
\end{equation}
{F}rom the definition of $\Omega_k$ (see (\ref{def:Omegak})), using
Lemma \ref{lm:nablaF} to bound the derivatives of $F_{mj}$, we find
\begin{equation}\label{eq:bound3}
\begin{split}
\int \, W^2 |\nabla_m \Omega_k | |\phi| |\nabla_m \phi| \leq \;& C
\ell^{-1} \int  \, W^2 |(\nabla^2 w)_{mj}| F_{mj}^{1/2} |\phi|
|\nabla_m \phi|  \\ &+ C\ell^{-2} \int  \, W^2 |(\nabla w)_{kj}|
F_{kj}^{1/2} (\theta_{mk} + \theta_{mj})
 |\phi|  |\nabla_m \phi|
\\ &+ C \ell^{-3}
\int  \, W^2 w_{kj} F_{kj}^{1/2} (\theta_{mk} + \theta_{mj}) |\phi|
|\nabla_m \phi| + O(\ell^{K-1-\eps}) \cF ( \phi) \; .
\end{split}
\end{equation}
We use that  $|(\nabla^2 w)_{mj}| \leq Ca \sigma_{mj}$, $|(\nabla
w)_{kj}| \leq Ca \lambda_{kj}$ and $w_{kj}\leq Ca\ell_1\lambda_{kj}$
(see (\ref{eq:wlambda})), and applying Lemma \ref{lm:err5} we obtain
\begin{equation}\label{eq:I1}
\int \, W^2 |\nabla_m \Omega_k | |\phi| |\nabla_m \phi| =o(1)
\cF(\phi) \; .
\end{equation}

For the other term in (\ref{eq:nablaOmegak}), using the definition
of $G_k$ (\ref{def:G}), we have
\begin{equation*}
\begin{split}
\int \, W^2 &|\Omega_k| \; |\nabla_m G_k | \; |\phi| \; |\nabla_m
\phi| \\ \leq \; &C \int  \, W^2 {\Bigg(} |(\nabla w)_{mj}| \;
|(\nabla w)_{mn}|\;  F_{mn} \;|\nabla_m F_{mj}| + |(\nabla w)_{kj}|
\; |(\nabla w)_{k m}| \;  F_{k m} \; |\nabla_k F_{kj}|
 \\ &+ |(\nabla w)_{mj}|\;  w_{m j} F_{m n} \; |\Delta_m F_{mj}|
+ |(\nabla w)_{k m}| \; w_{kj} F_{k m}\; |\Delta_k F_{kj}| \\
&+ |(\nabla w)_{kj}| \; w_{k n} \; |\nabla_m F_{k n}| \; |\nabla_k
F_{kj}| + w_{kj} w_{kn}\; |\Delta_k F_{kj}|\; |\nabla_m F_{kn} \; |
{\Bigg )} |\phi| |\nabla_m \phi| \; .
\end{split}
\end{equation*}
Taking as example the first term on the r.h.s. of the last equation,
we note that, unless $j = n$, the term is actually exponentially
small because, by Lemma \ref{lemma:noover}, $\| \wt \chi_{mj} \wt
\chi_{mn} F_{m n} \|_{\infty} \leq \exp (-c\ell^{-\e})$ if $j \neq
n$. Applying similar arguments to all other terms as well, and
applying Lemma \ref{lm:nablaF} to bound the derivatives of $F$, we
obtain
\begin{equation*}
\begin{split}
\int \, W^2 |\Omega_k| &|\nabla_m G_k | |\phi| |\nabla_m \phi| \\
\leq  &\, C  \int \, W^2 |\phi| |\nabla_m \phi| {\Big ( }\ell^{-1}
\, |(\nabla w)_{mn}|^2 F_{mn} + \left( \ell^{-2} \, |(\nabla
w)_{kj}| F_{kj}  + \ell^{-3} w_{kj}^2 F_{kj} \right) (\theta_{mk} +
\theta_{mj} ) {\Big )}
\\ &+ \left( O\left( e^{-c\ell^{-\e}}\right) + O(\ell^{K-2-\eps}) \right) \cF(\phi) \\
= \; &o(1)\cF(\phi)
\end{split}
\end{equation*}
following the estimate of the term (\ref{eq:bound3}) and using the
bounds $(\nabla w)^2\leq C a \sigma$ and $w^2\leq w$ from
(\ref{eq:wlambda}).

So, together with (\ref{eq:I1}) we find the following estimate for
the first term of (\ref{eq:nablaOmega}) \be \label{eq:I1+2} \int \,
W^2 \Big |\nabla_m \frac{\Omega_k}{2G_k} \Big| |\phi| |\nabla_m
\phi| \leq o(1)\cF(\phi) \; . \ee We consider next the second term
on the r.h.s. of (\ref{eq:nablaOmega}). Clearly we have
\begin{equation*}
\int  \, W^2 \Big |\nabla_m \frac{\Omega_{jk}}{2G_j}\Big| |\phi|
|\nabla_m \phi| \leq C \int  \, W^2 \left( |\nabla_m \Omega_{kj}| +
|\Omega_{kj}| |\nabla_m G_j| \right) |\phi| |\nabla_m \phi| \; .
\end{equation*}
Comparing $\Omega_{kj}$ with $\Omega_k$, it is clear that it only
remains to control the two terms
\begin{equation}\label{eq:II1}
\int \, W^2 w_{ij} |\Delta_k F_{ij}| |\nabla_m G_j| |\phi||\nabla_m
\phi|
\end{equation}
and
\begin{equation}\label{eq:II2}
\int \, W^2 |\nabla_m (w_{ij} \Delta_k F_{ij})| |\phi||\nabla_m
\phi| \; .
\end{equation}
We begin with (\ref{eq:II1}). The summation over $k$ is performed by
(\ref{eq:Fest}). We have
\begin{equation*}
\begin{split}
\int \, W^2 w_{ij} &|\Delta_k F_{ij}| |\nabla_m G_j| |\phi||\nabla_m
\phi| \leq \; C \ell^{-2}\int \, W^2 {\Big (} w_{ij} F_{ij}^{1/2}
|(\nabla w)_{jm}| F_{jm} \\ &+ w_{im} F_{im}^{1/2} |(\nabla w)_{m
n}| F_{m n} + w_{ij} F_{ij}^{1/2} w_{jn} |\nabla_m F_{jn}| {\Big )}
\, |\phi| |\nabla_m \phi| \; .
\end{split}
\end{equation*}
By Lemma \ref{lemma:noover} we know that, for example the first term
is exponentially small unless $i = m$. Similarly, apart from
exponentially small contributions, in the other two terms we can
consider only the case $i=n$. Hence we find
\begin{equation}\label{eq:II1bound}
\begin{split}
\int \, W^2 w_{ij} |\Delta_k F_{ij}| |\nabla_m G_j| |\phi||\nabla_m
\phi| \leq \; &C \ell^{-2}\int \, W^2 w_{mj} F_{mj}^{1/2} |(\nabla
w)_{mj}|\;  F_{mj}\;  |\phi|\; |\nabla_m \phi| \\ &+C \ell^{-2} \int
\, W^2 w_{jn}^2  F_{j n}^{1/2} |\nabla_m F_{jn}|\;  |\phi|\;
|\nabla_m \phi| \\ &+ O \left( \exp (-c \ell^{-\e})\right)
\cE(\phi)\; .
\end{split}
\end{equation}
On the right hand side  we estimate  $|\nabla w| \leq C a \lambda$
by (\ref{eq:wlambda}). In the first term, using $F\leq 1$, we have
\be
 C\ell^{-2} \int \, W^2 w_{mj}
F_{mj}^{1/2} |(\nabla w)_{mj}|\;  F_{mj}\;  |\phi|\; |\nabla_m \phi|
\leq Ca\ell^{-2}\int \, W^2 \lambda_{mj}\;  F_{mj}\; |\phi|\;
|\nabla_m \phi| \label{eq:wnabw} \ee that can be easily estimated by
$o(1) \cF(\phi)$, using the assumption $a \ll \ell^2$ and $a \ll
\ell_1 \ll \ell$.

In the second term of (\ref{eq:II1bound})  we estimate $w^2 \leq C
a^2 \lambda$. By Lemma \ref{lm:nablaF} we find
\begin{equation*}
\begin{split}
  C\ell^{-2} \int \, W^2 w_{jn}^2  & F_{j n}^{1/2}\;
  |\nabla_m F_{jn}|\;  |\phi|\;  |\nabla_m \phi|
   \leq Ca^2 \ell^{-2} \int W^2 \lambda_{jn} F_{jn}^{1/2}
(\theta_{mj} + \theta_{mn})
  |\phi|\; |\nabla_m \phi| = o(1) \cF(\phi) \; ,
\end{split}
\end{equation*}
where we also applied  Lemma \ref{lm:err5}.

We consider next (\ref{eq:II2}). We have
\begin{equation}\label{secc}
\begin{split}
\int \, W^2 |\nabla_m (w_{ij} \Delta_k F_{ij})| |\phi||\nabla_m
\phi| \leq \;& \int \, W^2 |(\nabla w)_{mj}| |\Delta_k F_{mj}|
|\phi||\nabla_m \phi| \\ &+ \int \, W^2 w_{ij} |\nabla_m \Delta_k
F_{ij}| |\phi||\nabla_m \phi|\; .
\end{split}
\end{equation}
After summation over $k$ using (\ref{eq:Fest}), the first term can
be treated as in (\ref{eq:wnabw}).

As for the second term in (\ref{secc}), we use that $w \leq C a
\ell_1 \lambda$ and Lemma \ref{lm:nablaF}. We find
$$
\int \, W^2 w_{ij} |\nabla_m \Delta_k F_{ij}| |\phi||\nabla_m \phi|
\leq a \ell_1 \ell^{-3}  \int W^2 \lambda_{ij} F^{1/2}_{ij}
(\theta_{mi} + \theta_{mj}) |\phi| |\nabla_m \phi|
$$
which can be estimated by $o(1)\cF (\phi)$ by Lemma \ref{lm:err5}.

Finally we control the third term on the r.h.s. of
(\ref{eq:nablaOmega}). After applying Lemma \ref{lemma:noover}
 we find, for any fixed $m$,
\be\label{eq:nablaGammanew}
\begin{split}
\Big | \nabla_m \Gamma \Big | \leq \;&C \Bigg[ |(\nabla^2w)_{mk}|\;
w_{jn} F_{mk}(\nabla_kF_{jn} + \nabla_m F_{jn})
+ | (\nabla w)_{ik}|\; |(\nabla w)_{jm}| \; F_{ik}\;  |\nabla_k F_{jm}| \\
& + | (\nabla w)_{ik}|\; w_{jn} \; |\nabla_m F_{ik}|\; |\nabla_k
F_{jn}|
+ | (\nabla w)_{ik}|\; w_{jn} \; |\nabla_m\nabla_k F_{jn}|\; F_{ik} \\
& + |(\nabla w)^2_{im}|\;  w_{jn} F_{im}^2\;  |\nabla_m F_{jn}|
+ |(\nabla w)^2_{km}| \; w_{jn} F_{km}^2 \; |\nabla_k F_{jn}|  \\
& + | (\nabla w)_{ik}|\; w_{ik} w_{jn} F_{ik}\; |\nabla_k F_{jn}| \;
|\nabla_m F_{ik}| + |(\nabla w)_{ik}|\; w_{jn}^2 F_{ik} \; |\nabla_k
F_{jn}|\;   |\nabla_m F_{jn}| \\ & + |(\nabla w)_{ik}|\; |(\nabla
w)_{jm}|\; w_{jm} F_{ik} F_{jm} \; |\nabla_k F_{jm}| + |(\nabla
w)_{mi}|\;  w_{jn} \; |\nabla_k  F_{mi}|\; | \nabla_k F_{jn}|
\\&+|(\nabla w)_{mi}| \; w_{mi} w_{jn} F_{mi} \; |\nabla_k F_{mi}|
|\nabla_k F_{jn}| + w_{in} w_{js}|\nabla_k \nabla_m F_{in}|
\; |\nabla_k F_{js}| \\
&+ w_{in}^2 w_{js} |\nabla_k F_{in}|\; |\nabla_m F_{in}|\; |\nabla_k
F_{js}| \Bigg] + O \left( e^{-c\ell^{-\e}}\right) \; .
\end{split}
\ee Using the estimates  (\ref{eq:Fest}), (\ref{eq:kfix}) and
(\ref{eq:kabfix}) from  Lemma \ref{lm:nablaF}  and the bounds
(\ref{eq:wlambda}), we obtain \be \label{eq:nabgam}
\begin{split}
\Big | \nabla_m \Gamma \Big | \leq C\Bigg[ &a \ell^{-1}
\sigma_{mk}F_{mk}^{1/2} + ( a \ell^{-2} + a \ell_1 \ell^{-3}+ a^2
\ell^{-3}) \lambda_{ik} F_{ik}^{1/2} (\theta_{im} + \theta_{km})
\\ &+ |(\nabla w)_{ik}|\; |(\nabla w)_{jm}| \; F_{ik} \; |\nabla_k F_{jm}|
+ |(\nabla w)_{ik}| \; w_{jn}^2 F_{ik}\;  |\nabla_m F_{jn}| \;
|\nabla_k F_{jn}|\Bigg] \, .
\end{split}
\ee The terms on the first line of the last equation can be bounded
using directly Lemma \ref{lm:err5}. Consider next the first term on
the second line. Its contribution, after two Schwarz inequalities is
given by
\begin{equation*}
\begin{split}
\int W^2 |(\nabla &w)_{ik}||(\nabla w)_{jm}| F_{ik} |\nabla_k
F_{jm}| |\phi||\nabla_m \phi| \\ \leq \, &\int W^2 \a|(\nabla
w)_{ik}|^2 \Big[ \; \beta |\phi|^2
+ \beta^{-1} |\nabla_m \phi|^2 \; \Big] \chi_{ik} F_{ik} \chi_{jm} |\nabla_k F_{jm}|\\
& + \int W^2  \a^{-1} |(\nabla w)_{jm}|^2  \Big[\; \gamma |\phi|^2 +
\gamma^{-1} |\nabla_m \phi|^2 \; \Big] \chi_{ik} F_{ik} \chi_{jm}
|\nabla_k F_{jm}|
\\ \leq \; & C\a \b a
\ell^{-1} \int W^2 \sigma_{ik} F_{ik} |\phi|^2 + C\a \b^{-1}
\ell^{-1}\int W^2 \lambda_{ik} F_{ik} |\nabla_m \phi|^2\\ &+
C\a^{-1}\gamma a \ell^{-1} \int W^2 \sigma_{jm} F_{jm}^{1/2}
|\phi|^2
+ C\a^{-1} \gamma^{-1} \ell^{-1} \int W^2 \lambda_{jm} F_{jm}^{1/2} |\nabla_m \phi|^2 \\
\leq \; &C \Big(\a\b a \ell^{-1} \log N + \a \b^{-1} \ell^{-1} (1 +
\ell_1 \ell^{-3})  + \a^{-1}\gamma a \ell^{-1} \log N \\ &+ \a^{-1}
\gamma^{-1} (1 + a \ell_1 \ell^{-3})\Big) \cF (\phi)\, .
\end{split}
\end{equation*}
Optimizing the choice of $\a$, $\b$ and $\gamma$,  we find that this
term is of order $o(1)\cF(\varphi)$ (using that $a \ll \ell^2$ and
$a \ell_1 \ll \ell^4$).

As for the last term on the r.h.s. of (\ref{eq:nabgam}) we use that
$w^2_{jn} \leq \chi_{jn}$. By Lemma \ref{lm:nablaF} and Lemma
\ref{lemma:combined} we get
\begin{equation*}
\begin{split}
\int W^2 |(\nabla w)_{ik}| \; w_{jn}^2 F_{ik} \; &|\nabla_m F_{jn}|
\; |\nabla_k F_{jn}|\; |\phi|\; |\nabla_m \phi| \\ \leq \; &C a
\ell^{-2} \int W^2
\lambda_{ik} F_{ik} (\a |\phi|^2 + \a^{-1} |\nabla_m \phi|^2) \\
\leq \; &C \left( \a N^{-1} (a\ell^{-2} + a \ell_1 \ell^{-5}) +
\a^{-1} (a\ell^{-2} + a \ell_1 \ell^{-5})\right) \cF (\phi) \\
\leq \; &C (a^{3/2} \ell^{-2} + a^{3/2} \ell_1 \ell^{-}) \cF (\phi)
= o(1) \cF (\phi).
\end{split}
\end{equation*}
{F}rom (\ref{eq:nabgam}), we find
$$
    \int W^2 \Big| \nabla_m \Gamma \Big | \, |\phi| \, |\nabla_m\phi|
    =o(1)\cF(\phi)\, .
$$
This completes the proof of Lemma \ref{lm:error4}.
\end{proof}

\section{A-priori Estimates on $U_{N,t}$}
\label{sec:apriori} \setcounter{equation}{0}

Recall that $\cH = L^2 (\Lambda, \rd x)$ is the one particle Hilbert
space, and that $\cH^{\otimes n}$ denotes the $n$-fold tensor
product $\cH^{\otimes n}$. We denote by $\cL^1_n := \cL^1
(\cH^{\otimes n})$ and by $\cK_n : = \cK (\cH^{\otimes n})$ the
space of trace class operators and the space of compact operators on
the Hilbert space $\cH^{\otimes n}$ respectively. We equip these
spaces with the trace norm,
 $\| \cdot \|_1:= \tr |\cdot |$,
 and  with the operator norm, $\|\cdot\|$, respectively.
 It is a well-known fact that
 $\cK_n^* = \cL_n^1$, that is
$\cL_n^1$ is the dual Banach space of $\cK_n$.

We define Sobolev-type norms on trace class operators. For
$\gamma^{(n)} \in \cL^1 (\cH^{\otimes n})$, we define the norm
\begin{equation*}
\| \gamma^{(n)} \|_{\cW_n}: = \left\{ \begin{array}{ll} \tr |S_1
\gamma^{(1)} S_1| &\quad \text{if } n=1 \\
\tr |S_1 S_2 \gamma^{(n)} S_1 S_2| &\quad \text{if }  n \geq 2
\end{array} \right.
\end{equation*}
where $S_j = (1 -\Delta_j)^{1/2}$, and the space $\cW_n = \{
\gamma^{(n)} \in \cL^1 (\cH^{\otimes n}) : \| \gamma^{(n)}
\|_{\cW_n} < \infty \}$.

Note that $\cW_n$ is the dual of the space
\[ \cA_n :=  \left\{ \begin{array}{ll}
\{ S_1 K^{(1)}  S_1 : K^{(1)} \in \cK_1 \} \quad &\text{if } n=1
\\ \{ S_1 S_2 K^{(n)} S_2 S_1 : K^{(n)} \in \cK_n
\} \quad &\text{if } n \geq 2 \end{array}\right.  \] equipped with
the norms
\[ \| T^{(1)} \|_{\cA_1} = \| S_1^{-1} T^{(1)} S_1^{-1} \|
\quad \text{and} \quad \| T^{(n)} \|_{\cA_n} := \| S_1^{-1} S_2^{-1}
T^{(n)} S_2^{-1} S_1^{-1} \| \quad \text{if } n \geq 2.
\]

\begin{lemma}\label{lm:BA} Suppose $U_{N,t} = \{ U_{N,t}^{(k)} \}_{k=1}^N$ is defined as in
(\ref{eq:UNt}), and that the assumptions of Theorem~\ref{thm:main}
are satisfied. Then there is $\nu >1$ large enough such that $\|
U_{N,t} \|_{H_-^{(\nu)}} \leq 1$. Moreover there exists a constant
$C >0$ such that
\begin{equation}\label{eq:b1}
\| U^{(k)}_{N,t} \|_{\cW_k} \leq C^k
\end{equation}
for any $t\in [0, T]$ and $k \leq N$.

Let $U_{\infty,t}$ be a limit point of  $U_{N,t}$  in $C ([0,T],
H_-^{(\nu)})$ with respect to the metric $\wh \rho$ (defined in
(\ref{def:varrho})). Then we also have $\| U_{\infty,t}
\|_{H_-^{(\nu)}} \leq 1$ and there is a version of $U_{\infty,t}$
such that \be\label{eq:limbound} \tr \, (1 -\Delta_1)
U^{(1)}_{\infty,t} \leq C \quad \quad \text{and } \quad
    \tr \, (1 -\Delta_i) ( 1-\Delta_j) U^{(k)}_{\infty,t}  \leq
    C^k\quad \text{if } \quad k \geq 2
\end{equation}
for every $k$, every $t \in [0,T]$, and every $i,j = 1, \dots ,k$
with $i \neq j$.
\end{lemma}
\begin{remark} This lemma does not yet prove the compactness
of the sequence $U_{N,t} \in C ([0,T], H_-^{(\nu)})$: to this end we
still need the equicontinuity of $U_{N,t}$, which will be proven in
Section \ref{sec:conti}. Here we prove that, if a limit
$U_{\infty,t} \in C ([0,T], H_-^{(\nu)})$ exists, then it satisfies
(\ref{eq:limbound}).
\end{remark}

\begin{proof} We have
\begin{equation*}
\| U^{(k)}_{N,t} \|^2 = \int \rd \bx_k \rd \bx'_k  |U^{(k)}_{N,t}
(\bx_k ; \bx'_k)|^2 \leq C_0^k \, \tr \, U^{(k)}_{N,t} \leq C_0^{2k}
\, .
\end{equation*}
Here we used that $U^{(k)}_{N,t}$ is the kernel of a positive
operator with trace bounded by $C_0^k$: this implies that also the
operator norm of $U_{N,t}^{(k)}$ is bounded by $C_0^k$. Choosing
$\nu > 2 C_0$ we immediately see that $\| U_{N,t} \|_{H_-^{(\nu)}}
\leq 1$. The bound $\| U_{\infty,t} \|_{H_-^{(\nu)}} \leq 1$ follows
because the norm can only drop when a weak limit is taken.

{F}rom Corollary \ref{cor:energy2} and (\ref{eq:WkW2}) we have
\begin{equation*}
\tr |S_1 U_{N,t}^{(1)} S_1| = \int \, |W^{(1)}|^2 \left(|\nabla_1
\phi_{N,t}|^2 + |\phi_{N,t}|^2 \right) \leq C \int \, W^2
\left(|\nabla_1 \phi_{N,t}|^2 + |\phi_{N,t}|^2 \right) \leq  C
\end{equation*}
and, for any $k \geq 2$,
\begin{equation}
\begin{split}
\tr \, |S_1 S_2 U^{(k)}_{N,t} S_1 S_2 | &= \int
 \, |W^{[k]}|^2 \Big[  |\nabla_1 \nabla_2 \phi_{N,t}|^2 +
 |\nabla_1 \phi_{N,t}|^2 + |\nabla_2 \phi_{N,t}|^2  +  | \phi_{N,t}|^2
\Big]\\
 &\leq C_0^k \int  \,
|W |^2 \Big[  |\nabla_1 \nabla_2 \phi_{N,t}|^2 +  |\nabla_1
\phi_{N,t}|^2 + |\nabla_2 \phi_{N,t}|^2  +  | \phi_{N,t}|^2
\Big] \\
& \leq C^k \;
\end{split}
\label{ssu}
\end{equation}
for some constant $ C$ independent of $k$. This proves
(\ref{eq:b1}).

For each fixed $t$, we can assume that $U_{N,t}$ converges to
$U_{\infty,t}$ in the weak$*$ topology of $H^{(\nu)}_-$ (otherwise
we choose an appropriate subsequence). This follows because $U_{N,t}
\in B_-^{(\nu)}$ (the unit ball of $H_-^{(\nu)}$) and because, on
$B_-^{(\nu)}$, the topology induced by the metric $\rho$ is
equivalent to the weak$*$ topology. This in particular implies that,
for every fixed $k$, $U^{(k)}_{N,t}$ converges to
$U^{(k)}_{\infty,t}$ in the weak topology of $L^2 (\Lambda^k \times
\Lambda^k)$. Note that $U_{\infty,t}^{(k)}$ is symmetric w.r.t.
permutations of the $k$ particles, because $U_{N,t}^{(k)}$ is
symmetric, and because the symmetry is clearly preserved in the
limiting process. By passing to a subsequence and using the
Alaoglu-Banach Theorem, we can assume that $U^{(k)}_{N,t}$ also
converges in the weak$*$ topology of $\cW_k$, and let $\wt
U_{\infty,t}$ be the limit. Testing these two limits against
operators with smooth kernels, one easily sees that $\wt
U^{(k)}_{\infty,t}=U^{(k)}_{\infty,t}$ as elements of $L^2
(\Lambda^k \times \Lambda^k)$, hence $U^{(k)}_{N,t}$ converges to
$U^{(k)}_{\infty,t}$ in both topologies. Then the estimate in
(\ref{eq:limbound}) for the version of $U_{\infty,t}^{(k)}$ given by
$\wt U^{(k)}_{\infty ,t}$ follow from (\ref{eq:b1}), from the
permutation symmetry of $U_{\infty,t}^{(k)}$, and from the fact that
the norm does not increase under weak$*$ limit.
\end{proof}

The next lemma will be used in Section \ref{sec:mainproof} to prove
part iii) of Theorem \ref{thm:main}.

\begin{lemma}\label{lm:trace}
Assume that $a \ll \ell_1 \ll \ell \ll 1$. Then, for any fixed $k$
and $t$, \begin{itemize} \item[i)] We have \begin{equation*} \tr
\left(U_{N,t}^{(k)} - \gamma_{N,t}^{(k)} \right) \to 0
\end{equation*}
as $N \to \infty$. \item[ii)] Assume that $U_{\infty,t} = \{
U_{\infty,t}^{(k)} \}_{k \geq 1} \in C([0,T], H_-^{(\nu)})$ is a
limit point of $U_{N,t} = \{ U_{N,t}^{(k)} \}_{k=1}^N$ with respect
to the metric $\wh \rho$. Then
\begin{equation}\label{eq:traceU12} \tr \, U_{\infty,t}^{(1)} =1 \,\quad \quad
\text{and } \quad \quad \tr \, U^{(2)}_{\infty,t} =1 \, .
\end{equation}
\end{itemize}
\end{lemma}
\begin{proof}
i) We have (with $\phi = \phi_{N,t} (\bx)$),
\begin{equation*}
\begin{split}
\tr& \left(U_{N,t}^{(k)} - \gamma_{N,t}^{(k)} \right) = \; \int \,
(W^2 - (W^{[k]})^2 ) |\phi|^2 \\ &= \int (1-\prod_{j=1}^k G_j )
\prod_{j=k+1}^N G_j |\phi|^2 + \sum_{m=k+1}^N \int
\prod_{j=k+1}^{m-1} G_j^{[k]} (G_m^{[k]} - G_m) \prod_{j=m+1}^N G_j
|\phi|^2
\end{split}
\end{equation*}
and hence, using (\ref{eq:GGk}),
\begin{equation*}
\begin{split}
\Big|\tr\left(U_{N,t}^{(k)} - \gamma_{N,t}^{(k)} \right)\Big| \leq
 &\; C^k \int \, (1 -\prod_{j=1}^k G_j) W^2 |\phi|^2 + C^k
\sum_{m=k+1}^N \int |G_m^{[k]} - G_m| W^2 |\phi|^2
\\ \leq &\; C^k \sum_{j \leq k} \sum_{m=1}^N \int W^2 w_{jm}
F_{jm} |\phi|^2 \\ &+ C^k \sum_{m,j >k}\sum_{r\leq k} \int W^2
w_{mj} F_{mj}^{[k]} \theta_{mr} |\phi|^2 + O(\ell^{K-\e}) .
\end{split}
\end{equation*}
So applying Lemma \ref{lemma:combined} for the first term  and part
(ii) of
 Lemma \ref{lm:sob} for the second term (with $U=\theta_{mr}$),
we find
\begin{equation*}
\begin{split}
\Big| \tr \Big(U_{N,t}^{(k)} - \gamma_{N,t}^{(k)} \Big) \Big| \leq
\; &C_k \left( a \ell_1 \sum_{j \leq k} \int W^2 |\nabla_j \phi|^2 +
a \ell_1 \ell^{-3} k \int W^2 |\phi|^2 \right)+ O(\ell^{K-\e})
\\ &+ C_k \left( a\ell_1 (\ell |\log \ell|)^2 \sum_{j>k}
\sum_{r<k} \int W^2 |\nabla_r \nabla_j \phi|^2 \right. \\ &\left. +
a \ell_1^2 \ell^{-3} (\ell |\log \ell|)^3 \sum_{m >k}\sum_{r \leq k}
\int W^2 (|\nabla_r \nabla_m \phi|^2 + |\nabla_r \phi|^2 + |\nabla_m
\phi|^2 + |\phi|^2) \right) \\ \leq \; &C_k ( a \ell_1 + a \ell_1^2
\ell^{-3} + \ell_1 \ell^2 |\log \ell|^2 + \ell_1^2 |\log \ell|^3) +
O(\ell^{K-\e}) = o(1)
\end{split}
\end{equation*}
for $N \to \infty$ using (\ref{apriori}).

ii) {F}rom part i) we have, for any fixed $k$ and $t$, $\tr\;
U^{(k)}_{N,t} \to 1$ as $N \to\infty$ using the normalization
(\ref{gammanorm}). By Lemma \ref{lm:BA}, the sequence
$U^{(1)}_{N,t}$ in $\cW_1$ is compact w.r.t. the weak$*$ topology of
$\cW_1$. By passing to a subsequence we can assume $U_{N,t}^{(1)}$
converges to $U_{\infty,t}^{(1)}$ in the weak$*$ topology of
$\cW_1$. In particular,
$$
  \tr \; U^{(1)}_{\infty,t} =\tr \; S_1^{-2} S_1 U^{(1)}_{\infty,t} S_1
  = \lim_{N\to\infty}\tr \; S_1^{-2} S_1 U^{(1)}_{N,t} S_1
  = \lim_{N\to\infty}\tr \; U^{(1)}_{N,t}=1
$$
since the operator $S_1^{-2}=(1-\Delta_1)^{-1}$ is compact on the
finite periodic box $\Lambda$. The proof of (\ref{eq:traceU12}) for
$U^{(2)}_{\infty,t}$ is similar.
\end{proof}

\section{Approximation of the Delta Function}\label{sec:approx}
\setcounter{equation}{0}

We consider the sequence of density matrices $U_{N,t} = \{
U_{N,t}^{(k)} \}_{k \geq 1}$. By viewing their kernels as
distributions, we will prove in Section \ref{sec:conti} that
$U_{N,t}$ is compact in $C ([0,T], H_-^{(\nu)})$. Assuming this
property for the moment, we denote a limit point by $U_{\infty,t}$.
Since for each fixed $t$ the kernel $U_{\infty,t}$ is defined only
as a distribution, the restriction of $U_{\infty,t}^{(k+1)}$ on the
diagonal $x_{k+1}=x_{k+1}'$, i.e., \be U_{\infty,t}^{(k+1)}(\bx_k,
x_{k+1}; \bx'_k, x_{k+1}) \label{rest} \ee has no a-priori meaning.
On the other hand, based upon Lemma \ref{lm:BA} we can view
$U_{N,t}$ and $U_{\infty,t}$ (or, more precisely, the families of
operators with kernels given by $U_{N,t}^{(k)}$ and
$U_{\infty,t}^{(k)}$) as elements of $\cW^{(\nu)}$, for every $t \in
[0,T]$.

We shall show in the next proposition that since $U_{\infty,t} \in
\cW^{(\nu)}$, the diagonal element (\ref{rest}) is well-defined. For
the rest of this section, we shall assume that $U^{(k+1)}$ is the
kernel of a density matrix with $\tr \; U^{(k+1)} < \infty$.

\begin{proposition}\label{prop:delta}
Suppose $\delta_{\beta} (x)$ is a radially symmetric function, with
$0 \leq \delta_{\beta} (x) \leq C \beta^{-3} \chi (|x| \leq \beta)$
and $\int \delta_{\beta} (x) \rd x = 1$ (for example $\delta_{\beta}
(x) = \beta^{-3} h (x/\beta)$, for a radially symmetric probability
density $h(x)$ supported in $\{ x : |x| \leq 1\}$). Then, for any
$J^{(k)} \in W^{1,\infty} (\Lambda^k \times \Lambda^k)$ and for any
smooth function $U^{(k+1)} (\bx_{k+1} ; \bx'_{k+1})$ corresponding
to a $(k+1)$-particle density matrix, we have, for any fixed $j \leq
k$,
\begin{equation}\label{eq:Uintbound}
\begin{split}
\Big| \int \rd {\bf x}_k \rd {\bf x}_k' &\rd x_{k+1} \rd x'_{k+1} \,
J^{(k)} (\bx_k ; \bx'_k) \, U^{(k+1)} (\bx_k, x_{k+1}; \bx'_k ,
x'_{k+1}) \\ &\times  \left(\delta_{\beta_1} (x'_{k+1} - x_{k+1})
\delta_{\beta_2} (x_j -x_{k+1}) - \delta (x'_{k+1} - x_{k+1}) \delta
(x_j - x_{k+1})\right) \Big|  \\ &\hspace{1.5cm} \leq C [\,
\|J\|_\infty + \|\nabla_j J \|_\infty \, \big ] \, (\beta_1 +
\sqrt{\beta_2} )\, \tr \, | S_j S_{k+1} U^{(k+1)} S_j S_{k+1}|\; .
\end{split}
\end{equation}
\end{proposition}
\begin{remark} {F}rom (\ref{eq:Uintbound}), using a standard
approximation argument, it also follows that
\begin{equation*}
\begin{split}
\Big| \int \rd {\bf x}_k \rd &{\bf x}_k' \rd x_{k+1} \rd x'_{k+1} \,
J^{(k)} (\bx_k ; \bx'_k) \, U^{(k+1)} (\bx_k, x_{k+1}; \bx'_k ,
x'_{k+1})\\  \times \big( \delta_{\beta_1} &(x'_{k+1} - x_{k+1})
\delta_{\beta_2} (x_j -x_{k+1}) - \delta_{\beta'_1}
(x'_{k+1} - x_{k+1}) \delta_{\beta'_2} (x_j - x_{k+1})\big)  \Big|  \\
\leq \; &C [\, \|J\|_\infty + \|\nabla_j J \|_\infty \, \big ] \,
\left(\beta_1 + \beta'_1 + \sqrt{\beta_2} + \sqrt{\beta_2'}
\right)\, \tr \, | S_j S_{k+1} U^{(k+1)} S_j S_{k+1}|
\end{split}
\end{equation*}
for any $U^{(k+1)}$ for which the r.h.s. is finite.
\end{remark}
\begin{proof}
It is enough to prove (\ref{eq:Uintbound}) for $U^{(k+1)}$ of the
form $U^{(k+1)} (\bx_{k+1} , \bx'_{k+1}) = f (\bx_{k+1})
\overline{f}(\bx'_{k+1})$ (in this section we use the notation
$\bx_{k+1} = (x_1 , \dots x_{k+1})$ and $\bx'_{k+1} = (x'_1, \dots
,x'_{k+1})$). Then, for a general density matrix $U^{(k+1)}$, the
proposition follows by considering the spectral decomposition:
\[
U^{(k+1)} (\bx_{k+1};\bx'_{k+1}) = \sum_n c_n f_n (\bx_{k+1})
\overline{f_n} (\bx_{k+1}')
\]
with  $c_n >0$ for all $n$ and $\| f_n\|_{L^2}=1$, and using the
fact that $\tr \;U^{(k+1)}= \sum c_n <\infty$. If $U^{(k+1)}
(\bx_{k+1}; \bx'_{k+1}) = f (\bx_{k+1}) \overline{f} (\bx'_{k+1})$
we can bound the l.h.s. of (\ref{eq:Uintbound}) by the sum
\begin{equation}\label{eq:delta0}
\begin{split}
\Big| &\int \rd {\bf x}_{k+1} \rd {\bf x}_{k+1}' \, J^{(k)} (\bx_k
;\bx'_k) \, \left(\delta_{\beta_1} (x'_{k+1} - x_{k+1}) - \delta
(x'_{k+1} - x_{k+1}) \right) \delta_{\beta_2} (x_j -x_{k+1}) \, f
(\bx_{k+1}) \overline{f} (\bx'_{k+1}) \Big|
\\ &+ \Big| \int \rd {\bf x}_{k+1} \rd {\bf x}_{k+1}' \, J^{(k)}
(\bx_k ; \bx'_k) \, \left( \delta_{\beta_2} (x_j -x_{k+1}) - \delta
(x_j - x_{k+1})\right) \delta (x'_{k+1} - x_{k+1}) f (\bx_{k+1})
\overline{f} (\bx'_{k+1}) \Big| \, .
\end{split}
\end{equation}
We next bound the first term by
\begin{equation}\label{eq:delta1}
\begin{split}
\Big| \int \rd {\bf x}_{k+1} \rd {\bf x}_{k+1}' \, J^{(k)} &(\bx_k ;
\bx'_k) \delta_{\beta_2} (x_j -x_{k+1}) 
\big(\delta_{\beta_1} (x'_{k+1} - x_{k+1}) - \delta (x'_{k+1} -
x_{k+1}) \big) \, f (\bx_{k+1}) \overline{f} (\bx'_{k+1}) \Big|
\\ \leq \; \| &J^{(k)} \|_{\infty} \, \int \rd \bx_{k+1} \rd \bx'_k \,
\delta_{\beta_2} (x_j -x_{k+1}) \, |f (\bx_{k+1})|
\\ &\times
\Big|f (\bx'_k , x_{k+1}) - \int \rd x'_{k+1} \,
\delta_{\beta_1} (x_{k+1} - x'_{k+1}) f (\bx'_k , x'_{k+1})\Big| \,
.
\end{split}
\end{equation}
A slight generalization of a standard Poincar\'e-type inequality
(see, e.g. Lemma 7.16 in \cite{GT}) yields that
\begin{equation}\label{pineq}
\begin{split}
\Big |  f (\bx'_k , x_{k+1}) - \int \rd x'_{k+1} \, &\delta_\beta
(x_{k+1} - x'_{k+1}) \, f (\bx'_k, x'_{k+1}) \Big| \le C \int_{|y|
\le \beta_1}
 \frac {|\nabla_{k+1} f (\bx'_k , x_{k+1} + y)|} { |y|^2  } \; \rd y
\end{split}
\end{equation} holds for any $\bx'_k$ and $x_{k+1}$. Inserting this
inequality on the r.h.s. of (\ref{eq:delta1}) and applying a Schwarz
inequality we get
\begin{equation*}
\begin{split}
\Big| \int \rd &{\bf x}_{k+1} \rd {\bf x}_{k+1}' \, J^{(k)} (\bx_k ;
\bx'_k) \, \left(\delta_{\beta_1} (x'_{k+1} - x_{k+1}) - \delta
(x'_{k+1} - x_{k+1}) \right)\delta_{\beta_2} (x_j -x_{k+1}) \, f
(\bx_{k+1})
\overline{f} (\bx'_{k+1}) \Big| \\
\leq \; & \| J^{(k)} \|_{\infty} \; \int \rd \bx_{k+1} \rd \bx'_k
 \rd y \, \frac{\chi (|y| \leq \beta_1)}{|y|^2} \,
\delta_{\beta_2} (x_j -x_{k+1}) \left( |f
(\bx_k ,x_{k+1})|^2 + |\nabla_{k+1} f (\bx'_k , x_{k+1} +y)|^2 \right) \\
\leq \; &C \| J^{(k)} \|_{\infty} \beta_1 \int \rd \bx_{k+1} \,
\delta_{\beta_2} (x_j -x_{k+1}) |f (\bx_k ,x_{k+1})|^2
\\ & + \| J^{(k)} \|_{\infty} \int \rd \bx'_k \rd x_{k+1}  \rd y \, \frac{\chi (|y| \leq
\beta_1)}{|y|^2} \, |\nabla_{k+1} f (\bx'_k , x_{k+1} +y)|^2
\end{split}
\end{equation*}
where in the first term we integrated over $\rd \bx'_k$ and over
$\rd y$ (the integration over $\rd y$ gives the factor $\beta_1$),
while in the second term we integrated over $\rd \bx_k$ (in
particular over $\rd x_j$). To control the first term on the r.h.s.
of the last equation we apply (\ref{eq:2delta}). In the second term
we shift the $x_{k+1}$ variable and then we integrate over $\rd y$.
We find, using that $U^{(k+1)} (\bx_{k+1} , \bx'_{k+1} )
=f(\bx_{k+1}) \overline{f} (\bx'_{k+1})$,
\begin{equation}\label{eq:delta8}
\begin{split}
\Big| \int \rd {\bf x}_{k+1} \rd {\bf x}_{k+1}' \, J^{(k)} (\bx_k ;
\bx'_k) &\left(\delta_{\beta_1} (x'_{k+1} - x_{k+1}) - \delta
(x'_{k+1} - x_{k+1}) \right) \delta_{\beta_2} (x_j -x_{k+1}) \, f
(\bx_{k+1})
\overline{f} (\bx'_{k+1}) \Big| \\
&\leq C \| J^{(k)} \|_{\infty} \, \beta_1 \, \tr \, (1 -
\Delta_{k+1}) ( 1 - \Delta_j) U^{(k+1)}
\end{split}
\end{equation}
uniformly in $\beta_2$. In order to control the second term on the
r.h.s. of (\ref{eq:delta0}), we use that
\begin{equation*}
\begin{split}
\int \rd {\bf x}_{k+1} &\rd {\bf x}_{k+1}' \, J^{(k)} (\bx_k ;
\bx'_k)  \delta (x'_{k+1} - x_{k+1}) \left( \delta_{\beta_2} (x_j
-x_{k+1}) - \delta (x_j - x_{k+1})\right)\, f (\bx_{k+1})
\overline{f} (\bx'_{k+1})   \\ &= \int \rd \bx_k \rd \bx'_k \rd
x_{k+1}  \, J^{(k)} (\bx_k ; \bx'_k) \left( \delta_{\beta_2} (x_j
-x_{k+1}) - \delta (x_j - x_{k+1})\right)\, f (\bx_{k}, x_{k+1})
\overline{f} (\bx'_{k}, x_{k+1})
\end{split}
\end{equation*}
and then we apply Lemma \ref{lm:sobsob} below, keeping all variables
fixed, apart from $x_j$ and $x_{k+1}$. This completes the proof of
the proposition.
\end{proof}

\begin{lemma}\label{lm:sobsob} Assume $\delta_{\beta}$ is defined as
in Proposition \ref{prop:delta}. Then we have
\begin{equation}\label{eq:intbound2}
\begin{split}
 \Big| \int \rd x \rd z &\rd x'\, J(x,x') (\delta_{\beta} (x-z) - \delta
(x -z)) f (x, z) \overline{f} (x', z) \Big|
\\
& \le C  \sqrt \beta \big [\, \|J \|_\infty + \|\nabla_x J \|_\infty
\, \big  ] \,  \Big |  \int \rd  x \rd z \overline{f}(x, z)
(1-\Delta_x) (1-\Delta_z) f(x, z)\Big |\; .
\end{split}
\end{equation}
\end{lemma}

\begin{proof} {F}rom the version \eqref{pineq} of the Poincar\'e inequality,
we can bound the left side of \eqref{eq:intbound2} by
\be\begin{split} C \int &\rd z \,  \rd x' \int \rd x \; 1(|x-z|\le
2\beta) \frac {|\nabla_x [J(x,x') f (x, z)]|} {|x-z|^2} |f (x',
z)| \\
&\le C \int \rd z  \, \rd x' \int \rd x \;  1(|x-z|\le2 \beta) \Big
[ \kappa \frac {|\nabla_x [J(x,x') f (x, z)]|^2} {|x-z|^2}
+\kappa^{-1} \frac {|f (x', z)|^2} {|x-z|^2} \Big ]\; .
\end{split}
\ee In the first term we drop the restriction $ 1(|x-z|\le2 \beta) $
 and apply the Hardy-type inequality (\ref{eq:hardy})
to the $z$ variable on the domain $\Lambda$. In the second term we
perform the $\rd x$ integration. Therefore the l.h.s. of
(\ref{eq:intbound2}) is controlled by
\begin{multline}
 C \int \rd z  \, \rd x' \int \rd x \;   \Big [ \kappa|
\nabla_x [J(x,x') \nabla_z f (x, z)]|^2 + \kappa | \nabla_x [J(x,x')
f (x, z)]|^2 + \kappa^{-1}\beta | f (x', z)|^2 \Big ] \; .
\end{multline}
The first two terms of the last expression are bounded by
$$
C \kappa  \big [\|J \|_\infty + \|\nabla_x J \|_\infty \big]^2
 \Big |  \int \overline{f} (x, z) (1-\Delta_x) (1-\Delta_z) f(x, z) \rd x \rd z\Big |\; ,
$$
while the last term is bounded by $C \beta \kappa^{-1} \| f\|_2^2$.
Optimizing the choice of $\kappa$, we obtain \eqref{eq:intbound2}.
\end{proof}

As a corollary of the proof of this lemma,  we can prove that for
all $0< \beta< 1$,
$$ \int \rd x \rd z \, \delta_{\beta} (x-z)  |f (x, z)|^2
\le C \,  \int \rd x \rd z \, \overline{f}(x, z) (1-\Delta_x)
(1-\Delta_z) f(x, z)
$$
with the constant $C$ independent of $\beta$. This reproves the
second part of Lemma \ref{lm:sob} (originally proven in Lemma 5.3 of
\cite{EY}) for the case of radially symmetric potential $U$ and
integration over a finite volume $\Lambda$.

\section{Proof of the Main Theorem}\label{sec:mainproof}
\setcounter{equation}{0}

The strategy for the proof of Theorem \ref{thm:main} is as follows.
First, in Section \ref{sec:regul}, we derive a hierarchy of equation
for the time evolution of the densities $U_{N,t}^{(k)}$, for large,
but finite $N$.

Using this Hierarchy we prove in Section \ref{sec:conti} that the
sequence $U_{N,t} = \{ U_{N,t}^{(k)} \}_{k =1}^N$ is equicontinuous
in $t$ with respect to the weak$*$ topology of $H_-^{(\nu)}$. This
result will be used to prove that $U_{N,t}$ is compact in the space
$C ([0,T], H_-^{(\nu)})$ with respect to the topology induced by the
metric $\wh \rho$ (defined in (\ref{def:varrho})).

In Section \ref{sec:compGamma} we prove then that any limit point of
$U_{N,t}$ in $C([0,T], H_-^{(\nu)})$ (w.r.t. the metric $\wh
\rho\,$) is also a limit point of the sequence $\Gamma_{N,t} = \{
\gamma_{N,t}^{(k)} \}_{k=1}^N$ (and that any limit point of
$\Gamma_{N,t}$ is also a limit point of $U_{N,t}$). This result
implies the compactness of $\Gamma_{N,t}$. Also part ii) and iii)
follow immediately from the fact that the limit points of $U_{N,t}$
and $\Gamma_{N,t}$ coincide, and from the results of Lemma
\ref{lm:BA} (in particular, the bound (\ref{eq:limbound})) and Lemma
\ref{lm:trace}. Having part ii), part iv) of Theorem \ref{thm:main}
follows easily from Proposition \ref{prop:delta}.

Finally, in Section \ref{sec:conv}, we complete the proof of Theorem
\ref{thm:main}, by proving part v): here we will start from the
hierarchy we are going to derive in Section \ref{sec:regul}, and we
will take the limit $N \to \infty$, using again Proposition
\ref{prop:delta}.

\subsection{Convergence to a Regularized Gross-Pitaevskii Hierarchy}
\label{sec:regul}

In this section we begin our analysis of the hierarchy of equations
governing the time evolution of the densities $U_{N,t}^{(k)}$, for
large $N$. Here and henceforth we use the pairing
\[
 \langle J^{(k)} , U^{(k)} \rangle = \int \rd \bx_k \rd \bx'_k
\, J^{(k)} (\bx_k; \bx'_k) U^{(k)} (\bx_k; \bx'_k) .
\]

The following lemma will be used in Sections \ref{sec:conti} and
\ref{sec:conv}, in order to prove the equicontinuity of
$U_{N,t}^{(k)}$ with respect to the weak$*$ topology of
$H^{(\nu)}_-$ and to prove that any limit point of $\Gamma_{N,t}$
satisfies the Gross-Pitaevskii Hierarchy (\ref{eq:GPH3}).

\begin{proposition}\label{lm:hierarchy}
Suppose $J^{(k)} \in W^{1,\infty} (\Lambda^k \times \Lambda^k)$. For
$0< \beta \le 1$ we choose a radially symmetric function
$\delta_{\beta} \in C_0^{\infty} (\bR^3)$ such that $0 \leq
\delta_{\beta} (y) \leq C \beta^{-3} \chi (|y| \leq \beta)$ and
$\int \rd y \, \delta_{\beta} (y) =1$. Then, for any $t
>0$,
\begin{equation}\label{eq:hierarchy}
\begin{split}
\langle J^{(k)}, U_{N,t}^{(k)}\rangle = \; &\langle J^{(k)},
U_{N,0}^{(k)} \rangle  -i \sum_{j=1}^k \int_0^t \rd s \, \int \rd
{\bf x}_k \rd {\bf x}_k' \, J^{(k)} ({\bf x}_k ; {\bf x}_k') \,
( -\Delta_j + \Delta_j') U^{(k)}_{N,s} ({\bf x}_k; {\bf x}_k') \\
&-8\pi i a_0 \sum_{j=1}^k \int_0^t \rd s \int \rd {\bf x}_k \rd {\bf
x}_k' J^{(k)} ({\bf x}_k ;{\bf x}_k')
\\ &\times \int \rd x_{k+1}
(\delta_{\beta} (x_j - x_{k+1}) - \delta_{\beta} (x_j' - x_{k+1}))
U_{N,s}^{(k+1)} ({\bf x}_k,x_{k+1}; {\bf x}_k',x_{k+1}) \\ &+ t \, O
(\beta^{1/2}) + t \, o (1)
\end{split}
\end{equation}
as $N \to \infty$.
\end{proposition}

\begin{proof}
Put $\phi_t = \phi_{N,t} (\bx)$. {F}rom
\[ i\partial_t \phi_t = L \phi_t + B \phi_t  \]
(see (\ref{def:L}) and (\ref{def:B})) it follows that
\begin{equation}\label{eq:conv1}
i\partial_t  U_{N,t}^{(k)}(\bx_k;\bx_k') = \int \rd {\bf x}_{N-k}
(W^{[k]})^2\Bigg[ \left( (L \phi_t)\overline{\phi'_t} -
(\overline{L' \phi'_t}) \phi_t\right) + (B-B') \phi_t
\overline{\phi'_t} \Bigg] \; ,
\end{equation}
where the superscript $'$ means that the coordinates $x_1,.., x_k$
are replaced by $x_1',.., x_k'$. Similarly, we will use $\nabla'_m:
= \nabla_{x_m'}$. In most cases we will not write out the arguments
of all functions fully, but it is understood that functions
$\phi_t$,  $B$, $\Omega$, etc.  with unspecified arguments depend on
the variables $x_1, x_2, \ldots, x_N$ and their primed versions,
$\phi_t'$,  $B'$, $\Omega'$, etc. depend on $x_1', x_2', \ldots,
x_k', x_{k+1}, \ldots , x_N$.

Next we note that for any fixed $k$
\begin{equation*}
\begin{split}
L  =\;&-\sum_{m=1}^{N} \Bigg( \Delta_m + 2 \nabla_m (\log W)
\nabla_m\Bigg)\\ = \; &-\sum_{m=1}^k \Delta_m -2 \sum_{m=1}^N\Big(
\nabla_m \log \frac{W}{W^{[k]}}\Big) \nabla_m - \sum_{m=k+1}^N
\Bigg(\Delta_m  + 2 \nabla_m (\log W^{[k]}) \nabla_m\Bigg)\;
\end{split}
\end{equation*}
since $W^{[k]}$ is independent of the first $k$ variables. The
contribution of the last term $\sum_m \Delta_m  + 2 \nabla_m (\log
W^{[k]}) \nabla_m $  cancels the analogous contribution from $L'$
 in (\ref{eq:conv1})  by
self adjointness on the space of the last $N-k$ variables. Moreover
we use the estimate (see (\ref{eq:wtOm})) \[ B- B' = (q_{jm} -
q'_{jm}) + (\wt \Omega - \wt \Omega') + O( \exp (-c\ell^{-\e})) \;
\] (with the summation convention). Thus, from (\ref{eq:conv1}) we
get
\begin{equation}\label{eq:conv2}
\begin{split}
i\partial_t &U_{N,t}^{(k)} ({\bf x}_k ; {\bf x}_k') \\ =
&\sum_{j=1}^k ( -\Delta_j + \Delta_j') U^{(k)}_{N,t} ({\bf x}_k ;
{\bf x}_k')
+ \sum_{m,j=1}^N\int \rd {\bf x}_{N-k} (W^{[k]})^2  (q_{jm}-q_{jm}') \phi_t \overline{\phi'_t}  \\
&-2 \sum_{m=1}^N \int \rd {\bf x}_{N-k} (W^{[k]})^2 \left[
 \Big(\nabla_m \log \frac{W}{W^{[k]}} \Big)\, (\nabla_{m} \phi_t)
\overline{\phi'_t} -\Big( \nabla'_m \log \frac{W'}{W^{[k]}}\Big) \,
(\nabla'_{m} \overline{\phi'_t} ) \phi_t \right] \\ &+\int \rd {\bf
x}_{N-k} (W^{[k]})^2 (\wt \Omega - \wt \Omega') \phi_t
\overline{\phi'_t} + O(\exp(-c\ell^{-\e})) \; .
\end{split}
\end{equation}
We can rewrite this equation in integral form
\begin{equation}\label{eq:intform}
\begin{split}
U_{N,t}^{(k)} ({\bf x}_k ; {\bf x}_k') = \, &U_{N,0}^{(k)} ({\bf
x}_k; {\bf x}'_k) -i \sum_{j=1}^k   \int_0^t \rd s \, ( -\Delta_j +
\Delta_j') U^{(k)}_{N,s} ({\bf x}_k ; {\bf x}_k')
\\ &-i\sum_{m,j=1}^N \int_0^t \rd s \int \rd {\bf x}_{N-k}
(W^{[k]})^2 (q_{jm} - q_{jm}') \phi_s
\overline{\phi'_s}  \\
&+2i \sum_{m=1}^N \int_0^t \rd s \int \rd {\bf x}_{N-k} (W^{[k]})^2
\\ &\hspace{1cm} \times\left[  \Big(\nabla_m  \log \frac{W}{W^{[k]}}
\Big)\, (\nabla_m \phi_s) \overline{\phi'_s} - \Big( \nabla'_m \log
\frac{W'}{W^{[k]}}\Big) \, (\nabla'_m \overline{\phi'_s} ) \phi_s
\right]\\ &-i \int_0^t \rd s \int \rd {\bf x}_{N-k} (W^{[k]})^2 (\wt
\Omega - \wt \Omega') \phi_s \overline{\phi'_s} + O(t
\exp(-c\ell^{-\e}))\; .
\end{split}
\end{equation}
Only the terms in the first three lines of last equation survive in
the limit $N \to \infty$. The other terms vanish as $N \to \infty$,
after they are tested against a function $J^{(k)}(\bx_k ; \bx_k')
\in W^{1,\infty} (\Lambda^k \times \Lambda^k)$: this is proven in
Lemma \ref{lm:conv1} and Lemma \ref{lm:conv2} in Section
\ref{sec:control}. Thus we are left with
\begin{equation}
\begin{split}
\langle J^{(k)}, U_{N,t}^{(k)}\rangle = \; &\langle J^{(k)},
U_{N,0}^{(k)}\rangle -i \sum_{j=1}^k \int_0^t \rd s \, \int \rd {\bf
x}_k \rd {\bf x}_k' \, J^{(k)} ( -\Delta_j + \Delta_j')
U^{(k)}_{N,s} \\ &-i \sum_{j,m=1}^N\int_0^t \rd s \int \rd {\bf x}_k
\rd {\bf x}_k' \rd {\bf x}_{N-k} \, J^{(k)} (W^{[k]})^2
(q_{jm}-q'_{jm}) \phi_s \overline{\phi_s}'
 + t\, o(1)
\end{split}
\label{la}
\end{equation}
as $N \to \infty$.

Next we note that $q_{jm} -q'_{jm} =0$ if $j,m
>k$. On the other hand if $j,m \leq k$, then we first use
 a Schwarz inequality to separate $\phi$ and $\phi'$, then we use
(\ref{eq:sob1}) with $U=\chi (|x_m -x_j| \leq 2 \ell_1) $ and we get
\begin{equation}\label{eq:jmleqk}
\begin{split}
\sum_{j,m \leq k} \int \rd {\bf x}_k & \rd\bx_k'\rd {\bf x}_{N-k}
|J^{(k)}|
(W^{[k]})^2 q_{mj} |\phi_s| \; |\overline{\phi}'_s | \\
\leq \; &a \ell_1^{-3}\| J^{(k)}\|_\infty \sum_{j,m \leq k} \int \rd
{\bf x}_k \rd\bx_k' \rd {\bf x}_{N-k} (W^{[k]})^2 \chi (|x_m -x_j|
\leq 2 \ell_1) \left(
|\phi_s|^2 + | \overline{\phi}'_s |^2 \right) \\
\leq \; & C k a \ell_1^{-3} \ell_1^2 \| J^{(k)}\|_\infty
\sum_{j=1}^k\int \rd {\bf x}_k \rd\bx_k' \rd {\bf x}_{N-k}
(W^{[k]})^2 \left(|\nabla_j \phi_s|^2 +  |\phi_s|^2 +
 | \overline{\phi}'_s |^2
\right) \\
\leq \; & C^k k  a \ell_1^{-1} \| J^{(k)}\|_\infty \sum_{j=1}^k
\Bigg[ \int \rd \bx \, W^2 (|\nabla_j \phi_s|^2 +  |\phi_s|^2) +
\int \rd \bx' \, (W')^2 |\phi'_s|^2\Bigg]\, .
\end{split}
\end{equation}
In the last line we used that the volume is finite and that
$W^{[k]}\leq C^k W'$ (by (\ref{lm:WkW})). Since $a \ll \ell_1$ by
assumption, the r.h.s. of (\ref{eq:jmleqk}) vanishes in the limit $N
\to \infty$.

The argument above implies that the factor $q_{jm} -q'_{jm}$ only
gives an important contribution to (\ref{la}) if $j \leq k < m$ or
$m \leq k < j$. Using the permutation symmetry of the wave function
we get
\begin{equation}\label{eq:conv6}
\begin{split}
\langle J^{(k)}, U_{N,t}^{(k)}\rangle  = \; & \langle J^{(k)},
U_{N,0}^{(k)}\rangle  -i \sum_{j=1}^k \int_0^t \rd s \, \int \rd
{\bf x}_k \rd {\bf x}_k' \, J^{(k)} ( -\Delta_j + \Delta_j')
U^{(k)}_{N,s} \\ &-2 i (N-k) \sum_{j=1}^k \int_0^t \rd s \int \rd
{\bf x}_k \rd {\bf x}_k' \rd {\bf x}_{N-k} \, J^{(k)}
(W^{[k]})^2 \\
&\times (q (x_j - x_{k+1}) - q (x_j' - x_{k+1})) \phi_s ({\bf x}_k,
{\bf x}_{N-k}) \overline{\phi_s} ({\bf x}_k', {\bf x}_{N-k})+ t \,
o(1)
\end{split}
\end{equation}
for $N \to \infty$.

In the next step we replace $W^{[k]}$ by $W^{[k+1]}$. To this end we
use that
\begin{equation}\label{eq:Wk3}
\begin{split}
(W^{[k]})^2 - (W^{[k+1]})^2 =\;& (G_{k+1}^{[k]} - 1) \prod_{m=k+2}^N
G^{[k]}_{m} \\ &+ \sum_{i=k+2}^N (G_i^{[k]} - G_i^{[k+1]})
\prod_{m=k+2}^{i-1} G_m^{[k+1]} \prod_{m=i+1}^N G_m^{[k]} \;
\end{split}
\end{equation} and estimate the effect of this difference in
(\ref{eq:conv6}). For the  first term in (\ref{eq:Wk3}) we use
\[
 G_{k+1}^{[k]} = 1 - \sum_{r > k+1} w_{k+1,r} F^{[k]}_{k+1,r}
\]
and its contribution to (\ref{eq:conv6})  for any fixed $j\leq k$
can be bounded by
\begin{equation*}
\begin{split}
(N-k) &\| J^{(k)}\|_\infty \sum_{r>k+1}\int \rd {\bf x}_k\rd\bx_k'
\rd {\bf x}_{N-k} \, w_{k+1 , r} F^{[k]}_{k+1,r}
\left(\prod_{m=k+2}^N G^{[k]}_{m}\right)
|q_{j,k+1}| |\phi_s| \; |\overline{\phi}'_s | \\
\leq \; &C a \ell_1^{-3}(N-k) \| J^{(k)}\|_\infty\sum_{r>k+1} \int
\rd {\bf x}_k\rd\bx_k'  \rd {\bf x}_{N-k} \, w_{k+1 , r}
F^{[k]}_{k+1,r} \left(\prod_{m=k+2}^N
G^{[k]}_{m}\right)  \\
&\times\chi (|x_j- x_{k+1}| \leq 2\ell_1)
\left( |\phi_s|^2 + |\overline{\phi}'_s |^2 \right) \\
\leq \; &C a \ell_1^{-1} (N-k)  \| J^{(k)}\|_\infty \sum_{r>k+1}
\int \rd {\bf x}_k \rd\bx_k'\rd {\bf x}_{N-k} w_{k+1 , r}
F^{[k]}_{k+1,r} \left(\prod_{m=k+2}^N G^{[k]}_{m}\right) \\
&\times \left( |\nabla_j \phi_s|^2 +  |\phi_s|^2 +
|\overline{\phi}'_s |^2 \right)\; ,
\end{split}
\end{equation*}
where  we used $\chi_{j, k+1}\leq C \ell_1^2 \lambda_{j, k+1}$ and
we applied the usual Hardy inequality (\ref{eq:hardyold}) for the
variable $x_j$. Next we use $w\leq Ca\ell_1\lambda$, $F^{[k]}_{k+1,
r}\leq CF_{k+1, r}$ and we estimate $\prod_m G^{[k]}_m \leq C^{k+1}
W^2$ (following from Lemma \ref{lm:WkW}),
 so that we can use (\ref{eq:nabw}) from Lemma \ref{lemma:combined} with respect to the
variable $x_{k+1}$. We find, for any fixed $j$,
\begin{equation}\label{eq:Gk}
\begin{split}
(N-k)  & \sum_{r>k+1}\int \rd {\bf x}_k \rd {\bf x}_{N-k} \, w_{k+1
, r} F^{[k]}_{k+1,r} \left(\prod_{m=k+2}^N G^{[k]}_{m}\right)
|q_{j,k+1}| |\phi_s| \; |  \overline{\phi}'_s |
\\ \leq \; &C  a \ell_1^{-1} (N-k)
\left\{ a \ell_1 \int \rd \bx \rd\bx_k'\rd  \, W^2
 \Big[|\nabla_{k+1} \nabla_j
\phi_s|^2 + |\nabla_{k+1} \phi_s|^2 + |\nabla_{k+1} \overline{\phi}'_s |^2 \Big] \right. \\
&\left. + a \ell_1^2 \ell^{-3} \int  \rd \bx \rd\bx_k' \; W^2 \Big[
|\nabla_j \phi_s|^2 + |\phi_s|^2 + |\overline{\phi}'_s |^2
\Big]\right\} \; .
\end{split}
\end{equation}
Finally we apply the estimate $W\leq C^k W'$ for the terms with
$\phi'$
 to obtain the energy norm.
Using Corollary \ref{cor:energy2} and the finiteness of the volume
we find (with $a\ell_1 \ll \ell^3$)
\begin{equation*}
\begin{split}
(N-k) \sum_{j < k} \sum_{r>k+1} \int &\rd {\bf x}_k \rd\bx_k'\rd
{\bf x}_{N-k} \, w_{k+1 , r} F^{[k]}_{k+1,r} \left(\prod_{m=k+2}^N
G^{[k]}_{m}\right) |q_{j,k+1}| |\phi_s| \;| \overline{\phi}'_s |
\\ &\leq C_ k( a + a \ell_1 \ell^{-3} ) \to 0 \qquad \mbox{as } \;
N\to\infty\; .
\end{split}
\end{equation*}

The other contributions coming from the second term of
(\ref{eq:Wk3}) can be bounded analogously. {F}rom (\ref{eq:conv6}),
and since
\begin{equation*}
\int \rd x_{k+2} \dots \rd x_N \, (W^{[k+1]})^2 \phi_s ({\bf x}_k,
{\bf x}_{N-k}) \overline{\phi_s} ({\bf x}_k', {\bf x}_{N-k}) =
U^{(k+1)}_{N,s} (\bx_k , x_{k+1} ; \bx'_k , x_{k+1})
\end{equation*}
it follows that
\begin{equation}\label{eq:conv7}
\begin{split}
\langle J^{(k)}, U_{N,t}^{(k)}\rangle  = \; &\langle J^{(k)},
 U_{N,0}^{(k)} \rangle
-i \sum_{j=1}^k \int_0^t \rd s \, \int \rd {\bf x}_k \rd {\bf x}_k'
\, J^{(k)}({\bf x}_k; {\bf x}_k') \, ( -\Delta_j + \Delta_j')
U^{(k)}_{N,s} ({\bf x}_k; {\bf x}_k') \\ &-2 i (N-k) \sum_{j=1}^k
\int_0^t \rd s \int \rd {\bf x}_k \rd
{\bf x}_k' \rd x_{k+1} \, J^{(k)}({\bf x}_k ;{\bf x}_k') \\
&\hspace{1cm} \times (q (x_j - x_{k+1}) - q (x_j' - x_{k+1}))
U_{N,s}^{(k+1)} ({\bf x}_k, x_{k+1} ; \bx'_k , x_{k+1}) \\ &+ t \,
o(1)\; .
\end{split}
\end{equation}
Note that the function $q$ depends on $N$, and that $Nq(x)$
approaches $4\pi a_0$ times a Dirac delta function as $N \to \infty$
(Lemma \ref{lm:w&q}). Using Proposition \ref{prop:delta} (with
$\beta_1 = 0$, that is $\delta_{\beta_1} (x) = \delta (x)$, and
$\beta_2 = 3/2 \ell_1$) we can replace $Nq(x)$ by $4\pi a_0$ times a
$\delta$-function. We can also replace $Nq(x)$ by $4\pi a_0$ times a
smoothed version of the $\delta$-function at some fixed length scale
$\beta$. More precisely, we choose a radially symmetric function
$\delta_{\beta} \in C_0^{\infty} (\bR^3)$ such that $0 \leq
\delta_{\beta} (x) \leq C \beta^{-3} \chi (|x| \leq \beta)$ and
$\int \rd x \, \delta_{\beta} (x) =1$. By the assumption $J^{(k)}
\in W^{1,\infty} (\Lambda^k \times \Lambda^k)$ and the a-priori
bound from Lemma \ref{lm:BA}, we have
\begin{multline*}
\Big| \int_0^t \rd s \int \rd \bx_k \rd \bx_k' \rd x_{k+1} \,
J^{(k)}(\bx_k;\bx_k') \left[ (N-k) q (x_j - x_{k+1}) - 4\pi a_0
\delta_{\beta} (x_j -x_{k+1})\right] \\ \times  U^{(k+1)}_{N,s}
(\bx_k , x_{k+1} ; \bx'_k , x_{k+1}) \Big| \leq C t (\ell_1^{1/2} +
\beta^{1/2})\; ,
\end{multline*}
for some constant $C$ depending on $k$ and $J^{(k)}$, but
independent of $N$ and $\beta$. Here we used  Lemma \ref{lm:sobsob}
twice, once for the function $q$ with lengthscale $\ell_1$ (Lemma
\ref{lm:w&q}) and once for the function $\delta_\beta$. {F}rom
(\ref{eq:conv7}) we find
\begin{equation*}
\begin{split}
\langle J^{(k)}, U_{N,t}^{(k)}\rangle = \; &\langle J^{(k)},
U_{N,0}^{(k)} \rangle  -i \sum_{j=1}^k \int_0^t \rd s \, \int \rd
{\bf x}_k \rd {\bf x}_k' \, J^{(k)}({\bf x}_k ; {\bf x}_k') \,
( -\Delta_j + \Delta_j') U^{(k)}_{N,s} ({\bf x}_k ; {\bf x}_k') \\
&-8\pi i a_0 \sum_{j=1}^k \int_0^t \rd s \int \rd {\bf x}_k \rd {\bf
x}_k' J^{(k)}({\bf x}_k ; {\bf x}_k') \\ &\times \int \rd x_{k+1}
(\delta_{\beta} (x_j - x_{k+1}) - \delta_{\beta} (x_j' - x_{k+1}))
U_{N,s}^{(k+1)} ({\bf x}_k,x_{k+1}; {\bf x}_k',x_{k+1})
\\ &+ t O (\beta^{1/2}) + t \, o(1)
\end{split}
\end{equation*}
as $N \to \infty$.
\end{proof}

\subsection{Compactness of $U_{N,t}$}
\label{sec:conti}

Using Proposition \ref{lm:hierarchy} we can now prove the
equicontinuity of $U_{N,t}^{(k)}$ in $t$ with respect to the metric
$\wh \rho$ on $C([0,T], H_-^{(\nu)})$, and thus the compactness of
$U_{N,t}$.

We recall the bound
\begin{equation}\label{eq:b2recall}
\| U_N \|_{C ([0, T] , H^{(\nu)}_-)} = \sup_{t \in [0,T]} \sum_{k
\geq 1} \nu^{-k} \| U_{N,t}^{(k)} \|_{2} \leq 1
\end{equation}
for some sufficiently large $\nu >1$ (Lemma \ref{lm:BA}) and we
recall the definition of the metric
 $\; \wh \rho\;$ from (\ref{def:varrho}). In
order to prove the compactness of the sequence $U_{N, t}$ with
respect to the topology induced on $C ([0,T], H^{(\nu)}_-)$ by the
metric $\wh \rho$, it is enough, by the Arzela-Ascoli Theorem, to
prove the equicontinuity of the sequence $U_{N,t}$.

\begin{lemma}\label{lm:equicont}
The sequence of families of density matrices $U_{N, t}=\{ U_{N,
t}^{(k)}\}_{k = 1}^N$, $N=1,2, \ldots$ satisfying
(\ref{eq:b2recall}) is equicontinuous on $H_-^{(\nu)}$ with respect
to the metric $\rho$ if and only if for every fixed $k \geq 1$, for
arbitrary $J^{(k)} \in W^{1,\infty} (\Lambda^k \times \Lambda^k)$
and for every $\eps >0$ there exists a $\delta
> 0$ such that
\begin{equation}\label{eq:equi02}
\Big| \langle J^{(k)} , U_{N,t}^{(k)} - U_{N,s}^{(k)} \rangle \Big|
\leq \eps
\end{equation}
whenever $|t -s| \leq \delta$.
\end{lemma}

\begin{proof}
Equicontinuity in the  metric $\rho$ means that, for any $\eps
> 0$ there exists $\delta
>0$ (independent of $N$), such that
\begin{equation}\label{eq:equi0}
\rho ( U_{N,t} , U_{N,s} ) = \sum_{j =1}^{\infty} 2^{-j} \Big|
\sum_{k \geq 1} \langle J_j^{(k)} , U_{N,t}^{(k)} - U_{N,s}^{(k)}
\rangle \Big| \leq \eps
\end{equation}
whenever $|t -s| \leq \delta$. Recall that $J_j = \{ J^{(k)}_j \}_{k
\geq 1}$, for $j \geq 1$ was chosen as a dense countable subset of
the unit ball of $H^{(\nu)}_+$. Using the uniform bound
(\ref{eq:b2recall}), one can
 approximate any given $J= (J^{(1)}, J^{(2)}, \ldots ) \in H^{(\nu)}_+$ by an
appropriate finite linear combinations of $J_j$ and  one can easily
prove that (\ref{eq:equi0}) implies (\ref{eq:equi02}).

On the other hand it is clear, by a standard approximation argument
(and because $W^{1,\infty} (\Lambda^k \times \Lambda^k)$ is dense in
$L^2 (\Lambda^k \times \Lambda^k)$), that (\ref{eq:equi02}) for all
$J^{(k)} \in W^{1,\infty} (\Lambda^k \times \Lambda^k)$ implies the
same bound for all $J^{(k)} \in L^2 (\Lambda^k \times \Lambda^k)$.
To prove that (\ref{eq:equi02}) for all $J^{(k)} \in L^2 (\Lambda^k
\times \Lambda^k)$ implies (\ref{eq:equi0}) one can proceed as
follow. Given $\eps >0$, we note that
\begin{equation*}
\begin{split}
\sum_{j > m} 2^{-j} \Big| \sum_{k \geq 1} \langle J_j^{(k)} ,
U_{N,t}^{(k)} - U_{N,s}^{(k)} \rangle \Big| &\leq \sum_{j
>m} 2^{-j} \| J_j \|_{H_+^{(\nu)}} \left( \| U_{N,t} \|_{H_-^{(\nu)}} +
\|U_{N,s} \|_{H_-^{(\nu)}} \right)\\ &\leq 2 \sum_{j >m} 2^{-j} \leq
\eps/3
\end{split}
\end{equation*}
if we choose $m$ sufficiently large. Hence
\begin{equation}\label{eq:eps1}
\sum_{j \geq 1} 2^{-j} \Big| \sum_{k \geq 1} \langle J_j^{(k)} ,
U_{N,t}^{(k)} - U_{N,s}^{(k)} \rangle \Big| \leq \eps/3 + \sum_{j
\leq m} 2^{-j} \Big| \sum_{k \geq 1} \langle J_j^{(k)} ,
U_{N,t}^{(k)} - U_{N,s}^{(k)} \rangle \Big|.
\end{equation}
Moreover we note that
\begin{equation*}
\begin{split}
\sum_{j \leq m} 2^{-j} \Big| \sum_{k > p} \langle J_j^{(k)} ,
U_{N,t}^{(k)} - U_{N,s}^{(k)} \rangle \Big| &\leq 2 \sup_{t \in
[0,T]} \|U_{N} (t)\|_{H^{(\nu)}_-} \sum_{j \leq m} 2^{-j}
\left(\sup_{k > p} \nu^{k}\|J_j^{(k)}\|_{k,+} \right) \\ &\leq \eps
/3
\end{split}
\end{equation*}
if we choose $p$ large enough, depending on $m$, because $\lim_{k
\to \infty} \nu^k \| J_j^{(k)} \|_{k,+} =0$ for every $j$. {F}rom
(\ref{eq:eps1}) we therefore have
\begin{equation}\label{eq:eps2}
\sum_{j \geq 1} 2^{-j} \Big| \sum_{k \geq 1} \langle J_j^{(k)} ,
U_{N,t}^{(k)} - U_{N,s}^{(k)} \rangle \Big| \leq 2\eps/3 + \sum_{j
\leq m} 2^{-j} \Big| \sum_{k \leq p} \langle J_j^{(k)} ,
U_{N,t}^{(k)} - U_{N,s}^{(k)} \rangle \Big|.
\end{equation}
Now, for every $j\leq m$ and $k \leq p$, we can find $\delta_{jk}
>0$ such that
\begin{equation*}
\Big| \langle J_j^{(k)} , U_{N,t}^{(k)} - U_{N,s}^{(k)} \rangle
\Big| \leq 2^{-k} \eps/3
\end{equation*}
if $|t -s| \leq \delta_{jk}$. Thus, for $|t-s| \leq \delta := \min
\{ \delta_{jk} \; : \;  j\leq m, \; k\leq p\}$, we have
\begin{equation*}
\sum_{j\geq 1} 2^{-j} \Big| \sum_{k \geq 1} \langle J_j^{(k)} ,
U_{N,t}^{(k)} - U_{N,s}^{(k)} \rangle \Big| \leq \eps.
\end{equation*}
This proves that (\ref{eq:equi02}) implies (\ref{eq:equi0}).
\end{proof}

\begin{lemma}\label{lm:compUN}
The sequence $U_{N,t} = \{ U^{(k)}_{N,t} \}_{k =1}^N \in C([0,T],
H_-^{(\nu)})$ is equicontinuous in $t$ with respect to the metric
$\rho$ (defined in \eqref{eq:rho}). In particular, by the
Arzela-Ascoli Theorem, the sequence $U_{N,t}$ is compact in
$C([0,T], H_-^{(\nu)})$ with respect to $\wh \rho$.
\end{lemma}

\begin{proof}
We prove (\ref{eq:equi02}) for all $J^{(k)} \in W^{1,\infty}
(\Lambda^k \times \Lambda^k)$. For such $J^{(k)}$ we can apply
Proposition \ref{lm:hierarchy} and we find
\begin{equation}\label{eq:equi1}
\begin{split}
\Big| \langle J^{(k)} , U_{N,t_1}^{(k)} &\rangle - \langle J^{(k)},
U_{N,t_2}^{(k)} \rangle \Big| \\ \leq \; &\sum_{j=1}^k
\int_{t_1}^{t_2} \rd s \, \Big| \langle J^{(k)} , (-\Delta_j +
\Delta_j') U^{(k)}_{N,s} \rangle \Big| \\ &+2 a_0 \sum_{j=1}^k
\int_{t_1}^{t_2} \rd s \Big| \int \rd {\bf x}_k \rd {\bf x}_k'
J^{(k)} ({\bf x}_k ;{\bf x}_k')
\\ &\times  \int \rd x_{k+1} (\delta_{\beta}
(x_j - x_{k+1}) - \delta_{\beta} (x_j' - x_{k+1})) U_{N,s}^{(k+1)}
({\bf x}_k,x_{k+1}; {\bf x}_k',x_{k+1})\Big| \\ &+ |t_1 - t_2| \, O
(\beta^{1/2}) + o(|t_1 - t_2|)\; .
\end{split}
\end{equation}
Next we bound
\[ \Big|\langle J^{(k)} , (-\Delta_j +
\Delta_j') U^{(k)}_{N,s} \rangle \Big| \leq \Big| \tr J^{(k)}
\Delta_j U^{(k)}_{N,s} \Big| + \Big| \tr \, J^{(k)} U^{(k)}_{N, s}
\Delta_j \Big| , \]
 and use
\begin{equation}\label{eq:JdeltaU}
\Big| \tr \, J^{(k)} \Delta_j U^{(k)}_{N,s} \Big| \leq \| S_j^{-1}
J^{(k)} S_j \| \| S_j^{-1} \Delta_j S_j^{-1} \| \tr \, \Big| S_j
U^{(k)}_{N,s} S_j \Big|\; .
\end{equation}
Using that $J^{(k)} \in W^{1,\infty} (\Lambda^k \times \Lambda^k)$
and the finiteness of the volume of $\Lambda$ we obtain
\[
\int \rd \bx_k \rd \bx'_k \, |\nabla_j J (\bx_k ; \bx'_k)|^2 <
\infty .
\]
It follows that $S_j J^{(k)}$ is a Hilbert-Schmidt operator, and
thus compact. This implies in particular that $S_j^{-1} J^{(k)} S_j$
is a bounded operator, and thus, by Lemma \ref{lm:BA}, the r.h.s. of
(\ref{eq:JdeltaU}) is bounded. Analogously, it also follows that
$|\tr \, J^{(k)} U^{(k)}_{N,s} \Delta_j |$ is bounded, uniformly in
$N$ and $s$.

As for the second term on the r.h.s. of (\ref{eq:equi1}), we use
that for fixed $\beta$, $\delta_{\beta}$ is bounded, $J^{(k)}(\bx_k
;\bx_k')$ is bounded and  $ \tr\; U^{(k+1)}_{N,s} \leq 1$: we obtain
\begin{multline}
\Big| \int \rd {\bf x}_k \rd {\bf x}_k' \rd x_{k+1} \, J^{(k)} ({\bf
x}_k ; {\bf x}_k') (\delta_{\beta} (x_j - x_{k+1}) - \delta_{\beta}
(x_j' - x_{k+1})) U_{N,s}^{(k+1)} ({\bf x}_k,x_{k+1}; {\bf
x}_k',x_{k+1})\Big| \leq C
\end{multline}
for a constant $C$, independent of $N$ and of $s$. It follows from
(\ref{eq:equi1}) that, for any fixed $J^{(k)} \in W^{1,\infty}
(\Lambda^k \times \Lambda^k)$,
\begin{equation*}
\Big| \langle J^{(k)} , U_{N,t_1}^{(k)} \rangle - \langle J^{(k)},
U_{N,t_2}^{(k)} \rangle \Big| \leq C |t_1 -t_2|
\end{equation*}
for a constant $C$ which depends on $k$ and on $J^{(k)}$, but is
independent of $N$ and of $t_1$ and $t_2$. This implies, by Lemma
\ref{lm:equicont}, that $U_N (t)$ is compact w.r.t. the topology
induced by the metric $\wh \rho$ on $C ([0,T], H_-^{(\nu)})$.
\end{proof}

\subsection{Compactness of $\Gamma_{N,t}$}
\label{sec:compGamma}

The aim of this section is to prove part i) of Theorem
\ref{thm:main}, stating the compactness of the sequence
$\Gamma_{N,t} = \{ \gamma^{(k)}_{N,t} \}_{k=1}^N$ in
$C([0,T],B_-^{(\nu)})$ with respect to the metric $\wh \rho$.

First of all, we note that, for any $\nu >2$,
\begin{equation*}
\| \Gamma_{N,t} \|_{H_-^{(\nu)}} = \sum_{k \geq 1} \nu^{-k} \|
\gamma_{N,t}^{(k)} \|_2 \leq 1
\end{equation*}
because $\| \gamma_{N,t}^{(k)} \|_2 \leq 1$ for every $k$. Thus
$\Gamma_{N,t} \in B_-^{(\nu)}$.

In order to prove the compactness of $\Gamma_{N,t}$ we use that, by
Lemma \ref{lm:compUN}, the sequence $U_{N,t}$ is compact. It only
remains to prove that limit points of $U_{N,t}$ are also limit
points of $\Gamma_{N,t}$. This is the aim of the next two lemmas.

Recall that
\[
\gamma^{(k)}_{N,t} ({\bf x}_k ; {\bf x}'_{k}) = \int \rd {\bf
x}_{N-k} \, W ({\bf x}_k , {\bf x}_{N-k}) W ({\bf x}'_k , {\bf
x}_{N-k}) \phi_{N,t} ({\bf x}_k , {\bf x}_{N-k})
\overline{\phi}_{N,t} ({\bf x}'_k , {\bf x}_{N-k})
\]
are the marginal densities corresponding to $\psi_{N,t} (x) = W(x)
\phi_{N,t} (x)$, while
\[ U^{(k)}_{N,t} ({\bf x}_k ; {\bf x}'_{k}) =
\int \rd {\bf x}_{N-k} \, W^{[k]} ({\bf x}_{N-k})^2 \phi_{N,t} ({\bf
x}_k , {\bf x}_{N-k}) \overline{\phi}_{N,t} ({\bf x}'_k , {\bf
x}_{N-k}). \] Recall that $W^{[k]} = W^{(1\dots k)}$ denotes the
wave function $W$ after removing its dependence on $\bx_k=(x_1,
\dots, x_k)$ (see (\ref{def:Wmanyk})).

\begin{lemma}\label{lm:L1bound}
Assume that $\Gamma_{N,t} = \{ \gamma_{N,t}^{(k)} \}_{k =1 }^N$ and
$U_{N,t} = \{ U_{N,t}^{(k)} \}_{k=1}^N$ are defined as above and
that the assumptions of Theorem \ref{thm:main} are satisfied. Then
we have, for every fixed $k \geq 1$ and $t \in [0,T]$,
\begin{equation*}
\int \rd \bx_k \rd \bx'_k \, \Big|\gamma^{(k)}_{N,t} (\bx_k ;
\bx'_k) - U_{N,t}^{(k)} (\bx_k ; \bx'_k)\Big| \to 0 \quad \quad
\text{as } N \to \infty.
\end{equation*}
\end{lemma}
\begin{proof}
In this proof  $k$ is considered fixed  and all constants may depend
on it. We have
\begin{equation*}
\begin{split}
\int \rd {\bf x}_k \rd {\bf x}'_k \, \Big| \gamma^{(k)}_{N,t} ({\bf
x}_k ; {\bf x}'_{k}) - U^{(k)}_{N,t} ({\bf x}_k ; {\bf x}'_{k})\Big|
\leq \; &\int \rd {\bf x}_k \rd {\bf x}'_k \rd {\bf x}_{N-k} \,
|\phi_{N,t} ({\bf x}_k , {\bf x}_{N-k})||\phi_{N,t} ({\bf x}'_k ,
{\bf x}_{N-k})| \\ &\times \Big| W ({\bf x}_k , {\bf x}_{N-k}) W
({\bf x}'_k , {\bf x}_{N-k}) - W^{[k]} ({\bf x}_{N-k})^2 \Big| \, .
\end{split}
\end{equation*}
Using the shorthand notation $\phi = \phi_{N,t} ({\bf x}_k , {\bf
x}_{N-k})$, $\phi' = \phi_{N,t} ({\bf x}'_k , {\bf x}_{N-k})$, and
$W=W({\bf x}_k , {\bf x}_{N-k})$, $W' = W ({\bf x}'_k , {\bf
x}_{N-k})$, and $W^{[k]} = W^{[k]} ({\bf x}_{N-k})$, we need to
bound
\begin{equation}\label{eq:diff1}
\int |W W' - (W^{[k]})^2||\phi||\phi'| \leq \int |W -W^{[k]}| W' |
\phi||\phi'| + \int |W' -W^{[k]}| W^{[k]} | \phi| |\phi'| \; .
\end{equation}
Using $W= \prod_{j=1}^N G_j^{1/2}$ and $W^{[k]} = \prod_{j=k+1}^N
(G_j^{[k]})^{1/2}$,
 we get
\begin{equation*}
\begin{split}
|W- W^{[k]}| \leq \; &\Big| 1 -\prod_{j=1}^k G_j^{1/2} \Big|
\prod_{j=k+1}^N G_j^{1/2} \\ &+ \sum_{m=k+1}^N
\left(\prod_{j=k+1}^{m-1} G_j^{1/2} \right)\, \Big|G_m^{1/2} -
(G_m^{[k]})^{1/2}\Big| \, \left( \prod_{j=m+1}^N (G_j^{[k]})^{1/2}
\right)\; .
\end{split}
\end{equation*}
Thus, the first term on the r.h.s. of (\ref{eq:diff1}) can be
bounded by
\begin{equation*}
\begin{split}
\int |W -W^{[k]}| W' | \phi||\phi'|  \leq \; &\int \Big|1
-\prod_{j=1}^k G_j^{1/2} \Big| \left(\prod_{j=k+1}^N G_j^{1/2}
\right) W' |\phi||\phi'| \\ &+ \sum_{m=k+1}^N \int
\left(\prod_{j=k+1}^{m-1} G_j^{1/2} \right) \Big|G_m^{1/2} -
(G_m^{[k]})^{1/2}\Big| \left(\prod_{j=m+1}^N (G_j^{[k]})^{1/2}
\right) W' |\phi||\phi'|\; .
\end{split}
\end{equation*}
Applying Schwarz inequality, with some  $\a >0$ which will be
specified later on, we find
\begin{equation}\label{eq:diff2}
\begin{split}
\int  \big |W -W^{[k]}\big | \; W' | \phi||\phi'| \leq \; &\a \int
\Big|1- \prod_{j=1}^k G_j^{1/2}\Big|^2 \left(\prod_{j=k+1}^N G_j
\right)|\phi|^2
\\ &+ \a \sum_{m=k+1}^N \int \left( \prod_{j=k+1}^{m-1} G_j \right) \Big|
G_m^{1/2} - (G_m^{[k]})^{1/2}\Big|^2 \left(\prod_{j=m+1}^N
G_j^{[k]} \right) |\phi|^2 \\ &+ C N \a^{-1} \int (W')^2 |\phi'|^2 \\
\leq \; &C \a \int \Big|1- \prod_{j=1}^k G_j^{1/2}\Big|^2 \, W^2
|\phi|^2
\\ &+ C\a \sum_{m=k+1}^N \int  \Big|G_m^{1/2}
- (G_m^{[k]})^{1/2}\Big|^2 W^2 |\phi|^2 + C N \a^{-1} \int (W')^2
|\phi'|^2\; .
\end{split}
\end{equation}
Next we note that, since $0< G_j \leq 1$, we have
\[ \Big|1-\prod_{j=1}^k G_j^{1/2} \Big| \leq \sum_{j \leq k} \sum_m w_{jm}
F_{jm}\;. \] Using (\ref{eq:nabw}) summed up for all $1\leq j\leq k$
with the help of (\ref{eq:overlapsum}), the first term on the r.h.s.
of (\ref{eq:diff2}) can be bounded by
\begin{equation*}
\begin{split}
\int \Big|1- \prod_{j=1}^k G_j^{1/2}\Big|^2 \, W^2 |\phi|^2 &\leq
\sum_{j \leq k} \sum_m \int w_{jm} F_{jm} W^2 |\phi|^2 \\ &\leq
a\ell_1 \sum_{j \leq k} \int W^2 |\nabla_j \phi|^2 + a \ell_1^2
\ell^{-3} k
\int W^2 |\phi|^2 \\
&\leq C_k (a \ell_1 + a \ell_1^2 \ell^{-3}) = o(N^{-1}
\ell^{\delta})
\end{split}
\end{equation*}
for $N \to \infty$ and for some $\delta >0$ (because $\ell_1 \ll
\ell^{3/2}$). As for the second term on the r.h.s. of
(\ref{eq:diff2}) we note that, since $G_j \geq c_1 >0$ (pointwise,
for $N$ large enough), we have for each fixed $m$
\begin{equation*}
\begin{split}
|G_m^{1/2} - (G_m^{[k]})^{1/2}| &\leq  C |G_m - G_m^{[k]}| \leq
\sum_{n\leq k} w_{mn} F_{mn} + \sum_{n >k} w_{mn} |F_{mn} -
F^{[k]}_{mn}| \\ &\leq \sum_{n\leq k} w_{mn} F_{mn} +
\ell^{-\e}\sum_{n >k} \sum_{r \leq k} w_{mn} \theta_{m r}
F^{[k]}_{mn} + O(\ell^{K-\e})
\end{split}
\end{equation*}
using (\ref{eq:FFk}). Thus, using Lemma \ref{lemma:noover} and the
fact that $w\leq \wt\chi$, we find,
\begin{equation*}
\begin{split}
\sum_{m>k} \int W^2 \Big|G_m^{1/2} - (G_m^{[k]})^{1/2}\Big|^2
|\phi|^2 \leq \; &C \sum_{m>k} \sum_{n_1, n_2 \leq k} \int W^2
w_{mn_1}w_{mn_2} F_{mn_1} F_{mn_2} |\phi|^2+ O(\ell^{K-\e})
 \\ &+ C \ell^{-2\e} \sum _{m, n_1,n_2 >k} \sum_{r_1, r_2 \leq k} \int
W^2 w_{mn_1} w_{mn_2} F^{[k]}_{mn_1} F^{[k]}_{mn_2} \theta_{mr_1}
\theta_{mr_2} |\phi|^2
 \\ \leq \; & C \sum_{m>k} \, \sum_{n\leq k} \int W^2 w_{mn}^2 F_{mn}
|\phi|^2 \\ &+ C \ell^{-2\e} \sum_{m, n >k} \sum_{r \leq k} \int W^2
w^2_{mn} F^{[k]}_{mn} \theta_{mr} |\phi|^2 + O(\ell^{K-\e}) \; .
\end{split}
\end{equation*}
Using the estimates (\ref{eq:wlambda}), Lemma \ref{lemma:combined}
and  (\ref{eq:sob1}) with $U=\theta_{mr}$, we obtain
\begin{equation*}
\begin{split}
\sum_{m=k+1}^N\int &W^2 \Big|G_m^{1/2} -(G_m^{[k]})^{1/2}\Big|^2
|\phi|^2 \\ \leq \; &C a^2 \sum_{n \leq k} \int W^2 |\nabla_n
\phi|^2 + C a^2 \ell_1 \ell^{-3} k \int W^2 |\phi|^2 \\ &+ C
\ell^{-2\e} (\ell |\log \ell|)^2 a^2 \sum_{m >k} \, \sum_{r\leq
k}\int W^2 ( |\nabla_m \nabla_r \phi|^2 + k |\nabla_m \phi|^2 )
\\& + C \ell^{-2\e} (\ell |\log \ell|)^2 a^2
\ell_1 \ell^{-3} N \sum_{r\leq k}\int W^2 ( |\nabla_r \phi|^2 +  |\phi|^2)+ O(\ell^{K-\e}) \\
\leq \; &C \Big(a^2 + a^2 \ell_1 \ell^{-3} +  a\ell^{-2\e} (\ell
|\log \ell|)^2 + a \ell_1 \ell^{-1-2\e} |\log \ell|^2\Big) +
O(\ell^{K-\e}) \\ = \; &o (N^{-1} \ell^{\delta} )
\end{split}
\end{equation*}
as $N \to \infty$, for some sufficiently small $\delta >0$ (here we
use (\ref{apriori}), that $ \ell_1 \ll \ell^{3/2}$, $a\ell_1 \ll
\ell^4$ and that $\e < 1/10$). So choosing $\a = N\ell^{-\delta}$
(for some $\delta$ small enough), we get, from (\ref{eq:diff2}),
\begin{equation*}
\int |W -W^{[k]}| W' | \phi||\phi'|  \leq o(1)
\end{equation*}
for $N \to \infty$. Analogously we can bound the second term on the
r.h.s. of (\ref{eq:diff1}).
\end{proof}

\begin{lemma}\label{lm:compGamma}
For any increasing subsequence $N_j$, the subsequence $U_{N_j,t}$
converges if and only if  $\Gamma_{N_j,t}$ converges (the
convergence is in $C([0,T], H_-^{(\nu)})$ with respect to the metric
$\wh \rho$). Moreover the limits coincide.
\end{lemma}
\begin{proof}
Suppose that, for a given subsequence $N_j$, $U_{N_j,t} \to
U_{\infty,t} = \{ U_{\infty,t}^{(k)} \}_{k \geq 1}$ as $j \to
\infty$, with respect to the metric $\wh \rho$. Then we prove that
$\Gamma_{N_j,t} \to U_{\infty,t}$ w.r.t. $\wh \rho\,$ for $j \to
\infty$. Since $\Gamma_{N_j,t} \in B_-^{(\nu)}$ (the unit ball of
$H_-^{(\nu)}$) it is enough to prove that for every fixed $k \geq 1$
and $t \in [0,T]$, and for all $J^{(k)}$ from any dense subset of
$L^2 (\Lambda^k \times \Lambda^k)$,
\begin{equation*}
\int \rd \bx_k \rd \bx'_k \, J^{(k)} (\bx_k ; \bx'_k) \left(
\gamma^{(k)}_{N_j,t} ( \bx_k ; \bx'_k) - U^{(k)}_{\infty,t} (\bx_k ;
\bx'_k) \right) \to 0
\end{equation*}
as $j \to \infty$. Assume now that $J^{(k)} \in W^{0, \infty}
(\Lambda^k \times \Lambda^k)$ (which is a dense subset of $L^2
(\Lambda^k \times \Lambda^k))$. Then we have
\begin{equation*}
\begin{split}
\Big| \int \rd \bx_k \rd \bx'_k \, J^{(k)} (\bx_k ; \bx'_k) (
&\gamma^{(k)}_{N_j,t} ( \bx_k ; \bx'_k) - U^{(k)}_{\infty,t} (\bx_k
; \bx'_k)) \Big|  \\ \leq \; &\int \rd \bx_k \rd \bx'_k \, \Big|
\gamma^{(k)}_{N_j,t} (\bx_k ; \bx'_k) - U^{(k)}_{N_j,t} (\bx_k ;
\bx'_k)\Big| \\ &+ \int \rd \bx_k \rd \bx'_k \, J^{(k)} (\bx_k ;
\bx'_k) \, \left( U^{(k)}_{N_j,t} (\bx_k ; \bx'_k) -
U^{(k)}_{\infty,t} (\bx_k ; \bx'_k) \right) \, .
\end{split}
\end{equation*}
The second term converges to zero, as $j \to \infty$, because we
assumed that $U_{N_j,t} \to U_{\infty,t}$ w.r.t. the metric $\wh
\rho$ as $j \to \infty$ (and because $U_{N_j,t} \in B_-^{(\nu)}$).
The first term converges to zero as $j \to \infty$, by Lemma
\ref{lm:L1bound}. This proves that $\Gamma_{N_j,t} \to U_{\infty,t}$
w.r.t. $\wh \rho$ as $j \to \infty$. Analogously one can prove that,
if $\Gamma_{N_j,t} \to \Gamma_{\infty,t}$, then also $U_{N_j,t} \to
\Gamma_{\infty,t}$.
\end{proof}

The last lemma, together with Lemma \ref{lm:compUN}, implies that
the sequence $\Gamma_{N,t}$ is compact, and completes the proof of
part i) of Theorem \ref{thm:main}.

Part ii) of Theorem \ref{thm:main} follows from Lemma
\ref{lm:compGamma} and from (\ref{eq:limbound}).

Part iii) of Theorem \ref{thm:main} follows on the other hand by
Lemmas \ref{lm:compGamma} and \ref{lm:trace}.

Part iv) follows by the remark after Proposition \ref{prop:delta},
using that $\Gamma_{\infty,t}$ satisfies the bound from part ii) of
Theorem \ref{thm:main}, $\tr \, (1 -\Delta_i) ( 1 -\Delta_j)
\gamma^{(k)}_{\infty ,t} \leq C^k$, for all $i \neq j$.

\subsection{Convergence to the Gross-Pitaevskii Hierarchy}
\label{sec:conv}

In this section we prove that the limit point $U_{\infty,t}$
satisfies the Gross-Pitaevskii Hierarchy, in the sense of
(\ref{eq:GPH3}).

\begin{proof}[Proof of part v) of Theorem \ref{thm:main}]
Using the assumption $J^{(k)} \in W^{2, \infty} (\Lambda^k \times
\Lambda^k)$ we can apply Proposition \ref{lm:hierarchy}. We find
\begin{equation}\label{eq:conv8}
\begin{split}
\langle J^{(k)}, U_{N,t}^{(k)}\rangle = \; &\langle J^{(k)},
U_{N,0}^{(k)} \rangle  -i \sum_{j=1}^k \int_0^t \rd s \, \int \rd
{\bf x}_k \rd {\bf x}_k' \, J^{(k)} ({\bf x}_k; {\bf x}_k') \,
( -\Delta_j + \Delta_j') U^{(k)}_{N,s} ({\bf x}_k; {\bf x}_k') \\
&-8\pi i a_0 \sum_{j=1}^k \int_0^t \rd s \int \rd {\bf x}_k \rd {\bf
x}_k' J^{(k)} ({\bf x}_k ;{\bf x}_k') \\ &\times \int \rd x_{k+1}
(\delta_{\beta} (x_j - x_{k+1}) - \delta_{\beta} (x_j' - x_{k+1}))
U_{N,s}^{(k+1)} ({\bf x}_k,x_{k+1}; {\bf x}_k',x_{k+1})
\\ &+ t O (\beta^{1/2}) + t o(1)
\end{split}
\end{equation}
as $N \to \infty$.

By passing to a subsequence and by Lemma \ref{lm:compGamma}, we can
assume here that $U_{N,t} \to \Gamma_{\infty,t} = \{
\gamma_{\infty,t}^{(k)} \}_{k \geq 1}$ with respect to the topology
induced by the metric $\wh \rho$ on $C ([0,T], H^{(\nu)}_-)$. This
in particular implies that $U_{N,t}^{(k)} \to \gamma_{\infty
,t}^{(k)}$ for every $k \geq 1$ and for every $t \in [0,T]$ w.r.t.
the weak topology of $L^2 (\Lambda^k \times \Lambda^k)$. Since
$J^{(k)} \in W^{2,\infty} (\Lambda^k \times \Lambda^k)$ and
$|\Lambda|<\infty$, we have $J^{(k)} (\bx_k ; \bx'_k) \in L^2
(\Lambda^k \times \Lambda^k)$. This means that
\begin{equation*}
\langle J^{(k)}, U_{N,t}^{(k)} - \gamma_{\infty,t}^{(k)} \rangle \to
0 \quad \text{and} \quad \langle J^{(k)}, U_{N,0}^{(k)} -
\gamma_{\infty,0}^{(k)} \rangle \to 0
\end{equation*}
as $N \to \infty$.

Moreover, since $J^{(k)} \in W^{2,\infty} (\Lambda^k \times
\Lambda^k)$, it also follows that $\Delta_{x_j} J^{(k)} (\bx_k;
\bx'_k)$ and $\Delta_{x'_j} J^{(k)} (\bx_k; \bx'_k)$ are elements of
$L^2 (\Lambda^k \times \Lambda^k)$. This and the fact that $U_{N,t}
\to \Gamma_{\infty,t}$ w.r.t. the topology induced by the metric
$\wh \rho$, imply that
\begin{equation*}
\sup_{s \in [0,T]} \sum_{j=1}^k \int \rd \bx_k \rd \bx'_k \,
\Delta_j J^{(k)} (\bx_k ; \bx'_k ) \left( U_{N,s}^{(k)} (\bx_k ;
\bx'_k) - \gamma_{\infty ,s}^{(k)} (\bx_k ; \bx'_k) \right) \to 0
\end{equation*}
as $N \to \infty$.

Finally we consider the limit $N \to \infty$ of the last term on the
r.h.s. of (\ref{eq:conv8}). {F}rom the proof of Proposition
\ref{prop:delta} (see in particular (\ref{eq:delta8})), we have
\begin{multline*}
\Big| \int \rd \bx_k \rd \bx'_k \rd x_{k+1} \rd x'_{k+1} J^{(k)}
(\bx_k ; \bx'_k) \delta_{\beta} (x_j - x_{k+1}) \\ \times \left(
\delta_{\eta} (x_{k+1} -x'_{k+1}) - \delta (x_{k+1} - x'_{k+1})
\right) U_{N,s}^{(k+1)} (\bx_k, x_{k+1} ; \bx'_k , x'_{k+1}) \,
\Big| \\  \leq C \, \eta \, \| J^{(k)}\|_\infty  \tr\; S_{k+1} S_j
U^{(k+1)}_{N,s} S_j S_{k+1}
\end{multline*}
for some finite constant $C$ independent of $s \in [0,T]$, of $N$,
and of $\beta$. In particular,
\begin{multline*}
\int \rd \bx_k \rd \bx'_k \rd x_{k+1} \, J^{(k)} (\bx_k ; \bx'_k)
\left(\delta_{\beta} (x_j - x_{k+1}) - \delta_{\beta} (x'_j -
x_{k+1}) \right) U_{N,s}^{(k+1)} (\bx_k , x_{k+1} ; \bx'_k, x_{k+1}
) \\ = \int \rd \bx_k \rd \bx'_k \rd x_{k+1}\rd x'_{k+1} \, J^{(k)}
(\bx_k ; \bx'_k) \left(\delta_{\beta} (x_j - x_{k+1}) -
\delta_{\beta} (x'_j -x_{k+1}) \right) \\ \times \delta_{\eta}
(x_{k+1} - x'_{k+1}) U_{N,s}^{(k+1)} (\bx_k , x_{k+1} ; \bx'_k,
x'_{k+1}) + O (\eta) \, .
\end{multline*}
On the r.h.s. of the last equation we can now let $N \to \infty$
keeping $\beta$ and $\eta$ fixed. By the assumptions on $J^{(k)}$
and by the choice of the functions $\delta_{\beta}$ and
$\delta_{\eta}$, it is easy to see that $J^{(k)} (\bx_k ; \bx'_k)
\delta_{\beta} (x_j -x_{k+1}) \delta_{\eta} (x_{k+1} -x'_{k+1})$ is
an element of $L^2 (\Lambda^{k+1} \times \Lambda^{k+1})$ for any
fixed $\beta$ and $\eta$. Hence
\begin{equation*}
\begin{split}
\int \rd \bx_k \rd \bx'_k \rd x_{k+1}\rd &x'_{k+1} \, J^{(k)} (\bx_k
; \bx'_k) \left(\delta_{\beta} (x_j - x_{k+1}) - \delta_{\beta}
(x'_j -x_{k+1}) \right) \delta_{\eta} (x_{k+1} - x'_{k+1}) \\
&\times \left( U_{N,s}^{(k+1)} (\bx_k , x_{k+1} ; \bx'_k, x'_{k+1})
- \gamma_{\infty,s}^{(k+1)} (\bx_k, x_{k+1} ; \bx'_k, x'_{k+1})
\right) \to 0
\end{split}
\end{equation*}
for $N \to \infty$, uniformly in $s$. So, after taking the limit $N
\to \infty$, (\ref{eq:conv8}) becomes
\begin{equation*}
\begin{split}
\langle J^{(k)} , \gamma_{\infty,t}^{(k)} \rangle = \; &\langle
J^{(k)}, \gamma_{\infty,0}^{(k)} \rangle -i \sum_{j=1}^k \int_0^t
\rd s \, \int \rd {\bf x}_k \rd {\bf x}_k' \, J^{(k)} ({\bf x}_k;
{\bf x}_k') \, ( -\Delta_j + \Delta_j') \gamma^{(k)}_{\infty,s}
({\bf x}_k; {\bf x}_k') \\
&-8\pi i a_0 \sum_{j=1}^k \int_0^t \rd s \int \rd {\bf x}_k \rd {\bf
x}_k' \rd x_{k+1} \rd x'_{k+1} \, J^{(k)} ({\bf x}_k ;{\bf x}_k')
\delta_{\eta} (x_{k+1} -x'_{k+1})
\\ &\hspace{1cm}
\times (\delta_{\beta} (x_j - x_{k+1}) - \delta_{\beta} (x_j' -
x_{k+1})) \gamma^{(k+1)}_{\infty,s} (\bx_{k}, x_{k+1}; \bx'_{k},
x_{k+1}') \\ &+ O(\beta^{1/2}) + O (\eta)
\end{split}
\end{equation*}
for any fixed $t$ and $k$. Next we apply Proposition
\ref{prop:delta} to replace $\delta_{\eta} (x_{k+1} -x'_{k+1})$ by
$\delta (x_{k+1} -x_{k+1}')$ and $\delta_{\beta} (x_j - x_{k+1})$
(respectively, $\delta_{\beta} (x'_j - x_{k+1})$) by $\delta (x_j -
x_{k+1})$ (respectively, by $\delta (x'_j - x_{k+1})$). The error
here is of order $\beta^{1/2} + \eta$. Hence, letting $\eta \to 0$
and $\beta \to 0$ we find
\begin{equation*}
\begin{split}
\langle J^{(k)} , \gamma_{\infty,t}^{(k)} \rangle = \; &\langle
J^{(k)}, \gamma_{\infty,0}^{(k)} \rangle -i \sum_{j=1}^k \int_0^t
\rd s \, \int \rd {\bf x}_k \rd {\bf x}_k' \, J^{(k)} ({\bf x}_k ;
{\bf x}_k') \, ( -\Delta_j + \Delta_j') \gamma^{(k)}_{\infty,s}
({\bf x}_k ; {\bf x}_k') \\
&-8\pi i a_0 \sum_{j=1}^k \int_0^t \rd s \int \rd {\bf x}_k \rd {\bf
x}_k' \rd x_{k+1} \, J^{(k)} ({\bf x}_k ; {\bf x}_k')
\\ &\times
(\delta (x_j - x_{k+1}) - \delta (x_j' - x_{k+1}))
\gamma^{(k+1)}_{\infty,s} (x_1, \dots, x_{k+1}; x_1',\dots, x_{k}',
x_{k+1})\; .
\end{split}
\end{equation*}
\end{proof}

\section{Control of Some Error Terms}\label{sec:control}
\setcounter{equation}{0}

In this section we use the notation \[ \cD (\phi) = \int \rd x \,
W^2 \,\left( |\nabla_1 \nabla_2 \phi|^2 + N^{-1} |\nabla_1^2 \phi|^2
+ |\nabla_1 \phi|^2 + |\phi|^2 \right). \] By Corollary
\ref{cor:energy2}, $\cD (\phi) \leq C$ if $(W \phi, \wt H W \phi)
\leq C_1 N$ and $(W \phi, \wt H^2 W \phi) \leq C_2 N^2$.

\begin{lemma}\label{lm:conv1}
Assume $a \ll \ell_1 \ll \ell \ll 1$, $a \ell_1 \ll \ell^4$, $\ell_1
\ll \ell^{3/2}$. Then for any fixed $k$ and any $J^{(k)}\in
W^{0,\infty}(\Lambda^k\times\Lambda^k)$, and for every wave function
$\phi$ symmetric w.r.t. permutation, we have
\begin{equation}\label{eq:conv4}
\Big| \int \rd {\bf x}_k \rd {\bf x}_k' \rd {\bf x}_{N-k} \,
J^{(k)}({\bf x}_k,{\bf x}_k') (W^{[k]})^2 ( \wt \Omega - \wt \Omega'
) \phi ({\bf x}_k , {\bf x}_{N-k}) \, \overline{\phi} ({\bf x}_k' ,
{\bf x}_{N-k})\Big| \leq o(1) \cD (\phi)
\end{equation}
as $N \to \infty$
\end{lemma}
\begin{proof}
We recall that
\begin{equation*}
\wt \Omega = \frac{\Omega_p}{2 G_p} + \frac{\Omega_{pj}}{2 G_j} +
\Gamma .
\end{equation*}
Using the definitions of $\Omega_p$, $\Omega_{pj}$ and $\Gamma$ from
the beginning of Section \ref{sec:energy} and the estimates from the
end of Appendix \ref{app:remove}, we find
\begin{equation}\label{eq:conv5}
\begin{split}
|\wt \Omega - \wt \Omega'|  \leq \; &C \ell^{-1} \sum_i\sum_{j \leq
k} \left( |(\nabla w)_{ij}| F_{ij}^{1/2} + |(\nabla w')_{ij}|
(F')_{ij}^{1/2} \right) \\ &+ C \ell^{-2} \sum_i \sum_{j \leq k}
\left( w_{ij} F_{ij}^{1/2} + w_{ij}' (F_{ij}')^{1/2} \right)
\\ &+ C \ell^{-1-\eps} \sum_{i,j > k} \sum_{r,r' =1}^k |(\nabla w)_{ij}|
(F_{ij}^{[k]})^{1/2} ( \theta_{ri} + \theta_{r' i}) \\ &+ C
\ell^{-2-\eps} \sum_{i,j > k} \sum_{r,r'=1}^k w_{ij} (\theta_{r i} +
\theta_{r' i}) (F_{ij}^{[k]})^{1/2} \\ &+ O(N^4 \ell^{K-3-\e})\,.
\end{split}
\end{equation}
This inequality relies on the fact that if both indices $i,j > k$
then there is a  cancellation between $F_{ij}$ and $F_{ij}'$.

Using $|w|\leq Ca\ell_1 \lambda \ll Ca\ell \lambda$ and $|\nabla
w|\leq Ca\lambda$, it is sufficient to control the first and the
third terms on the r.h.s of (\ref{eq:conv5}) with $|(\nabla
w)_{ij}|$ replaced with $Ca\lambda_{ij}$.

Inserting the first term on the r.h.s. of the last equation into
(\ref{eq:conv4}), taking the absolute value, and estimating
$J^{(k)}$ by its sup-norm, we find that this contribution is bounded
by
\begin{equation*}
\begin{split}
C a\ell^{-1} \sum_i \sum_{j \leq k} \int &\rd {\bf x}_k \rd \bx'_k
\rd {\bf x}_{N-k} (W^{[k]})^2 \lambda_{ij} F_{ij}^{1/2} \, |\phi
({\bf x}'_k, {\bf x}_{N-k})| \, |\phi ({\bf x}_k, {\bf x}_{N-k})|
\\ \leq \; &Ca \ell^{-1} \sum_i \sum_{j \leq k} \int \rd {\bf x}_k \rd \bx'_k
\rd {\bf x}_{N-k} (W^{[k]})^2 \lambda_{ij} F_{ij}^{1/2} \left(
|\phi|^2 + | \phi'|^2 \right)
\\  \leq \; &C\sum_i \sum_{j \leq k} \left[ a\ell^{-1} \int \rd {\bf x}_k
\rd {\bf x}_{N-k} (W^{[k]})^2 \wt \chi_{ij} F_{ij}^{1/8} \,
|\nabla_j \phi|^2  \right.
\\ &+ a\ell_1 \ell^{-4} \int \rd {\bf x}_k \rd \bx'_k \left. \rd {\bf
x}_{N-k} \, (W^{[k]})^2 \wt \chi_{ij} F_{ij}^{1/8} \left(
|\phi|^2 + |\phi'|^2 \right) \right] \\
\leq \; &C^k \sum_{j \leq k}  a \ell^{-1} \int \rd \bx \, W^2 \,
|\nabla_j \phi|^2 \\ &+ C^k k a \ell_1 \ell^{-4} \left( \int \rd \bx
\, W^2 \, |\phi|^2 + \int \rd {\bf x}' \, (W')^2 \, |\phi'|^2
\right) \\ \leq \; & C_k ( a \ell^{-1} + a\ell_1 \ell^{-4}) \cD
(\phi) \\ =\; &o(1)\cD (\phi)
\end{split}
\end{equation*}
because $a \ell_1 \ell^{-4} \ll 1$. Here $\bx' = (\bx'_k,
\bx_{N-k})$.

The contribution from the third term on the r.h.s. of
(\ref{eq:conv5}) can be bounded as follows. Also here we estimate
$J^{(k)}$ by its sup-norm.
\begin{equation*}
\begin{split}
C a\ell^{-1-\eps} &\sum_{ i,j>k} \sum_{r \leq k} \int \rd {\bf x}_k
\rd \bx'_k \rd {\bf x}_{N-k} (W^{[k]})^2 \lambda_{ij} F_{ij}^{[k]}
\theta_{ri} \, |\phi'|\,  |\phi| \\ \leq \; &Ca \ell^{-1-\eps}\sum_{
i,j>k} \sum_{r \leq k} \int \rd {\bf x}_k \rd \bx'_k \rd {\bf
x}_{N-k} \, (W^{[k]})^2\lambda_{ij} F_{ij}^{[k]} \theta_{ri} \left(
|\phi|^2 + |\phi'|^2 \right)
\\  \leq \;& C a \ell^{-1-\eps} \sum_{ i,j>k} \sum_{r \leq k}\int \rd \bx_k
\rd \bx'_k \rd {\bf x}_{N-k} (W^{[k]})^2 \wt \chi_{ij}
(F_{ij}^{[k]})^{1/4} \theta_{ir} \left( |\nabla_j \phi|^2  + |
\nabla_j \phi' |^2 \right)
\\ &+ C a \ell_1 \ell^{-4-\eps} \sum_{ i,j>k} \sum_{r \leq k} \int \rd \bx_k
\rd \bx'_k \rd {\bf x}_{N-k} \, (W^{[k]})^2 \wt \chi_{ij}
(F_{ij}^{[k]})^{1/4} \theta_{ir} \left( |\phi|^2 + | \phi'|^2
\right)\,.
\end{split}
\end{equation*}
Next we apply Lemma \ref{lm:sob}. We find
\begin{equation*}
\begin{split}
C \ell^{-1-\eps} &\sum_{i,j>k} \sum_{r \leq k} \int \rd {\bf x}_k
\rd {\bf x}_{N-k} (W^{[k]})^2 \lambda_{ij} F_{ij}^{[k]} \theta_{ri} |\phi' | |\phi| \\
\leq \; &C \sum_{i,j>k} \sum_{r \leq k} \left\{ a \ell^{-1-\eps}
\ell^2 |\log \ell|^2 \int \rd \bx (W^{[k]})^2 \wt \chi_{ij} \,
(F_{ij}^{[k]})^{1/4} \left( |\nabla_r \nabla_j \phi|^2 + |\nabla_j
\phi|^2 \right) \right. \\ & + a \ell^{-4-\eps} \ell_1 (\ell |\log
\ell|)^3 \\ &\left.\hspace{1cm}\times\int \rd \bx\, (W^{[k]})^2 \,
\wt \chi_{ij} \, (F_{ij}^{[k]})^{1/4} \, \left(|\nabla_i \nabla_r
\phi|^2 + |\nabla_r \phi|^2 +  |\nabla_i \phi|^2 + |\phi|^2 \right)
\right\}
\\ &+ C k \sum_{i,j >k} \left\{ a \ell^{-1-\eps} \ell^2 |\log
\ell|^2 \int \rd \bx'
 (W^{[k]})^2 \wt \chi_{ij} \, (F_{ij}^{[k]})^{1/4} \,
|\nabla_j \phi'|^2 \right. \\ &\left. +a \ell^{-4-\eps} \ell_1 (\ell
|\log \ell|)^3 \int \rd \bx' \, (W^{[k]})^2 \, \wt\chi_{ij} \,
(F_{ij}^{[k]})^{1/4} \, \left( | \nabla_i \phi' |^2 + |\phi' |^2
\right) \right\}
\\ \leq \; &C_k (\ell^{1-\eps} |\log \ell|^2 + \ell_1
\ell^{-1-\eps} |\log \ell|^3 )\cD (\phi) = o(1) \cD (\phi)\,.
\end{split}
\end{equation*}
We used the assumption $\ell_1 \ll \ell^{3/2}$ (and $0< \eps <
1/4$).
\end{proof}

\begin{lemma}\label{lm:conv2}
Assume $a \ll \ell_1 \ll \ell$, $\ell_1 \ll \ell^{3/2}$, $N \ell_1^3
\ll \ell^{9/4}$, and $N^2 \ell^5 \ll 1$. Then for any fixed $k$, any
$J^{(k)}\in W^{0,\infty}(\Lambda^k\times\Lambda^k)$, and any wave
function $\phi$ symmetric w.r.t. permutations, we have
\begin{equation}\label{97}
\Big| \sum_{m=1}^N \int \rd {\bf x}_k \rd {\bf x}_k' \rd {\bf
x}_{N-k} J^{(k)}({\bf x}_k, {\bf x}_k') \, (W^{[k]})^2  \left(
\nabla_m \log \frac{W}{W^{[k]}} \right) \, (\nabla_m \phi)  \,
\overline{\phi}' \Big| \leq o(1) \cD (\phi)
\end{equation}
as $N \to \infty$.
\end{lemma}

\begin{proof}
We have
\begin{equation*}
\nabla_m \log \frac{W}{W^{[k]}} = \frac{1}{2} \sum_{j \leq k}
\frac{\nabla_m G_j}{G_j} + \frac{1}{2} \sum_{j > k }\Bigg(
\frac{\nabla_m G_j}{G_j} - \frac{\nabla_m G_j^{[k]}}{G_j^{[k]}}
\Bigg).
\end{equation*}
Using that for each fixed $j$
\[ G_j = 1 -w_{jn} F_{jn} \quad \quad  \text{and }
\quad \quad G_j^{[k]} = 1 -w_{jn} F_{jn}^{[k]} \chi (n
> k) \qquad \mbox{for} \;\; j>k
\]
we have, for each fixed $j>k$ and $m$,
$$
   | G_j -G_j^{[k]}| \leq \sum_{n\leq k} w_{jn} F_{jn} + \sum_{n>k}  w_{jn} | F_{jn} - F^{[k]}_{jn}|
$$
and \begin{equation*}\begin{split}
   | \nabla_m(G_j -G_j^{[k]})| \leq \; &|(\nabla w)_{jm}|\; | F_{jm} - F^{[k]}_{jm}|
   + w_{jn} |\nabla_m( F_{jn} - F^{[k]}_{jn})|
   \\ &+  |(\nabla w)_{jm}| F_{jm} \chi(m\le k) \; .
\end{split}
\end{equation*}
We also use that $w_{jn}$ is supported on $|x_j-x_n|\leq 2\ell_1$,
and
\[
\chi ( |x_j -x_{n}| \leq 2 \ell_1) \, F_{jn} \, G_j = \chi ( |x_j
-x_{n}| \leq 2 \ell_1) \, F_{jn} \, (1 -w_{jn}F_{jn}) + O (\exp
(-c\ell^{-\e})). \]

Estimating $J^{(k)}$ by its sup-norm, we find
\begin{equation}\label{eq:conv3}
\begin{split}
\sum_m \Big| \int \rd {\bf x}_k &\rd {\bf x}_k' \rd {\bf x}_{N-k}
J^{(k)}({\bf x}_k, {\bf x}_k') \, (W^{[k]})^2 \left( \nabla_m \log
\frac{W}{W^{[k]}}\right) \, (\nabla_m \phi) \, \overline{\phi}' \Big| \\
\leq \; &\sum_m \sum_{j \leq k} \int \rd {\bf x}_k \rd \bx'_k \rd
{\bf x}_{N-k} \, (W^{[k]})^2 \left( | (\nabla w)_{j m}| F_{jm} +
w_{jn} |\nabla_m F_{jn}| \right) |\nabla_m \phi| \, |\phi'|
\\ &+ \sum_m \sum_{n ,j >k} \int \rd {\bf x}_k \rd \bx'_k
\rd {\bf x}_{N-k} \, (W^{[k]})^2 w_{jn} |\nabla_m ( F_{jn} -
F_{jn}^{[k]})| |\nabla_m \phi| \, |\phi'|\\
 &+ \sum_m \sum_{j >k} \int \rd {\bf x}_k \rd \bx'_k
\rd {\bf x}_{N-k} \, (W^{[k]})^2 |(\nabla w)_{jm}| | ( F_{jm} -
F_{jm}^{[k]})| |\nabla_m \phi| \, |\phi'| \\
&+ O(\exp (-c\ell^{-\e})) \cD (\phi) \; .
\end{split}
\end{equation}
Applying a Schwarz inequality we can bound the first term on the
r.h.s of the  last equation by (using the summation convention for
$m$)
\begin{equation*}
\begin{split}
\sum_{j \leq k} \int \rd {\bf x}_k \rd \bx'_k \rd {\bf x}_{N-k} \,
&(W^{[k]})^2 \, | (\nabla w)_{j m}| F_{jm} |\nabla_m \phi| | \phi' |
\\ \leq \; &a \a \sum_{j \leq k} \int \rd {\bf x}_k \rd \bx'_k \rd
{\bf x}_{N-k} \, (W^{[k]})^2 \, \lambda_{j m} F_{jm} |\nabla_m
\phi|^2 \\ &+ a \a^{-1}\sum_{j \leq k} \int \rd {\bf x}_k \rd \bx'_k
\rd {\bf x}_{N-k} \, (W^{[k]})^2 \, \lambda_{j m} F_{j m} |\phi'|^2.
\end{split}
\end{equation*}
To bound the first term on the r.h.s. of the last equation we use
Lemma \ref{lemma:combined}; to bound the second one, we estimate
$F_{jm} \leq 1$, and, for each fixed $m$, we integrate over the
variable $x_j$ using that $\phi'$ doesn't depend on it. We get
\begin{equation*}
\begin{split}
\sum_{j \leq k} \int \rd {\bf x}_k \rd \bx'_k \rd &{\bf x}_{N-k} \,
(W^{[k]})^2 \, | (\nabla w)_{j m}| F_{jm} |\nabla_m \phi| | \phi' |
\\ \leq \; & C a \a \sum_{j\leq k}\int \rd \bx_k \rd \bx'_k \rd
\bx_{N-k} \, (W^{[k]})^2 \wt \chi_{j m} F^{1/4}_{jm} |\nabla_j
\nabla_m \phi|^2 \\ &+ C a \a \ell_1 \ell^{-3} \sum_{j\leq k} \int
\rd \bx_k \rd \bx'_k \rd \bx_{N-k} (W^{[k]})^2 \wt \chi_{j m}
F^{1/4}_{jm}  |\nabla_m \phi|^2 \\ &+ C a \a^{-1} \ell_1 N \int \rd
\bx' \, (W^{[k]})^2 |\phi' |^2 \\ \leq \;& C_k \left(\a a N + \a a
\ell_1 \ell^{-3} N  + \a^{-1} a \ell_1 N \right) \cD (\phi) = o(1)
\cD (\phi)
\end{split}
\end{equation*}
where we used $\a = \ell^{3/2}$ and that $\ell_1 \ll \ell^{3/2}$.
Next we consider the second term on the r.h.s. of (\ref{eq:conv3}).
We have, applying first Lemma \ref{lemma:combined}, and then Eq.
(\ref{eq:sob1}) from Lemma \ref{lm:sob}, (using the summation
convention for the indices $m,n$)
\begin{equation*}
\begin{split}
\sum_{j \leq k} \int \rd &{\bf x}_k \rd \bx'_k \rd {\bf x}_{N-k} \,
(W^{[k]})^2  w_{jn} |\nabla_m F_{jn}| \,  |\nabla_m \phi| | \phi' |
 \\ \leq \; &Ca \ell_1 \sum_{j \leq k} \int \rd {\bf x}_k \rd
 \bx'_k \rd {\bf x}_{N-k} (W^{[k]})^2 \lambda_{jn} |\nabla_m F_{jn}|
\left( |\nabla_m \phi|^2 + |\phi'|^2 \right)
\\ \leq \; &Ca\ell_1\ell^{-1} \sum_{j \leq k} \int \rd \bx (W^{[k]})^2 F_{jn}^{1/2} \wt \chi_{jn} |\nabla_j
\nabla_m \phi|^2 \\ &+ Ca \ell_1^2 \ell^{-4} \sum_{j\leq k} \int \rd
\bx_k \rd \bx'_k \rd \bx_{N-k} (W^{[k]})^2 \theta_{jm} F_{jn}^{1/2}
\wt \chi_{jn} \left( |\nabla_m \phi|^2 + |\phi' |^2
\right) \\
\leq \; &C_k ( \ell_1\ell^{-1} + \ell_1^2 \ell^{-4} \ell^2 |\log
\ell|^2) \cD (\phi) = o(1) \cD (\phi)
\end{split}
\end{equation*}
for $N \to \infty$, because $\ell_1 \ll \ell$.

We consider now the second term on the r.h.s. of (\ref{eq:conv3}).
The bound (\ref{eq:FFkder}) implies that (here we are summing over
$n ,j > k$ and $ r \leq k$)
\begin{equation*}
\begin{split}
\int \rd {\bf x}_k \bx'_k \rd &{\bf x}_{N-k} \, (W^{[k]})^2 w_{jn}
|\nabla_m ( F_{jn} - F_{jn}^{[k]})| |\nabla_m \phi| |
\phi' | \\
\leq \; &C \ell^{-1-\e} \int \rd \bx_k \rd \bx'_k \rd \bx_{N-k} \,
(W^{[k]})^2 w_{jn} F_{jn}^{1/2} \theta_{jr} \left( \a |\nabla_m
\phi|^2 + \a^{-1} |\phi' |^2 \right) +O(\ell^{K-1-\e} )\\
\leq \; &C k \a a \ell_1 \ell^{-1-\e} \int \rd \bx \, (W^{[k]})^2
\wt \chi_{jn} F_{jn}^{1/8} |\nabla_j \nabla_m \phi|^2 \\ &+ C \a a
\ell_1^2 \ell^{-4-\e} \int \rd \bx \, (W^{[k]})^2 \wt
\chi_{jn} F_{jn}^{1/8} \theta_{jr} |\nabla_m \phi|^2 \\
&+ C\a^{-1} a \ell_1 \ell^{-1-\e} \int \rd \bx' \rd \bx_k \,
(W^{[k]})^2 \wt \chi_{jn} F_{jn}^{1/8} \theta_{jr} |\nabla_j \phi'|^2 \\
&+ C \a^{-1} a \ell_1^2 \ell^{-4-\e} \int \rd \bx' \rd \bx_k \,
(W^{[k]})^2 \wt \chi_{jn} F_{jn}^{1/8} \theta_{jr}|\phi' \Big|^2
+O(\ell^{K-1-\e} )\cD (\phi) \; .
\end{split}
\end{equation*}
Applying Lemma \ref{lm:sob}, we find
\begin{equation*}
\begin{split}
\int \rd {\bf x}_k &\rd {\bf x}_{N-k} \, (W^{[k]})^2 w_{jn}
|\nabla_m ( F_{jn} - F_{jn}^{[k]})| |\nabla_m \phi| |\phi'|\\
\leq \; &C\a a \ell_1 \ell^{-1-\e} k \int \rd \bx \, (W^{[k]})^2
|\nabla_j \nabla_m \phi|^2 \\ &+ C\a a \ell_1^2 \ell^{-4-\e} \ell^2
|\log \ell|^2 k \int \rd \bx \, (W^{[k]})^2 \left(
|\nabla_j \nabla_m \phi|^2 + N |\nabla_m \phi|^2 \right)  \\
&+ C k \a^{-1} a \ell_1 \ell^{-1-\e} \ell^3 |\log \ell|^3 \int \rd
\bx' (W^{[k]})^2 |\nabla_j \phi'|^2
\\ &+ C k\a^{-1} a \ell_1^2 \ell^{-4-\e} \ell^3 |\log \ell|^3
\int \rd \bx \, (W^{[k]})^2  \left(|\nabla_j \phi'|^2 + N | \phi'|^2
\right) + O(\ell^{K-1-\e} )\\ \leq \; &C_k |\log \ell|^3 (\a a\ell_1
\ell^{-1-\e} N^2 +\a a\ell_1^2 \ell^{-2-\e} N^2 + \a^{-1} a\ell_1
\ell^{2-\e}  N \\ &+
\a^{-1} a \ell_1^2 \ell^{-1-\e} N + \ell^{K-1-\e}) \cD (\phi) \\
\leq \; &C_k (\a N \ell_1 \ell^{-1-\e} + \a^{-1} \ell_1^2
\ell^{-1-\e})\cD(\phi)  \leq C_k \ell_1^{3/2} N^{1/2} \ell^{-1-\e}
\cD(\phi)= o(1) \cD(\phi)
\end{split}
\end{equation*}
for $N\to \infty$. Here we first used that $\ell_1 \ell^2 \ll
\ell_1^2 \ell^{-1}$ (as follows from $N^2 \ell^5 \ll 1$ and $a \ll
\ell_1 \ll \ell \ll 1$), then we optimized the choice of $\a$ and we
used that $N \ell_1^3 \ll \ell^{\, 9/4}$ (and that $\e < 1/8$).

Finally we consider the third term on the r.h.s. of
(\ref{eq:conv3}). Using (\ref{eq:FFk}) we find (using the summation
convention for $m$)
\begin{equation*}
\begin{split}
 \sum_{j > k} \int \rd {\bf x}_k \rd \bx'_k &\rd {\bf
x}_{N-k} \, (W^{[k]})^2 |(\nabla w)_{jm}| | ( F_{jm} -
F_{jm}^{[k]})| |\nabla_m \phi| \, |\phi'| \\ &\leq a \sum_{j >k}
\sum_{r \leq k} \int \rd \bx_k \rd \bx'_k \rd {\bf x}_{N-k} \,
(W^{[k]})^2 \lambda_{jm} \theta_{mr} (F_{jm}^{[k]})^{1/2} |\nabla_m
\phi| \, |\phi'|.
\end{split}
\end{equation*}
Next we apply a weighted Schwarz inequality. We find
\begin{equation*}
\begin{split}
\sum_{j > k} \int \rd {\bf x}_k \rd \bx'_k \rd &{\bf x}_{N-k} \,
(W^{[k]})^2 |(\nabla w)_{jm}| | ( F_{jm} - F_{jm}^{[k]})| |\nabla_m
\phi| \, |\phi'| \\  \leq \; &\a \sum_{j >k} \sum_{r \leq k} \int
\rd \bx_k \rd \bx'_k \rd {\bf x}_{N-k} \, (W^{[k]})^2
\, \theta_{mr} \wt \chi_{jm} (F_{jm}^{[k]})^{1/2} |\nabla_m \phi|^2 \\
&+ \a^{-1} a \sum_{j >k} \sum_{r \leq k} \int \rd \bx_k \rd \bx'_k
\rd {\bf x}_{N-k} \, (W^{[k]})^2 \sigma_{mj} \theta_{mr}
(F_{jm}^{[k]})^{1/2} |\phi'|^2
\end{split}
\end{equation*}
where we used that $a \lambda_{jm}^{1/2} \leq \wt \chi_{jm}$ and
that $\lambda_{jm}^{3/2} \leq \sigma_{mj}$ (recall the definition of
$\sigma_{jm}$ from (\ref{eq:lambda-sigma})). In the first term on
the r.h.s. of the last equation we can sum over the index $j$ (using
Lemma \ref{lemma:noover}) and then we can apply Lemma \ref{lm:sob}
for the integration over the variable $x_r$. As for the second term
on the r.h.s. of the last equation we can first integrate over the
variable $x_r$ (using that $\phi'$ is independent of $x_r$) and then
we can apply (\ref{eq:sigmaest}). We find (using the summation
convention for the index $m$),
\begin{equation*}
\begin{split}
\sum_{j > k} \int \rd {\bf x}_k \rd \bx'_k \rd {\bf x}_{N-k} \,
&(W^{[k]})^2 |(\nabla w)_{jm}| | ( F_{jm} - F_{jm}^{[k]})| |\nabla_m
\phi| \, |\phi'| \\ \leq \; &C \a \ell^2 |\log \ell|^2 \sum_{r \leq
k} \int \rd \bx \, (W^{[k]})^2 |\nabla_m \nabla_r \phi|^2 \\ &+ C
\a^{-1} a \ell^3 (\log N) \sum_{j >k} \int \rd \bx' \, (W^{[k]})^2
|\nabla_j \nabla_m \phi'|^2 \\ \leq \; &C_k (\log N) (\a N \ell^2 +
\a^{-1} \ell^3 N) \cD (\phi) = o(1) \cD(\phi)
\end{split}
\end{equation*}
where we chose $\a = \ell^{1/2}$ and we used that $N^2 \ell^5 \ll
1$. This completes the proof of the lemma.
\end{proof}




\appendix


\section{Properties of the Two Body Problem}
\label{app:2body} \setcounter{equation}{0}

 We set $r=|x|$ for
$x\in\bR^3$ and for a radial function $g(x)$ on $\bR^3$ we use the
notation $g(r)$ for the function  $r\mapsto  g(x)|_{|x|=r}$. Let
$g'(r):=\partial_rg(r)$ be the derivative of this radial function
$g$ w.r.t. $r$.

Consider the (unique) solution to the
 zero energy problem
\[
 \fh \omega =0, \quad  \fh := -\Delta  + (1/2) V\; ,
\]
on $\bR^3$, where $\omega$ is radially symmetric and satisfies the
condition $\lim_{r \to \infty} \omega (r) =1$. It is known that
$\om$ is always non-negative and by Harnack inequality actually
$\om>0$ since $V$ is regular.
 By the maximum principle, the $C^2$ norm of $\om$ is
also bounded. By writing $\omega(r)= f(r) / r$, the scattering
length, $a$,  of a potential $V$ is defined  by \be\label{sl} a :=
\lim_{r \to \infty} r - f(r)\, . \ee
 When $V$ has a
compact support, that is,  $V(x)=0$ for $|x| \ge R$, then $f(r)=
r-a$ for $r \ge R$. It is known that $a$ is bounded by
\be\label{abound} a \le \frac{1}{8 \pi} \int V \rd x \; \ee and also
$a< R$ since $f>0$.

\begin{lemma}\label{lm:E&g}
Let $V$ be a  smooth, positive, spherical symmetric function such
that $\supp V \subset \{ x \in \bR^3 : |x| \leq R \}$, for some $R
>0$. Let $\varrho := (8\pi)^{-1} (\| V \|_{\infty} + \|V
\|_{L_1})$ and denote the scattering length of $V$ by $a$. Let
$\varphi$ be the ground state of the Neumann problem \be\label{A2}
(-\Delta + \sfrac 1 2  V) \varphi = E \varphi \ee on the sphere of
radius $L$, with the boundary condition
$$
\varphi(L) =1, \quad   \partial_r \varphi (L)=0 \, .
$$
Then, if $L$ is sufficiently large, we have
\begin{itemize}
\item[i)] \be\label{1}
    E = 3 a L^{-3} (1 + O(1/L))\;  \quad \mbox{as} \quad L\to \infty \, .
\ee \item[ii)] There is a constant $0< c_0 < 1$ such that for all $
|x|\le L $ we have \be\label{2}
   c_0 \leq \varphi(x) \leq 1, \quad
   1-\varphi (x) \leq   \frac{C}{|x|}
\ee where $C$ is a constant, depending only on the potential.
\item[iii)] For all $ |x|\le L $ we also have the following
bounds:
\begin{equation}
|\nabla \varphi (x)| \leq  \frac{C \varrho}{|x|^2 + R^2}, \quad
\text{and}  \quad  |\nabla^2 \varphi (x)| \leq \frac{C \varrho
}{|x|^3+R^3} \, .\label{eq:gradbound}
\end{equation}
\end{itemize}
Note that, in ii) and iii), the constant $C$ is independent of $L$,
if $L$ is large enough.
\end{lemma}

\begin{proof}
i) We first prove an upper bound for the energy $E$. We write the
zero energy state $\omega (r)$ as $ \omega(r)= f(r) /r$ and let \be
\label{eq:psik} \psi (r) := \sin (k f(r))/r . \ee Note that
\[
\partial_r \psi (r) = \frac{ k f'(r) r \cos (kf(r)) - \sin (k
f(r))}{r^2} \; .
\]
Therefore, assuming $L \geq R$, $\psi$ satisfies Neumann boundary
conditions at $r=L$ if and only if \be k L = \tan (k (L -a))
\label{keq} \ee where $a$ is the scattering length defined in
\eqref{sl}. We define $k$ to be the smallest positive real number
satisfying equation (\ref{keq}), in particular $\psi>0$. It is easy
to check that there are constants $C_1$ and $C_2$ such that
\begin{equation}\label{eq:k2}
\frac{3a}{L^3} \Big(1 - \frac{C_1R}{ L}\Big ) \leq k^2 \leq
\frac{3a}{L^3}\Big(1 +\frac{C_1R}{ L}\Big )\; .
\end{equation}
Let $\tfh =  -\partial_r^2  + (1/2) V $, then $\tfh f = r \fh\om
=0$, i.e. \be\label{feq} -f''(r) + \frac{1}{2} V(r) f(r) =0 \; . \ee
In particular, $f$ is linear for $r>R$ and by the normalization
$\lim_{r\to\infty} \om (r)=1$, we have $f'(r)=1$ for $r>R$.
Moreover, $f$ satisfies  uniform bounds $f(r)\leq C(1+r)$,
$\|f'\|_\infty , \|f''\|_\infty\leq C$ by the boundedness of $\om$
and $f$ does not vanish.

With the help of the identity \be - \big [\sin (k f(r)) \big ] '' =
k^2 f' (r)^2 \sin (k f (r)) - k f'' (r) \cos (kf(r))\; ,
\label{eq:iden} \ee we compute
$$
\psi\fh \psi  =  k^2 \psi^2+
   k^2 (f'-1)
\Big(\frac{\sin kf}{r}\Big)^2  +
   \frac{1}{r^2}\Big[ -kf'' (\sin kf)(\cos kf) + \frac{1}{2} V
   (\sin kf)^2 \Big] \; .
$$
The last two terms are supported on $|x|\leq R$. Using the equation
(\ref{feq}) and the boundedness properties of $f$, we see that the
$O(k^2)$ terms cancel in the square bracket: \be \psi\fh \psi  = k^2
\psi^2+ r^{-2}\chi(r\leq R) O(k^3) \label{php} \ee and therefore
$$
    \langle \psi , \fh \psi \rangle =  k^2  \langle \psi , \psi \rangle  + O(k^3) \;,
$$
where $\langle \cdot, \cdot \rangle$ denotes the scalar product on
$L^2(|x|\leq L)$.

 Using the estimate
\be \sin k f(r) \ge C k f(r) \ge Ckr \label{lowerb} \ee $r\ge 2R$,
we also get the lower bound
$$
    \langle \psi , \psi \rangle \ge 4\pi \int_0^L (\sin k f(r))^2 \rd r
    \ge C \int_{2R}^L  k^2 r^2 \rd r = C k^2 L^3 \; .
$$
So we can use $\psi$ as a trial function for the upper bound on the
energy
\begin{equation*}
E \leq \frac{\langle \psi , \fh \psi \rangle}{\langle \psi , \psi
\rangle} \leq k^2 + O(kL^{-3}) = \frac{3a}{L^3} (1 + O(L^{-1})) \; .
\end{equation*}
This proves the upper bound in \eqref{1}. Before proving the lower
bound in \eqref{1}, we prove parts ii) and iii) of the lemma.

ii), iii) We now prove  \eqref{2} and \eqref{eq:gradbound}. We set
$m(r):= r\varphi (r)$ and we rewrite the eigenfunction equation
\eqref{A2} as \be \tfh m= (-\partial_r^2 + \sfrac 1 2  V) m = E m \;
. \ee
 Since $V$ has compact support, we can explicitly write $m (r)$
for $r \in (R,L]$. Let $\lambda:= \sqrt E$. {F}rom the boundary
conditions on $\varphi$, we have \be\label{A3} m(r)=
\lambda^{-1}\sin(\lambda (r-L))+ L \cos (\lambda (r-L)), \quad R < r
\le L\; . \ee
 {F}rom i) we have $\lambda \le
CL^{-3/2}$. This allows us to expand $m(r)$ up to $O(\lambda^4)$, we
find \be\label{A4} m(r) = r - a + O(1/L) \ee uniformly in $r$, for
$r \in (R,L]$. Using that $a < R$, this proves that $\varphi (r) =
m(r) /r$ satisfies the inequalities in \eqref{2} for $r \in [R,L]$,
if $L$ is large enough. The properties (\ref{eq:gradbound}) on the
interval $[R,L]$ can be easily proved using the explicit formula
(\ref{A3}) (and using that $a <R$).

We now consider the region $r \le R$. {F}rom the Harnack inequality,
the ratio between the supremum and the infimum of $\varphi$ in a
given ball is bounded. Since $\varphi$  satisfies \eqref{2} for $r
\le R$, it follows that $\varphi(r)$ is bounded away from zero for
$r \le R$. This proves the lower bound of $\varphi(r)$ in \eqref{2}.

To prove the bound $\varphi(x) \le 1$ for $r \le R$, consider the
ball $B_R$ of radius $R$ about the origin. On the boundary of this
domain $\varphi(x)\leq 1-\delta$ with some $\delta>0$, uniformly for
all sufficiently large $L$ from (\ref{A4}). Inside this ball,
$$
  -\Delta \varphi \leq -\Delta \varphi + \frac{1}{2}V\varphi = E\varphi \leq CL^{-3}
$$
since $\varphi$ is bounded by Harnack principle, i.e. $ -\Delta
(\varphi+ CL^{-3} x^2) \leq 0$\; . By maximum principle, $\varphi(x)
\leq 1-\delta + CL^{-3}R^2 <1$ for large enough $L$ for all $x\in
B_R$.

Finally we have to prove  \eqref{eq:gradbound} for $r <R$. Since
$\varphi(0)$ is bounded, we have $m(0) =0$. {F}rom the equation of
$m$ inside $r \le R$, $|m''(r)| \le C \varrho$. By integrating $m'$
from $0$ to $r$ we obtain $m'(r) = m'(0) + O(\varrho r )$.
Integrating once more yields
$$
      m(r) = m'(0) r + O(\varrho r^2 )\; ,
$$
hence the bound $|\varphi'| = O(\varrho)$ follows. Differentiating
the equation  (\ref{A2}), we obtain $|m'''|\leq C\varrho$ for $r\in
[0, R]$ and integrating three times we obtain
\begin{equation*}
\begin{split}
    m''(r) &= m''(0)  + O(\varrho r ), \qquad
    m'(r) = m'(0) + m''(0) r + O(\varrho r^2), \\
    m(r) &= m'(0)r + \frac{m''(0) r^2}{2} + O(\varrho r^3) \; ,
\end{split}
\end{equation*}
and the bound on $\varphi''$ follows.

Next we prove the lower bound in (\ref{1}). Given any wave function
$\phi$ satisfying the Neumann boundary condition at $|x| =L$, we can
write it as $\phi (x) = g(x) \psi (x)$, where $\psi$ is given in
(\ref{eq:psik}), and $g>0$ satisfies Neumann boundary condition at
$|x| = L$ as well. {F}rom the identity $\fh \phi = (\fh \psi) g -
(\Delta g) \psi -2 \nabla g \nabla \psi$, we have
\begin{equation*}
\int_{|x|\leq L} \rd x \, \overline{\phi} \fh \phi = \;
\int_{|x|\leq L} \rd x |\nabla g|^2 \psi^2 + \int_{|x|\leq L} \rd x
\, |g|^2 \psi \fh \psi
\end{equation*}
and from (\ref{php})
$$
\int_{|x|\leq L} \rd x \, \overline{\phi} \fh \phi \ge \; k^2
\|\phi\|^2 + \int_{|x|\leq L} \rd x |\nabla g|^2 \psi^2
-Ck^3\int_{|x|\leq R} \rd x \, \frac{|g(x)|^2}{|x|^2}\; .
$$
Using (\ref{lowerb}) and that $f$ does not vanish, we have
$\psi(r)\ge ck$ with some positive constant $c$, thus
$$
\int_{|x|\leq L} \rd x \, \overline{\phi} \fh \phi \ge \; k^2
\|\phi\|^2 + ck^2 \int_{|x|\leq L} \rd x |\nabla g|^2
-Ck^3\int_{|x|\leq R} \rd x \, \frac{g^2(x)}{|x|^2}\; .
$$
Using Hardy's inequality (Lemma \ref{lm:hardy})
$$
    k^3\int_{|x|\leq R}
\frac{g^2(x)}{|x|^2}\; \rd x \leq k^3 \int_{|x|\leq L} \rd x \,
|\nabla g|^2 + \frac{k^3R}{L^3} \int_{|x|\leq L} \rd x \, |g|^2
$$
and by $\psi(r)\ge ck$ the last term can be controlled by
$$
   \frac{k^3R}{L^3} \int_{|x|\leq L} \rd x \, |g|^2 \leq \frac{k R}{c^2L^3}
   \int_{|x|\leq L} \rd x \, |g|^2 \psi^2
   \leq Ck^2L^{-3/2} \|\phi\|^2 \; .
$$
This implies the lower bound
$$
\int_{|y|\leq L} \rd y \, \overline{\phi} \fh \phi \ge \; k^2( 1 +
O(L^{-3/2})) \|\phi\|^2 \geq \frac{3a}{L^3} ( 1 + O(L^{-1})) \| \phi
\|^2
$$
in (\ref{1}).
\end{proof}

Next we apply last lemma to prove some important properties of the
function $w(y)$ and $q(y)$, defined in Section \ref{sec:2body}.

\begin{lemma}\label{lm:w&q}
Assume $V_a (x) = (a_0/a)^2 V( (a_0/a) x) = N^2 V(Nx)$ (because
$a=a_0/N$), where $V \geq 0$ is a smooth, spherical symmetric
potential with scattering length $a_0$ and with $\supp V \subset \{
x \in \bR^3 : |x| \leq R_0 \}$. Let $ \varrho = (8\pi)^{-1}\, (\|V
\|_1 + \| V \|_{\infty})$. Suppose that the functions $w(x)$ and
$q(x)$ are defined as in (\ref{eq:w}) and (\ref{eq:q}), with $a \ll
\ell_1 \ll 1$.
\begin{itemize}
\item[i)] There is a constant $c_0>0$ such that \be
   c_0 \leq 1- w(x) \leq 1
\ee for all $x \in \bR^3$. Moreover \be w(x) \leq C a \frac{\chi
(|x| \leq 3\ell_1 /2)}{|x|}, \ee for some constant $C$ independent
of $N$. \item[ii)] For $x \in \bR^3$ we have the following bounds on
the derivatives of $w(x)$:
\begin{equation}
\begin{split}
|\nabla w(x)| &\leq C a \frac{\chi (|x| \leq 3\ell_1/2)}{|x|^2 +
a^2}, \\
|\nabla^2 w (x)| &\leq C \varrho \, a \frac{\chi (|x| \leq
3\ell_1/2)}{|x|^3 + a^3} \leq C \varrho \frac{\chi (|x| \leq
3\ell_1/2)}{|x|^2} \, .
 \label{eq:derbound}
 \end{split}
\end{equation}
\item[iii)] We have $\supp q \subset \{ x \in \bR^3: |x|\leq 3
\ell_1 /2 \}$. Moreover \[ 0\leq q (x) \leq C a\ell_1^{-3} \chi (|x|
\leq 3 \ell_1 /2)\] and \[ | \nabla q (x)| \leq C
\frac{a}{|x|\ell_1^{3}} \chi (|x| \leq 3 \ell_1 /2) \] for all $x
\in \bR^3$. \item[iv)] The $L^1$ norm of $q(x)$ is given by
\[ \int_{\bR^3} q(x) \rd x = 4 \pi a(1+o(1)) \; , \qquad a\to 0 \; .\]
\end{itemize}
The constant $C$ in i),ii) and iii) is independent of $N$, if $N$ is
large enough.
\end{lemma}
\begin{remark} Using the functions $\lambda (x)$ and $\sigma (x)$
introduced in (\ref{eq:lambda-sigma}), this lemma in particular
proves all the bounds in (\ref{eq:wlambda}). The bound $|\nabla w|^2
\leq C a \sigma$ follows because $a (|x|^2 + a^2)^2 \leq C (|x|^3 +
a^3)^{-1}$, for all $x \in \bR^3$.
\end{remark}
\begin{proof}
Let $e_{\kappa}$ and $\psi_{\kappa} (x) = 1 -w_{\kappa} (x)$ be the
lowest Neumann eigenvalue and eigenfunction on the ball $\{ |x| \leq
\kappa \}$, that is
\begin{equation*}
(-\Delta + \frac{1}{2} V_a (x) ) \psi_{\kappa} (x) = e_{\kappa}
\psi_{\kappa} (x)
\end{equation*}
with the condition that $\psi_{\kappa} (x) =1$ if $|x| = \kappa$.
$\psi_{\kappa} (x)$ is then extended to be one, for $|x| \geq
\kappa$. We define $\phi_{\kappa} (x) := \psi_{\kappa} ((a/a_0) x)$.
Then we have
\begin{equation*}
(-\Delta + \frac{1}{2} V (y) ) \phi_{\kappa} (y) = (a/a_0)^2
e_{\kappa} \phi_{\kappa} (y)
\end{equation*}
for $|y| \leq (a_0 /a) \kappa =: L$, and with $\phi_{\kappa} (L)
=1$. By Lemma \ref{lm:E&g}, part i), we have
\begin{equation*}
(a/a_0)^2 e_{\kappa} = 3 a_0 L^{-3} (1 + o(1))
\end{equation*}
and thus
\begin{equation}\label{eq:ekappa}
e_{\kappa} = 3 a \kappa^{-3} (1 + o(1)).
\end{equation}
{F}rom Lemma \ref{lm:E&g}, part ii) we get immediately
\begin{equation}\label{eq:psikappa}
c_0 \leq \psi_{\kappa} (x) \leq 1 \quad \text{and } \quad w_{\kappa}
(x) \leq C a \frac{\chi (|x| \leq \kappa)}{|x|}
\end{equation}
where the constants $c_0$ and $C$ are independent of $\kappa$ and
$a$ (they depend only on the properties of the unscaled potential
$V(x)$). Moreover from Lemma \ref{lm:E&g}, part iii) we find
\begin{equation}\label{eq:wkappa}
|\nabla w_{\kappa} (x)| \leq C \varrho \, a \frac{\chi (|x| \leq
\kappa)}{|x|^2 + a^2} \quad \text{and } \quad |\nabla^2 w_{\kappa}
(x)| \leq C \varrho \, a \frac{\chi (|x| \leq \kappa)}{|x|^3 + a^3}
\end{equation}
where $C$ is independent of $\kappa$ and $a$.

{F}rom (\ref{eq:psikappa}), taking the average over $\kappa$ w.r.t.
the probability measure $\mu$, part i) follows trivially. As for
part ii), from (\ref{eq:wkappa}) we have
\begin{equation}\label{eq:nabla-w}
\begin{split}
|\nabla w (x)| &\leq \int |\nabla w_{\kappa} (x)| \mu (\rd \kappa)
\leq \frac{C\varrho\, a}{|x|^2+a^2} \int \chi ( |x| \leq \kappa) \mu
(\rd \kappa) \\ &\leq C \varrho \, a \frac{\chi (|x| \leq
3\ell_1/2)}{|x|^2 + a^2}
\end{split}
\end{equation}
because the measure $\mu$ is supported on $[\ell_1 /2 , 3\ell_1
/2]$. The bound for $|\nabla^2 w (x)|$ in (\ref{eq:derbound}) can be
proven analogously, using (\ref{eq:wkappa}).

In order to prove iii) and iv) recall that $q(x)$ was defined by
\[ q(x) = \frac{ \int e_{\kappa} \chi (|x| \leq \kappa)
\psi_{\kappa} (x) \mu (\rd \kappa)}{\int \psi_{\kappa} (x) \mu (\rd
\kappa)} = \frac{ \int e_{\kappa} \chi (|x| \leq \kappa)
\psi_{\kappa} (x) \mu (\rd \kappa)}{1-w(x)}\, .\] {F}rom
(\ref{eq:ekappa}) and since the measure $\mu$ is supported on
$[\ell_1 /2 , 3\ell_1/2]$, we get $q(x) \leq C a\ell_1^{-3}$. The
gradient of $q$ can be estimated by
\begin{equation}\label{eq:nabla-q}
\begin{split}
|\nabla q (x)| \leq  \; &\frac{\int e_{\kappa} \kappa^{-1} \delta
(|x| -\kappa) \mu (\rd \kappa)}{1 - w(x)} + \frac{ \int e_{\kappa}
\chi (|x| \leq \kappa) |\nabla_x w_{\kappa} (x)| \mu (\rd \kappa)}{1
-w(x)} \\ &+ \frac{|\nabla w(x)|}{1-w(x)} q(x) \, .
\end{split}
\end{equation} The first term on the r.h.s. of the last
equation (where we already used the condition $\psi_{\kappa} (x) =1$
if $|x| = \kappa$) is different from zero only if $|x| \in [\ell_1
/2 , 3 \ell_1 /2]$. Since the delta function forces $\kappa =|x|$ we
find, because of (\ref{eq:ekappa}) and because $1 -w(x) \geq c_0$,
\begin{equation*}
\frac{\int e_{\kappa} \delta (|x| -\kappa) \mu (\rd \kappa)}{1 -
w(x)} \leq C \frac{a}{\ell_1^3}\,  \chi (|x| \leq 3\ell_1 /2) \, .
\end{equation*}
The second term on the r.h.s. of (\ref{eq:nabla-q}) can be
controlled using (\ref{eq:ekappa}) and (\ref{eq:wkappa}). The third
term on the r.h.s. of (\ref{eq:nabla-q}) can be estimated using the
bounds (\ref{eq:nabla-w}) and $q(x) \leq C a \ell_1^{-3}$. This
completes the proof of iii).

Finally, to prove the estimate for the $L^1$ norm of $q (x)$, we
note that
\begin{equation}\label{eq:L1normofq}
\begin{split}
\int \rd x \, q(x) &= \int \rd x \, \frac{\int e_{\kappa} \chi (|x|
\leq \kappa) \psi_{\kappa} (x) \mu (\rd \kappa)}{\int \psi_{\kappa}
(x) \mu (\rd \kappa)} = \int \rd x \, \frac{\int e_{\kappa} \chi
(|x| \leq \kappa) \mu (\rd \kappa)}{\int \psi_{\kappa} (x) \mu (\rd
\kappa)} + a \, o(1)
\end{split}
\end{equation}
for $a \to 0$, because $\psi_{\kappa} (x) = 1 - w_{\kappa} (x)$ and
\begin{equation*}
\begin{split}
\Big|\int \rd x \int e_{\kappa} \chi (|x| \leq \kappa ) w_{\kappa}
(x) \mu (\rd \kappa) \Big| &\leq C a \int e_{\kappa} \left( \int \rd
x \frac{\chi (|x|\leq \kappa)}{|x|} \right) \mu (\rd \kappa)
\\ &\leq C a^2 \int \kappa^{-1} \mu (\rd \kappa) \leq C a^2
\ell_1^{-1} = a O(a / \ell_1) \, .
\end{split}
\end{equation*}
Analogously we can estimate the contribution of $w_{\kappa} (x)$ in
the denominator on the r.h.s. of (\ref{eq:L1normofq}). We get
\begin{equation*}
\begin{split}
\int \rd x \, q(x) &=  \int \rd x \, \int e_{\kappa} \chi (|x| \leq
\kappa) \mu (\rd \kappa) + a \, o(1) = \int e_{\kappa} \left( \int
\rd x \, \chi (|x| \leq \kappa) \right) \mu (\rd \kappa) + a \, o(1)
\\ &= \frac{4\pi}{3} \int e_{\kappa} \kappa^3 \mu (\rd \kappa) + a
\, o(1)
\end{split}
\end{equation*}
and thus, with (\ref{eq:ekappa}),
\begin{equation*}
\int \rd x \, q(x) = 4 \pi a (1 + o(1))
\end{equation*}
for $a \to 0$, which proves iv).
\end{proof}

\section{Properties of the Triple Cutoff Function $F_{ij}$}
\label{app:cutoff} \setcounter{equation}{0}

In this appendix we collect some properties of the function
$F_{ij}$, defined in Section \ref{sec:rem}.

\begin{lemma}[No overlap]\label{lemma:noover}
For any exponent $q>0$ and any fixed $i$ and $j\neq j'$ \be
   \| \wt\chi_{ij}\wt\chi_{ij'}F_{ij}^q\|_\infty \leq e^{-cq/\ell^\e} \; .
\label{eq:overlap} \ee
 Moreover, for any  fixed $i$,
 \be \sum_j \wt\chi_{ij}
F_{ij}^q \le c_q \; .\label{eq:overlapsum} \ee
\end{lemma}
\begin{proof}
Clearly $|F|\leq 1$. Suppose there is an overlap between two
functions $\wt\chi_{ij}$, $\wt\chi_{ij'}$, i.e. for some $j\neq j'$
we have $\wt\chi_{ij}\neq 0$ and $\wt \chi_{ij'}\neq0$. Then
$|x_i-x_j|, |x_i-x_{j'}|\leq \ell$, so
$$
    F_{ij} \leq e^{-\ell^{-\e} h(x_i-x_{j'})}\leq
    e^{-c\ell^{-\e}}
$$
so any overlap forces $F_{ij}$ to be exponentially small. This
proves (\ref{eq:overlap}) and also (\ref{eq:overlapsum}).
\end{proof}

\bigskip

We have analogous controls on the derivatives of $F_{ij}$ as well.
We define the sum norm of a vector $\bx =(x_1, \ldots x_N)\in
\bR^{3N}$ as
$$
   \thi \bx \thi := \sum_{k=1}^N |x_k|
$$
where $|x|$ is the Euclidean length in $\bR^3$. Similarly, if $\bA =
(A_1, \ldots , A_N)$, where $A_j$ is an $n$-tensor on $\bR^3$ for
each $j$, then
$$
 \thi \bA \thi : = \sum_{k=1}^N | A_k|_{(\bR^3)^{\otimes n}} \; ,
$$
where $| A_k| =| A_k|_{(\bR^3)^{\otimes n}}$ denotes the tensor norm
on $(\bR^3)^{\otimes n}$ derived from the Euclidean norm. In
particular, if $\bnabla =(\nabla_1, \ldots , \nabla_N)$ denotes the
$3N$ dimensional derivative, and $\a \in \bN$, we have
$$
    \thi \bnabla^\alpha f \thi: = \sum_{k_1}\sum_{k_2}\ldots \sum_{k_\alpha}
     |\nabla_{k_1}\ldots\nabla_{k_\a} f|_{(\bR^3)^{\otimes \a}}
$$
where
$$
|\nabla_{k_1}\ldots\nabla_{k_\a} f|_{(\bR^3)^{\otimes \a}} = \sup
\Big\{ \Big | (\nabla_{k_1} \ldots \nabla_{k_{\a}} f )(v_1\otimes
v_2 \otimes \ldots \otimes v_\alpha)\Big| \, : \, v_j\in \bR^3\; ,
\;|v_j| =1\Big\}
$$
with
$$
    (\nabla_{k_1} \ldots \nabla_{k_{\a}} f )(v_1\otimes v_2 \otimes \ldots \otimes v_\alpha)
    = \sum_{\beta_1, \ldots \beta_\a=1}^3 \Big( \prod_{j=1}^\alpha v_{j\beta_j}\Big)
    \Big( \prod_{j=1}^\alpha \frac{\partial }{\partial x_{j\beta_j}}\Big)f
$$
with $v_j= (v_{j1}, v_{j2}, v_{j3})$.

\begin{lemma}[Control of the derivatives of $F_{ij}$] \label{lm:nablaF}
Let $\ell\ll 1$, i.e. $\ell\leq N^{-\kappa}$ for some $\kappa>0$.
For sufficiently large $N$ the following pointwise bounds hold:

\begin{itemize}
\item[i)] Let $\a\in\bN$ and $q>0$. Then for any fixed $i,j$ \be
   \thi \bnabla^\a F_{ij}^q \thi \leq c_\a\ell^{-\a} F_{ij}^{q/2} \; .
\label{eq:Fest} \ee \item[ii)] Let $K>0$ and recall the definition
of $\theta_{kj}$ from (\ref{def:theta}). For any  $\alpha\in \bN$
and for $k\neq i,j$, and arbitrary $m$
\begin{equation}
|\nabla_k^\alpha F_{ij}^q| \leq c \ell^{-\a} ( \theta_{ik}
+\theta_{jk}) F_{ij}^{q/2} + O (\ell^{K-\a-\e}), \label{eq:koverlap}
\end{equation}
\begin{equation}
|\nabla_k\nabla_m F_{ij}^q| \leq c \ell^{-2} ( \theta_{ik}
+\theta_{jk})
 F_{ij}^{q/2} + O
(\ell^{K-2-\e}) \; .\label{eq:koverlap1}
\end{equation}
\item[iii)] For any fixed $\alpha\in \bN$ and index $k$ \be
\sum_{ij} |\nabla_k^\alpha F_{ij}^q|\wt{\chi}_{ij} \leq c_{q,\a}
\ell^{-\alpha} \; .\label{eq:kfix} \ee \item[iv)] For any fixed
$\alpha,\beta \in \bN$ and index $k$ \be \sum_{ijm} |\nabla_k^\alpha
\nabla_m^\beta F_{ij}^q| \wt{\chi}_{ij} \leq c_{q,\a,\beta}
 \ell^{-\alpha-\beta} \; .
\label{eq:kabfix} \ee
\end{itemize}
\end{lemma}

\begin{proof} For i) note that
$$
      |\nabla^\a h(x)| \leq c_\alpha\ell^{-\alpha} h(x)
$$
for $\a\in \bN$. Then
\begin{equation*}
     \thi \bnabla F_{ij}^q\thi=
     |\nabla_i F_{ij}^q| + |\nabla_j F_{ij}^q|+
     \sum_{k\neq i,j} |\nabla_k F_{ij}^q|
    \leq c q \ell^{-1}
    \ell^{-\e}\sum_{k\neq i,j} [ h_{ki}+h_{kj}] F_{ij}^q
    \leq c\ell^{-1} F^{q/2}_{ij}
\end{equation*}
using that $ze^{-z}\leq ce^{-z/2}$ with $z= c q N\ell^{-\e}$. For
higher derivatives the proof is similar:
$$
     \thi \bnabla^\a F_{ij}^q\thi
    \leq c\ell^{-\a}
    \Big[\ell^{-\e}\sum_{k\neq i,j} [ h_{ki}+h_{kj}] \Big]^\a F_{ij}^q
    \leq c_\a\ell^{-\a} F^{q/2}_{ij}
$$
using that $z^\a e^{-z}\leq c_\a e^{-z/2}$ with $z=\cN\ell^{-\e}$.

For the proof of ii) we note that  for $k\neq i,j$ the derivative
$\nabla_k F_{ij}$ is smaller than $\ell^{K-\e}$ unless $|x_i-x_k|$
or $|x_j-x_k|$  is smaller than $K\ell |\log\ell|$, therefore
\begin{equation*}
\begin{split}
|\nabla_k F_{ij}^q|&\leq c\ell^{-1-\e}\Big[ \theta_{ki} +
\theta_{kj} \Big] \, \left(h_{kj} + h_{ki}\right) F_{ij}^q
+\ell^{K-1-\e} \\ &\leq c \ell^{-1} \Big[ \theta_{ki} + \theta_{kj}
\Big] F_{ij}^{q/2} + \ell^{K-1-\e} \qquad k\neq i,j \; .
\end{split}
\end{equation*}
The proof for higher derivatives is similar.

As for iii), we use the previous estimate when $i, j\neq k$
$$
\sum_{ i , j \neq k} |\nabla_k^\a F_{ij}^{q} | \wt{\chi}_{ij} \leq
c\ell^{-\a}\sum_{ i , j \neq k} \theta_{kj} F_{ij}^{q/2}
\wt{\chi}_{ij} + O(N^2\ell^{K-\alpha-\e}) \; .
$$
 We consider the set $S_k=\{ j \; : \; |x_j-x_k|\leq
K\ell|\log\ell|\}$. If $j,j'\in S_k$, and $|x_j-x_{j'}|\leq
\frac{\e}{2}\ell|\log\ell|$, then $F_{ij}\leq \exp(-\ell^{-\e/2})$.
Therefore, apart from exponentially small error, the cardinality of
$S_k$ is $(2K/\e)^3$, since if there are more $j$'s in the set
$S_k$, then $|x_j - x_{j'}| \leq (\e/2) \ell | \log \ell|$, at least
for two indices $j,j' \in S_k$. The same argument holds for the $i$
indices, showing that the $i$ summation is only over a finite set.
This implies that
\begin{equation*}
\sum_{ i , j \neq k} \theta_{kj} F_{ij}^{q/2}  \wt{\chi}_{ij} \leq
\sum_{j \in S_k, j\neq k}\;  \sum_{i\in S_k  i \neq k} F_{ij}^{q/2}
\wt{\chi}_{ij} \leq c_q (K/\e)^6 \;.
\end{equation*}
Choosing $K$ sufficiently large we obtain the bound (\ref{eq:kfix})
for $i,j\neq k$. For the terms with $k=i$ or $k=j$ we use \be
   |\nabla_k^\a F_{kj}^q|\leq |\bnabla^\a F_{kj}^q|\leq c\ell^{-\a}
   F_{kj}^{q/2}
\label{kkj} \ee and the $j$ summation is finite by
(\ref{eq:overlapsum}).

Finally, the proof of iv) is similar. If $k=i$,  then we have
$$
\sum_{jm} |\nabla_k^\alpha \nabla_m^\beta F_{kj}^q| \wt{\chi}_{kj}
\leq c_{\a,\beta}\ell^{-\a-\beta} \sum_j  F_{kj}^{q/2}
\wt{\chi}_{kj} =  c_{q,\a,\beta}\ell^{-\a-\beta}
$$
using (\ref{eq:Fest}) and (\ref{eq:overlapsum}) and the same
estimate holds if $k=j$. If $i,j\neq k$, then again the cardinality
of the index set $S_k$ for $i$ and $j$ is bounded, apart from an
exponentially small error. Therefore
$$
\sum_{ijm \atop i,j\neq k} |\nabla_k^\alpha \nabla_m^\beta F_{ij}^q|
\wt{\chi}_{ij}
  \leq \sum_{i, j\in S_k\atop i,j\neq k}
  \sum_m |\nabla_k^\alpha \nabla_m^\beta F_{ij}^q|
  \leq c_{\a,\beta}\ell^{-\a-\beta} (K/\e)^6  + O(N^3\ell^{K-\a-\beta-2\e})
$$
using (\ref{eq:Fest}).
\end{proof}

\section{Local Structure after Particle Removal}
\label{app:remove} \setcounter{equation}{0}

\begin{lemma}\label{lm:WkW}
There exists a constant $C_0>0$ such that for any $k$ \be
        C_0^{-1} \leq \frac{W^{(k)}}{W} \leq C_0
\label{eq:Wk} \ee pointwise for sufficiently large $N$. Moreover,
for any $\a\in \bN$ and sufficiently large $N$
 we have the pointwise estimate
\begin{equation}\label{eq:WkW2}
C_0^{-\a} \leq \frac{W^{(k_1 \dots  k_\a)}}{W} \leq C_0^\a
\end{equation}
 uniformly for any family of indices $k_1, \dots,
k_\a$.
\end{lemma}

{\it Proof.} Similarly to the proof of  (\ref{eq:Gsep}), we see that
for any fixed $k$ and sufficiently big $N\ge N(k)$, all
 $G_i^{(k)}$ are  separated away from zero uniformly in $i$.
Therefore it is sufficient to show that \be
    \sum_{i\neq k} |G_i^{(k)}- G_i|
   \leq  \sum_{i\neq k} \Big[ w_{ik} F_{ik} +\sum_{ m\neq i,k} w_{im}|F^{(k)}_{im}
    -F_{im}|\Big]
\label{Fk} \ee is uniformly bounded for any fixed $k$. The
boundedness of the first term follows from (\ref{eq:overlapsum}). As
for the second term, note that
$$
|F^{(k)}_{im}   -F_{im}| \leq \Big| 1 -
e^{-\ell^{-\e}[h_{ik}+h_{mk}]}\Big| F^{(k)}_{im} \leq
\ell^{-\e}[h_{ik}+h_{mk}] F^{(k)}_{im} \; ,
$$
so
$$
 \sum_{i\neq k} \sum_{ m\neq i,k} w_{im}|F^{(k)}_{im}
    -F_{im}| \leq 2\ell^{-\e} \sum_{i,m\neq k} \chi_{im} h_{ik}F^{(k)}_{im} \;
$$
since $w_{im}\leq \chi_{im}$ by the support of $w$. If $|x_i -
x_k|\ge K\ell|\log\ell|$ for some $i$, then
 $ h_{ik} \leq \ell^{K}$ and and if $K$ is a
large constant, then $\ell^{K-\e}N^2\to 0$, so this term is
negligible even after the summation. Now we look at the set
$$
    S_k: =\{ i  \; : \; i\neq k,  |x_k-x_i|\le K\ell|\log\ell| \} \, .
$$
If $i, i'\in S_k$, $i\neq i'$ and $|x_i-x_{i'}|\leq \frac{\e}{2}
\ell |\log\ell|$ then $F^{(k)}_{im} \leq \exp( -\ell^{-\e} h_{ii'})
\leq \exp (-\ell^{-\e/2})$ that is exponentially small. Therefore,
modulo exponentially small errors, $|S_k|\leq (K/(\e/2))^3$, which
guarantees that the summation over $i$ in (\ref{Fk})  is finite. For
each fixed $i$ the summation over $m$ is finite by
(\ref{eq:overlapsum}). The second bound (\ref{eq:WkW2}) follows
easily by induction. $\;\;\Box$

The same proof immediately gives the following two bounds that are
used in the proof. Here $k$ is fixed and the constants may depend on
$k$. \be
      |F_{im}^{[k]} - F_{im}| \leq C\ell^{-\e} \sum_{r\leq k} (\theta_{ir}+\theta_{mr}) F_{im}^{[k]}
      + O(\ell^{K-\e})
\label{eq:FFk} \ee and \be
     \sum_j |\nabla_j^\alpha(F_{im}^{[k]} - F_{im})|
     \leq C\ell^{-\alpha-\e} \sum_{r\leq k} (\theta_{ir}+\theta_{mr})
    (F_{im}^{[k]})^{1/2}
      + O(\ell^{K-\alpha-\e}) \; .
\label{eq:FFkder} \ee We will often multiply these inequalities by
$w_{im}$, then on the support of $w_{im}$ we can use that
$\theta_{ir}$ and $\theta_{mr}$ are comparable. In particular, we
also obtain for each $m$ \be
      |G_m^{[k]}- G_m|\leq C\ell^{-\e}\sum_{j>k}\sum_{r\leq k} w_{mj} F^{[k]}_{mj}\theta_{mr}
      + O(\ell^{K-\e}) \; .
\label{eq:GGk} \ee

\section{Upper bound for $\wt H^2$}
\setcounter{equation}{0}

The aim of this appendix is to prove that the assumptions of Theorem
\ref{thm:main} about the energy distribution of the initial data
$\psi_{N,0} = W \phi_{N,0}$ are satisfied for a large class of
$\phi_{N,0}$. In particular, in the next lemma we prove that
$(W\phi_{N,0}, \wt H^2 W\phi_{N,0}) \leq C N^2$ is satisfied, if the
function $\phi_{N,0}$ is sufficiently smooth. The proof of the
inequality $(W \phi_{N,0}, \wt H W \phi_{N,0}) \leq C N$ for
sufficiently smooth $\phi_{N,0}$ is similar (but much easier) and
therefore omitted.

\begin{lemma}\label{lm:h2bound}
Assume $a \ll \ell_1 \ll \ell \ll 1$, $a \ell_1 \ll \ell^4$ and
$\ell_1^3\ll a\ell^{5/2}$. Suppose moreover that $\phi$ satisfies
\begin{equation}\label{eq:assphi}
\sum_{i,j,k,m} \int \overline{\phi} (1 -
\Delta_i)(1-\Delta_j)(1-\Delta_k)(1-\Delta_m) \phi \leq D N^4
\end{equation}
for some $D >0$. Then there is a constant $C >0$ such that
\begin{equation*}
(W \phi, \wt H^2 \, W\phi) \leq C N^2\; .
\end{equation*}
\end{lemma}
\begin{proof}
Following the steps from (\ref{eq:HWphi}) to (\ref{eq:HWphi2}) (but
this time without neglecting the positive contributions), we find
(using the summation convention)
\begin{equation*}
\begin{split}
(W \phi, \wt H^2 \, W\phi) = \; &\int |\wt H W \phi|^2 \\ \leq \;
&\int \, W^2 |\nabla_i \nabla_j \phi|^2 + C \int \, W^2 \left(
|\nabla_i \nabla_j G_{\ell}| + |\nabla_i G_{\ell}| |\nabla_j
G_{\ell}|\right) \, |\nabla_i \phi| |\nabla_j \phi|  \\ &+ 2 \int \,
W^2 |B| |\nabla_m \phi|^2 + 2  \int \, W^2 |\nabla_m B| |\phi|
|\nabla_m \phi| + \int W^2 B^2 |\phi|^2 ,
\end{split}
\end{equation*}
with $B =q_{kj} + \Omega = q_{kj} + \wt \Omega + O \left( e^{-C
\ell^{-\e}} \right)$. Using Lemmas \ref{lm:error1} -
\ref{lm:error4}, we get
\begin{equation*}
\begin{split}
\int |\wt H W \phi|^2 \leq \; &C \int W^2 \left( |\nabla_i \nabla_j
\phi|^2 + N |\nabla_i \phi|^2 + N^2 |\phi|^2 \right) + C
\int W^2 q_{ij} |\nabla_m \phi|^2 \\
&+ C \int \, W^2 |\nabla_i G_{\ell}| |\nabla_j G_{\ell}| |\nabla_i
\phi| |\nabla_j \phi| + \int W^2 B^2 |\phi|^2 \, .
\end{split}
\end{equation*}
Since $W \leq 1$, and by (\ref{eq:assphi}), we find
\begin{equation}\label{eq:D1}
\begin{split}
\int |\wt H W \phi|^2 \leq \; &C N^2 + C \int |\nabla_i G_{\ell}|
|\nabla_j G_{\ell}| |\nabla_i \phi|^2 + C \int q_{ij} |\nabla_m
\phi|^2 \\ &+ C \int q_{ij} q_{km} |\phi|^2 + C \int \wt \Omega^2
|\phi|^2\; .
\end{split}
\end{equation}
We start by considering the second term on the r.h.s. of the last
equation. Using Lemma \ref{lemma:noover}, we find
\begin{equation*}
\begin{split}
\int |\nabla_i G_{\ell}| |\nabla_j G_{\ell}| |\nabla_i \phi|^2 \leq
\; &C \int \Bigg( |(\nabla w)_{mi}|^2 F_{mi} + |(\nabla w)_{im}|
w_{im} F_{im} |\nabla_j F_{im}| \\ &+ |(\nabla w)_{jm}| w_{jm}
F_{jm} |\nabla_i F_{jm}| + w_{jm}^2 |\nabla_i F_{jm}| |\nabla_n
F_{jm}| \Bigg) |\nabla_i \phi|^2 \\ &+ O(e^{-c \ell^{-\e}})\; .
\end{split}
\end{equation*}
Using $|\nabla w|^2 \leq C\lambda$, $|\nabla w|\leq Ca\lambda$,
Lemma \ref{lm:nablaF}, Lemma \ref{lemma:combined} (with $W\equiv 1$)
and (\ref{eq:wlambda}) we find
\begin{equation*}
\begin{split}
\int |\nabla_i G_{\ell}| |\nabla_j G_{\ell}| |\nabla_i \phi|^2 \leq
\; &C \int \lambda_{mi} F_{mi}^{1/2} |\nabla_i \phi|^2 + C
a\ell^{-1} \int \lambda_{mj} F_{mj} |\nabla_i \phi|^2 \\ &+
C\ell^{-2} \int |\nabla_i \phi|^2 + O(e^{-c \ell^{-\e}})
 \\ \leq \; &C \int \wt
\chi_{mi} F_{mi}^{1/2} |\nabla_m \nabla_i \phi|^2 + \ell_1 \ell^{-3}
\int  \wt \chi_{mi} F_{mi}^{1/2} |\nabla_i \phi|^2 \\ &+ C a
\ell^{-1}\int \wt \chi_{mj} F_{mj}^{1/2} |\nabla_j \nabla_i \phi|^2
\\ &+ C a \ell_1 \ell^{-4} \int \wt \chi_{mj} F_{mj}^{1/2} |\nabla_i
\phi|^2+ O(e^{-c \ell^{-\e}})
\\ \leq \; &C(1+a\ell^{-1}) \int |\nabla_m \nabla_i \phi|^2 +
\ell_1 \ell^{-4} \int |\nabla_i \phi|^2 \\ &+ O(e^{-c
\ell^{-\e}})\;,
\end{split}
\end{equation*}
which is bounded by $CN^2$, because $a \ell_1 \ell^{-4}\ll 1$, and
because of \eqref{eq:assphi}.

Next we consider the third term in (\ref{eq:D1}); we obtain
\begin{equation*}
\begin{split}
\int q_{ij} |\nabla_m \phi|^2 &\leq C a \ell_1^{-3} \int \chi_{ij}
|\nabla_m \phi|^2 \\ &\leq C a \int \left( |\nabla_i \nabla_j
\nabla_m \phi|^2 + N |\nabla_i \nabla_m \phi|^2 + N^2 |\nabla_m
\phi|^2 \right) \\ &\leq C N^2
\end{split}
\end{equation*}
using (\ref{eq:2delta}) and  (\ref{eq:assphi}). As for the fourth
term on the r.h.s. of (\ref{eq:D1}), we have
\begin{equation*}
\begin{split}
\int q_{ij} q_{km} |\phi|^2 \leq \; &C a^2 \ell_1^{-6} \Big (
\sum_{(i,j) \neq (k,m)} \int \chi_{ij} \chi_{km} |\phi|^2 +
\sum_{i,j} \int \chi_{ij} |\phi|^2 \Big) \\ \leq \; &C a^2
\sum_{(i,j) \neq (k,m)} \int \overline{\phi} (1 -\Delta_i) (1
-\Delta_j) (1- \Delta_k) (1 -\Delta_m) \phi \\ &+ C a^2
\ell_1^{-3} \int \overline{\phi} (1-\Delta_i)(1- \Delta_j) \phi \\
\leq \; &C a^2 N^4 + a^2 \ell_1^{-3} N^2 \leq C N^2
\end{split}
\end{equation*}
because of $a \ll \ell_1^{3/2}$ and of (\ref{eq:assphi}).

Finally we consider the last term on the r.h.s. of (\ref{eq:D1}). We
use that, by (\ref{def:Omtilde}), $\wt \Omega^2 \leq C \Omega_{kj}
\Omega_{im} + C \Gamma^2$. By Lemma \ref{lm:nablaF} and by the
estimate $w\leq a\ell\lambda$, $|\nabla w|\leq a\lambda$, we obtain
\begin{equation}\label{eq:D4}
\begin{split}
\int \Omega_{kj} \Omega_{im} |\phi|^2 \leq \; &C \int \Bigg( w_{sj}
w_{rm} |\Delta_k F_{sj}||\Delta_i F_{rm}| \\ &+ |(\nabla w)_{kj}|
|(\nabla w)_{im}| |\nabla_k F_{kj}| |\nabla_i F_{im}| \Bigg)
|\phi|^2
\\ \leq \; & C a^2\ell^{-2} \int \lambda_{kj}\lambda_{im}
F_{kj}^{1/2} F_{im}^{1/2} |\phi|^2 \; .
\end{split}
\end{equation}
We note that, if, for example $k=i$, then $j$ is forced to be equal
to $m$, up to exponentially small errors, due to the strong
non-overlapping properties of the functions $F_{kj}$. Hence, up to
errors which are exponentially small in $N$ and using Lemma
\ref{lemma:combined},  we  can estimate the second term on the r.h.s
of (\ref{eq:D4}) by
\begin{equation*}
\begin{split}
\int \Omega_{kj} \Omega_{im} |\phi|^2 \leq \;  & C a^2\ell^{-2}
\sum_{kj} \int \lambda_{kj}^2 F_{kj} |\phi|^2 + C a^2\ell^{-2}
\sum_{(k,j) \neq (i,m)} \int \lambda_{kj}\lambda_{im} F_{kj}^{1/2}
F_{im}^{1/2} |\phi|^2 \\ \leq \; & C \ell^{-2} a \int \sigma_{kj}
|\phi|^2 + C \ell^{-2} a^2 \sum_{(k,j) \neq (i,m)} \int \lambda_{kj}
\lambda_{im} F_{kj}^{1/2} F_{im}^{1/2} |\phi|^2
\\ \leq \; &C a \ell^{-2} (\log N )\int \overline{\phi} ( 1
-\Delta_k) (1-\Delta_j) \phi \\ &+ C a^2 \ell^{-2} \sum_{(k,j) \neq
(i,m)} \left( \int \lambda_{im} F_{im}^{1/2} |\nabla_j \phi|^2 +
\ell_1 \ell^{-3} N \int \lambda_{im} F_{im}^{1/2} |\phi|^2 \right)
\\ \leq &\; o(N^2) + C a^2 \ell^{-2} \int |\nabla_i \nabla_j \phi|^2
+ C a^2 \ell^{-5} \ell_1 N \int |\nabla_j \phi|^2 \\ &+ Ca^2
\ell^{-8} \ell_1^2 N^2 \int |\phi|^2 \; ,
\end{split}
\end{equation*}
which is $o(N^2)$ because $a\ell_1 \ll \ell^4$. We still have to
control the contribution from $\Gamma^2$. Using Lemma
\ref{lm:nablaF} we have
\begin{equation}\label{eq:D6}
\begin{split}
\int \Gamma^2 |\phi|^2 \leq  \; &C \int \Bigg( |(\nabla w)_{ik}|
|(\nabla w)_{mr}| w_{jp} w_{ns} F_{ik} F_{mr} |\nabla_k F_{jp}|
|\nabla_r F_{ns}| \\
& + w_{ik} w_{mr} w_{jp} w_{ns} |\nabla_q F_{ik}| |\nabla_q F_{mr}|
|\nabla_d F_{jp}| |\nabla_d
F_{ns}| \Bigg) |\phi|^2 \\
\leq \; &C \int \left( \ell^{-2} |(\nabla w)_{ik}| |(\nabla w)_{mr}|
F_{ik} F_{mr} + \ell^{-4}
w_{ik} w_{mr} F_{ik} F_{mr}\right) |\phi|^2 \\
\leq \; &Ca^2 \ell^{-2}\int\lambda_{ik}\lambda_{mr} F_{ik} F_{mr}
|\phi|^2
\end{split}
\end{equation}
which can be bounded by $CN^2$ in the same way we bounded the r.h.s.
of (\ref{eq:D4}). This completes the proof of the lemma.
\end{proof}

{\bf Acknowledgements}:
We would like to thank J. Yngvason for communicating this problem
to us and for helpful discussions.

\end{document}